中山大学博士学位论文

# 高能物理和量子信息视角的强关联电子系统有效场论和数值研究

**Effective field theory and numerical study of strongly correlated electron systems from the perspective of high energy physics and quantum information**

专业：　　理论物理

研究方向：　强关联电子理论

博士生：　　马嘉政

指导教师：　姚道新教授

答辩委员会　（签名）

主席:

委员:

中山大学物理学院●中国广州

2024 年 11 月 8 日

## 论文原创性声明

本人郑重声明：所呈交的学位论文，是本人在导师的指导下，独立进行研究工作所取得的成果。除文中已经注明引用的内容外，本论文不包含任何其他个人或集体已经发表或撰写过的作品成果。对本文的研究作出重要贡献的个人和集体，均已在文中以明确方式标明。本人完全意识到本声明的法律结果由本人承担。

学位论文作者签名：

日期：　　年　月　日

## 学位论文使用授权声明

本人完全了解中山大学有关保留、使用学位论文的规定，即：学校有权保留学位论文并向国家主管部门或其指定机构送交论文的电子版和纸质版，有权将学位论文用于非赢利目的的少量复制并允许论文进入学校图书馆、院系资料室被查阅，有权将学位论文的内容编入有关数据库进行检索，可以采用复印、缩印或其他方法保存学位论文。

保密论文保密期满后，适用本声明。

学位论文作者签名：　　　　　　导师签名：

日期：　　年　月　日　　　　日期：　　年　月　日

# 高能物理和量子信息视角的强关联电子系统有效场论和数值研究

专业：理论物理

博士生：马嘉政

指导教师：姚道新教授

## 摘要

凝聚态物理是一个传统但持续富有活力的领域，而强关联电子系统的性质研究是其中极具挑战性的分支。通常强关联系统简化抽象而来的有效格点模型看起来很简单，如高温超导的哈伯德模型，磁性杂质巡游电子相互作用的安德森模型等，但是通常它们除了极少数特殊参数点和低维情况，一般是不能严格求解的。因此学者们开发了许多有效理论和数值方法研究它们，并得到了非常丰富的相图。近年来拓扑相，拓扑序，摩尔材料，新型高温超导关联相等新奇物相的发现又给凝聚态有效理论和数值方法的研究带来了新的挑战和机遇。拓扑相以及拓扑相变不能用通常的朗道范式的局域序参量描述，而是由一些全局的性质所决定，例如陈数，纠缠谱等。在类似转角电子学系统以及摩尔超晶格中，因为超晶格常数在小转角达到数个纳米的量级，单胞中的原子数通常达到了 $10^4$ 量级，使得实空间的精确对角化以及第一性原理计算变得十分困难。因此迫切需要开发新的理论，数值方法。

俗话说他山之石，可以攻玉。虽然凝聚态物理,高能物理以及量子信息是研究方法看起来十分不同的领域，但实际它们之间有不少共通之处。本文将考虑高能物理以及量子信息在凝聚态物理问题中的应用，其中第二，三章的工作主要是高能物理在凝聚态演生论观点中的启示，第四章主要是高能物理在还原论中的启示。

在第一个工作中，我们基于几何理论从实空间出发提出适用于一般转角的扭转双层石墨烯有效模型。在传统的扭转双层石墨烯理论中，麦当劳模型在小转角下做单能谷近似，并把层间耦合处理为微扰，摩尔条纹对应的布里渊区折叠为更小的所谓摩尔布里渊区(mBZ)，层间耦合导致同一个能谷两层的狄拉克锥(小能谷)发生隧

穿，能很好地解释小转角处的电子局域化，以及所谓一系列魔角处出现的近平带。转角双层石墨烯的奇异物理几乎都源于平带不同填充下的关联行为。但是正如之前所说麦当劳连续模型只适用于小转角，中等转角在加上能谷间耦合或者公度近似也能较好处理，但对于接近$30^0$左右的TBG，此时系统没有平移对称性，不能统一到麦当劳模型的理论框架。相应地，在$30^0$转角TBG中具有特殊的12重旋转对称性，可以构造相应的准晶模型。如何对一般转角提出统一的理论仍然鲜有研究。我们从实空间出发，将层间刚性扭转视为一种形变，进而这种形变会诱导出等效的弯曲时空，因此我们猜想一般转角下的理论可以看成狄拉克费米子在这个等效弯曲时空下做出的响应，即时空标架作为序参量与狄拉克费米子耦合。这个理论模型在高能物理领域和广义相对论中已经被广泛研究。基于这一理论我们得出了第一魔角$\theta = 1.05^0$处的最低近平带，和麦当劳模型给出的带宽相近。我们模型不仅适用于小转角和公度的情况，也适用于非公度情况例如$30^0$转角的TBG准晶，基于这个理论的$30^0$准晶的准能带也和纳米角分辨光电子能谱(nanoARPES)实验结果定性符合。更重要的是我们的理论能预言一类新的体系的性质，即转动双层石墨烯。这种体系是之前理论无法解释和预言的，因为之前的理论并没有统一的一般转角理论，而转动双层石墨烯也不同于一般的弗洛凯周期光场驱动系统，是一种结构上的周期驱动。基于我们的理论框架，可以预言对于给定有限尺寸区域在双层相对旋转过程中一个周期内的电荷泵浦，它和博特指数联系在一起，是量子化的。同样电荷泵浦会依赖于驱动频率，在频域发生拓扑相变，转动双层石墨烯的这一现象可能用特定输运实验予以验证。相关内容已发表于Physical Review B，详见博士期间发表论文。

在第二个工作中，我们模拟了一个相互作用扭转双层棋盘格(TBCB)模型。基于类BM模型，我们得出了具有二次型能量交点的最低近平带。在求解贝里曲率以及陈数的过程中把微分形式的积分问题化成一个格点规范问题，使得不需要固定不同参数点波函数规范，这是其中一来自高能的启示。此外我们还计算了平带的富比尼-斯塔迪量子几何张量，其虚部的反对称组合也能给出贝里曲率分布，实部则称为量子度规，反映波函数在动量空间中平移时保真度的变化，相比之前所用的黎曼度规具有更丰富的复结构。基于能带带宽，贝里曲率在mBZ的分布标准差，和理想量子几何偏离值这3大判据，可以判定在相应参数下TBCB是否可能具有分数陈绝缘体(FCI)相，FCI作为分数量子霍尔效应（一种拓扑序）的格点版本，可

能在弱得多的磁场甚至零磁场下实现，因此在理论和实验上都备受关注。根据动量空间中对称分辨的精确对角化(ED)结果，可以得到涡旋附加条件所预期的基态简并度。除此之外，量子信息同样在人们理解FCI拓扑序中给予了重要启示，从现代理论的观点看，拓扑序是一类具有强量子纠缠的多体物态。借用量子信息中的概念，拓扑序的特征能用纠缠熵和纠缠谱描述，并且能从中揭示拓扑和相互作用的丰富联系。基于之前的ED多体能谱，我们可以进一步求出FCI系统在磁通下的多体谱流动，可以初步判断出拓扑简并基态子空间和简并度，基于这些简并态构造的密度矩阵，可以求出简并混态的粒子纠缠谱(PES)，PES在纠缠能隙以下的态对应着FCI中普适的准空穴激发，受到高能中色单态的启示，可以验证它们的计数服从相应填充下Halperin赝自旋单态的广义泡利原理，而这种计数是相应的电荷密度波(CDW)竞争相所没有的，因此PES的准空穴谱计数也成为FCI相的指纹。基于PES判据我们求解了TBCB中C=2平带在魔角$\theta \approx 1.608^{o}$下的准空穴计数，证实TBCB处于FCI相，并有可能在人工量子模拟器平台实现这类C=2的FCI相。相关内容已被Physical Review B接收。

在第三个工作中，我们首先用修正自旋波理论(MSWT)以及简易密度泛函+动力学平均场模型研究了近藤半金属$CeFe_2Al_{10}$的磁激发，一定程度上重复出了中子散射实验得到的激发谱，物理图像上可以用所谓的自旋激子理解。然而从更为还原论的角度出发，即高能物理角度看，我们可以把这样的凝聚态或量子化学系统看成是具有特定空间群对称性的量子电动力学(QED)-费米子物质场相互作用问题，通过自洽求解物质场以及电磁场来求解系统基态。当电磁场作为相互作用的媒介时，猜测不同于通常的量子光学问题，电磁场可以当成准经典的，这可以大大简化原来的高能物理第一性原理问题，即格点规范问题。基于这个假设我们开发了一种经典蒙特卡洛规范场抽样的固定规范算法，即使得抽样的规范场构型满足规范固定条件，比高能物理领域固定规范的引入鬼场和优化求解最大树算法更为简便。

**关键词:** 弯曲时空标架，狄拉克有效场论，扭转(转动)双层石墨烯，分数陈绝缘体，量子几何，精确对角化，粒子纠缠谱，格点规范

# Effective field theory and numerical study of strongly correlated electron systems from the perspective of high energy physics and quantum information

Marjor:   Theoretical physics

Name:   Jia-Zheng Ma

Supervisor:   Prof. Dao-Xin Yao

## ABSTRACT

Condense matter physics is a traditional but energetic area. Among it, the properties of strong correlated electron system is a very challenging aspect. Usually, an abstract effective lattice model for strong correlated system is simple. For example, Hubbard model for high Tc superconductivity, Anderson model for interaction between magnetic impurity and itinerant electron. However most of them are not exactly solvable except under certain low dimension and special parameters. So researchers develop many effective theories and numerical methods to investigate them and get very rich phase diagram. However, the occurence of topological phase ,topological order, twistronics, new kind of high Tc superconductive correlated phases and so on in recent decades have brought new challenge and chance for effective theories and numerical methods. On one hand, topological phase and the transition between them can not be characterized by local order parameter, which is beyond Landau paradigm. Instead, they are characterized by some global properties like chern number, entanglement entropy and so on. On the other hand, in twistronics system and moire superlattice, since under small twisted angle the superlattice constant can reach the magnitude of few nanometers. The atoms in unit (super) cell can reach the magnitude of $10^4$, which gives rise huge difficulties in real space exact diagonalization and ab initio calculation. These difficulties call for improved theoretical and numerical methods.

As the old saying, the stone from another mountain can break the jade unex-pectly. Althought condense matter physics, high energy physics as well as quantum information seem to be very different areas. Actually they share much common. This thesis will discuss how to apply high energy physics and quantum information methods in traditional condense matter problems. In Chapters 2 and 3, we mainly introduce the high energy physics inspiration in condensed matter via emergent point of view while in Chapter 4 we use the reductionism in high energy community.

In the first work, we hope to raise an effective model applicable for general twisted angle based on real space geometry theory. In traditional TBG theory, under small twisted angle, one can do the single valley approximation for BM model and consider the interlayer tunneling as perturbation. The twist produce the moire pattern. Accordingly, BZ is folded into mBZ. The interlayer coupling induces the two Dirac cones (mini-valley) in two layer tunel with each other within the same valley. This formalism can well explain the eletron localization at small twisted angle and the occurence of the almost flat bands at a series magic angles. The exotic physics in TBG is nearly all emergent from the correlation behaviors of the flat bands under different filling. However as mentioned above, BM continuous model is only applicable for small twisted angle. For moderate twisted angle, considering intervalley coupling and using commensurate approximation can also charaterize TBG well. However, for the TBG around $30^o$ twisted angle. There is no translation symmetry in such system. So it can not be unified under the theoretical framework of BM model. Instead, $30^o$ TBG hosts special 12 fold rotation symmetry. One can construct corresponding quasi-crystal model. How to build a universal theory for general twisted angle is still few researched. We start from real space and consider the interlayer rigid twist as a kind of deformation. This kind of deformation will induce effective curved space-time. Hence we conjecture that the theory under general twisted angle can be considered as the response of Dirac fermion under such curved space-time. That is the space-time vierbein acts as order parameter and couples with Dirac fermion. This theory and model have been widely investigated in high energy and general relativity community. Based on this theory we explain the lowest almost flat bands near the first magic angle $\theta = 1.05^o$. And we find that the band width is nearly the same as the result given by BM model. Our model is not only applicable for small twisted angle and commensurate case, but also incommensurate case like $30^o$ TBG. Our theory can give the quasi-energy band for $30^o$ TBG which met the experimental result of nanoARPES qualitatively. More importantly, our theory can predict the property of a new kind of system. We dub it as rotating bilayer graphene (RBG). This kind of system can not be explained and predicted by previous theories since there is no universal accepted theory for general twisted angle at past. And RBG is also very different from ordinary optical periodic driven floquet system. It's a periodic evolution in structure. Based on our framework, one can predict the charge pumping during one rotation period for given finite size region, which is related to Bott index and ought to be quantized. And the charge pumping will depend on the driven frequency and show topological phase transition in frequency domain. This phenomenon in RBG may be examined by certain transport experiment. This work have published in Physical Review B. One can check the detail in publication list during phd.

In the second work, we model a interacting twisted bilayer checkerboard lattice(TBCB) model. Based on BM-like model, we get the lowest flat bands with quadratic band touching. When figuring out the Berry curvature and Chern number, the first inspiration from high energy physics is that, one can transform the integral of differential form into a lattice gauge problem. This transformation make us do not have to fix the gauge of wave function under different parameter. In addition, we get the Fubini-Study quantum geometric tensor for flat bands. Its anti-symmetric combination of imaginary part can also give out the Berry curvature. While the real part is the quantum metric, which reflects the variety of wave function fidelity when tranlate it in momentum space. Compared with the previous Riemannian metric. The quantum geometric tensor hosts richer complex structure. One can consider following three indicator to diagnose whether there's fractional chern insulator (FCI) phase in TBCB. Band width, the standard deviation of Berry curvature over mBZ and failure of the trace condition (The deviation of ideal quantum geometry). FCI acts as a lattice version of fractional quantum hall effect (FQHE , a kind of topological order) and it may be realized at weak magnetic field even zero field. So FCI is concerned both in theory and experiment. According the result of symmetry resolved exact diagonalization (ED) in momentum space, one can get the TBCB ground state degeneracy predicted by vortex attachment theory. In addition, quantum information also gives us important inspiration in interpreting FCI topological order. In the modern point of view, topological order is a kind of many body matter with strong quantum entanglement. One can borrow the concept from quantum information and charaterize the topological order with entanglement entropy and entanglement spectrum, which can reveal the very rich interpaly between topology and interaction. Based on the ED many body energy spectrum mentioned above, we can solve the many body spectrum flow of FCI system under flux insertion and diagnose the topological degenerate ground state substate as well as the degree of degeneracy roughly. By using such degenerate states one can construct a density matrix and figure out its particle entanglement spectrum(PES). The state under PES gap correspond to the universal quasi-hole excitation in FCI. Inspired by the color singlet in high energy physics, one can examine that the quasi-hole counting obey the generalized Pauli principle based on Halperin pseudo spin singlet ansatz under corresponding filling. Such counting feature is absent in FCI's competing candidate, CDW phase. In conclusion, the quasi-hole counting in PES can be considered as the fingerprint of FCI. Based on PES criterion we solve the quasi-hole counting in TBCB C=2 topological flat bands under magic angle $\theta \approx 1.608^o$. We examine that TBCB indeed locates in FCI phase. Moreover, it's possible to realize such C=2 FCI phase in artificial quantum simulator platform. This work is accepted by Physical Review B.

In the third work, we first use disorder spin self-consistent mean field effective model along with poor man DFT+DMFT to investigate the magnetic excitation of Kondo semimetal like $CeFe_2Al_{10}$. We interpret the magnetic excitation spectrum measured in inelastic neutron scattering (INS) experiment to some extend. In terms of physical picture, one can interpret the result with spin exciton picture. However, on the another hand, if one interpret from the perspective of recovery theory. That is one can start from high energy physics, one can consider such condense matter system or quantum chemistry system as quantum electrodynamics (QED)-Fermion matter interaction problem under certain space group symmetry. One has to solve the matter field and electromagnetic field self-consistently. When the electromagnetic field act as the media of interaction, one can conjecture that electromagnetic field can act as quasi-classical object unlike the case in quantum optics. This can largely simplify the original high energy ab initio problem. That is the lattice gauge theory (LGT). Based on this assumption, we develop a kind of gauge fixing algorithm for Monte-Carlo sampling quasi- classical gauge field. That is forcing the sampling configuration to satisfy the gauge condition. At least in our problem, our algorithm is simpler than introducing ghost field and optimizing the maximal tree in gauge configuration.

**Key Words:** curved space-time vierbein, Dirac effective field theory, twisted (ratationg) bilayer graphene, fractional Chern insulator, quantum geometry, exact diagonalization, particle entanglement spectrum, lattice gauge theory

# 目录

# 第1章 前言

关联电子体系是凝聚态古老而传统的研究领域，近年来也面临着飞速发展与许多的机遇挑战。过去的理论和数值方法虽然一定程度解决了强关联的困难。但近年来重费米子变价化合物，摩尔材料，分数陈绝缘体拓扑序候选者等的出现也对理论数值的发展提出新的要求和挑战，这就要求凝聚态研究者不能拘泥于传统的凝聚态方法，而需要向其他的领域借鉴方法来解决，其中高能物理也是凝聚态强关联物理中经常借鉴的领域，本文将讨论一些强关联电子系统中来自高能物理的启示，并于一些传统凝聚态方法进行比较研究。

本文将讨论以下主题，第一个主题是一般转角下的扭转双层石墨烯(TBG)的有效模型，最初的扭转石墨烯麦当劳(BM)模型，只适用于小转角，公度转角的情况，受到高能中弯曲时空量子场论(QFT)的启示，我们提出了适用于一般转角的TBG有效模型，并复现了第一魔角的平带，30°准晶TBG的准能带。更重要的是这个理论预言了匀速转动双层石墨烯的拓扑性质，并有望通过特定的输运实验予以验证。第二个主题是分数陈绝缘体潜在候选模型扭转双层棋盘格模型(TBCB)的量子化学与拓扑性质研究，其中求解贝里曲率和陈数过程应用了高能的格点规范思想，在判断模型是否具有分数陈绝缘体相时，需要用到量子几何张量，以此定义参数空间态之间的距离和相位关系，具有比相对论中的黎曼度规更丰富的复结构。通过精确对角化(ED)多体能谱，我们得到了TBCB在特定填充数下的拓扑基简并度。进一步在量子信息的启示下，考虑这些拓扑简并基态候选者的磁通泵浦以及粒子纠缠谱，相应纠缠谱准空穴计数将提供摩尔系统是否存在分数陈绝缘体(FCI)的关键证据。第三个主题是近藤半金属$CeFe_2Al_{10}$的凝聚态第一性原理和高能物理的第一性原理的计算方案，与前两个主题不同，这里我们希望从还原论角度实现高能物理对应的算法，凝聚态方法即密度泛函+动力学平均场（DFT+DMFT）虽然能解释出更多特征，但多多少少依赖于一些先验实验数据，并且存在一定的唯象性。传统DMFT也难以处理低温的强关联材料，以及剩余自能的取法也存在一定争议。而利用高能中

的第一性原理即格点规范则相对只需要更少的先验数据和经验参数，原则上只需要晶格结构原子种类就能进行模拟，可以推广到其他材料，即具有更强的扩展性。本文主要讨论规范固定的一种简便算法，实践仍需要进一步研究。

本文结构安排如下：

1）第二章，我们介绍一般转角扭转双层石墨烯的有效弯曲时空理论，并用它来解决一般转角下的TBG的能带和准能带问题，相应的启示来自弯曲时空的QFT，标架作为序参量与TBG的狄拉克费米子发生最小耦合。原则上可以解析导出一般转角下的形变场和相应诱导的度规。这个理论并不受限于小转角和公度转角，具有更多的普适性。这个理论定性复现了实验和BM模型的魔角平带和30°准晶TBG准能带。并可以相当自然地推广到转动双层石墨烯（RBG），只需要把标架替换成对应的匀速旋转时空标架。RBG是一种非典型的弗洛凯系统，一般具有瞬时的空间准周期性，因此我们用准晶中的实空间拓扑指标-博特指标描述，它可以与实空间电荷泵浦联系，即在输运实验上提出了验证这个理论的可能性。

2）第三章，我们介绍TBCB的类BM理论和相应数值结果，非相互作用TBCB的类BM同样能给出最低平带，与TBG不同的是此时单层在费米能级的色散是二次型的，并且陈数为2，其平带部分填充时可能存在分数陈绝缘体（FCI）相，和陈数1的FCI不同，$C = 2$的情况没有简单的连续空间朗道能级对应。当把屏蔽库伦相互作用投影到平带，通过对相互作用模型做ED，对导带分数填充$\nu = 1/5$的魔角扭转双层棋盘格（MATBCB）可以得到10个近简并基态。它们在磁通泵浦下相互转化而不会跑出简并子空间。为了进一步确定其中是FCI相还是CDW相，可以通过量子信息启示，考虑拓扑简并基态候选者组成的混态的粒子纠缠谱（PES），PES能隙以下的准空穴态对于FCI满足广义泡利原理计数。对于$C > 1$的情况，类比高能物理语言，还需要额外满足Halperin色单态约束，对$C = 2$的特例约化为携带自旋情况下的广义泡利原理。我们发现MATBCB中FCI相候选者的PES准空穴计数符合上述规则，说明MATBCB很可能实现$C = 2$ FCI。然而由于保护能隙较小，猜测很可能需要外磁场进行稳定化。

3）第四章，我们介绍近藤半金属材料$CeFe_2Al_{10}$的相关实验现象与理论解释，我们的核心是关注它的磁激发，因为它不同于同类金属间化合物，相比$CeRu_2Al_{10}, CeOs_2Al_{10}$它是磁无序的，中子散射给出一般自旋波理论无法解释的分立点状信号。本章首先讨论用唯象修正自旋波理论自洽地从平均场水平模拟该

系统，定性地解释了自旋关联函数的分布，它正比于中子散射的散射振幅。图像上可以用自旋激子理论理解。然而上述模型未考虑巡游电子，基于安德森模型以及DMFT，我们也能定性得到磁无序的激发谱。最后，从更一般的角度考虑，原则上可以对一个一般的给定晶格常数和原子种类及分布晶体材料进行类似高能物理中的第一性原理模拟，即基于多体量子电动力学(QED)的凝聚态模拟方案，本章解决的一个关键问题是其中的电磁场抽样的规范固定问题。

4）第五章，我们本文的主要结论和对未来研究的展望。

# 第2章 扭转双层石墨烯的新视角-弯曲时空量子场论的应用

本章我们讨论的是扭转双层石墨烯(TBG)系统，一种人造的摩尔(moiré)超晶格系统，也是近几年来实验和理论学者非常关注的一个领域。由此衍生出相当多的新概念，例如魔角(magic angle)，转角电子学(twistronics)，摩尔量子模拟器等，转角摩尔系统相当于一个新奇关联量子物态的游乐园，很多强关联物态都可以通过调节转角，应变，填充等手段进行调控，因此在实验，理论，应用方面均具有广阔前景。本章开创性地提出了TBG的弯曲时空有效理论，从演生论角度看TBG这类凝聚态系统在长波低能下演生出来的类似高能理论的行为。下面先做一个简单的背景介绍。

## 2.1 研究背景和意义

转角摩尔系统就是把一些二维材料按照一些特定取向进行堆垛，类似量子版本的乐高积木，整体通过应变或者相对扭转可以产生比单层晶格尺度大得多的超晶格，相应的动量空间布里渊区(BZ)也会被折叠到更小的摩尔布里渊区(mBZ)。对TBG系统最早的理论工作出现于2011-2012年，首先是毕斯楚瑟-麦当劳(Bistrizer-MacDonald,BM)模型的出现，在小转角下，公度和非公度的差别并不大，可以把层间耦合当成微扰处理，对于TBG，层间耦合会诱导3个主导的动量转移方向，这样就能构造出BM模型，即TBG在动量空间的有效连续模型 [1]。比较新奇的是，MacDoanald等学者发现，TBG在某些特定转角下，重整化的费米速度会变成接近0，费米面附近也会出现带宽极窄的平带，意味着在这些特定转角下，TBG电子强烈局域化。此外，Guinea等学者将这种层间隧穿诱导的摩尔势能看成一种非阿贝尔规范场，从赝朗道能级的角度自然解释了魔角附近的电子局域化 [2]。而Castro等研究者对于所有公度转角下的TBG发展了基于紧束缚模型的系统连续理论 [3]。遗憾的是，当时TBG并没有引起实验学者足够重视，理论研究也

相对较少进而停滞。

直到2018年，Jarillo，曹原等实验学者在输运实验中发现TBG在特定填充存在关联绝缘相和超导相 [4, 5]，才掀起TBG，摩尔电子学的研究热潮。学者们在TBG以及相应的多层推广中发现的新奇物象包括但不限于，轨道磁性相 [6]，陈绝缘体 [7]，超导 [5]，脆弱拓扑相 [8, 9]，高阶拓扑绝缘体 [10]，高阶拓扑超导 [11]等。除此之外，基于TBG的异质结构还在超导干涉器件 [12]，Majorana零能模的实现 [13]等方面有所应用。

本章主要从理论层面分析TBG，希望能建立一种适用于一般转角的TBG理论。下面先回顾一下TBG的理论进展，自TBG中发现了魔角附近的新奇关联绝缘相，理论学家发展了很多理论试图理解魔角的出现以及物理图像。戴希等学者从赝朗道能级的图像出发，通过摩尔等效磁场的磁长度对应的磁面积和摩尔超胞的适配性，可以简单定出魔角条件 [14]。Ashvin等学者则从另外的角度出发，考虑所谓手征极限(chiral limit)下的BM模型 [15]，忽略层间相同子格(sublattice)之间的耦合，发现相应的BM哈密顿具有块状形式，只有非对角块是非平庸的，波函数此时是精确可解的，并且是(反)全纯函数，即波函数只是 $z = x + iy$ 或 $\bar{z} = x - iy$ 的函数。根据手征极限下的零能模方程可以定出相应的费米速度为0的点，对应魔角序列。牛谦等学者则从半经典近似角度出发 [16]，发现TBG的半经典波函数能表示成艾里(Airy)函数的线性组合，但波函数并非单值，而是存在斯托克斯割线，在不同的区域分片单值。根据不同区域之间的转移条件，要求转移一圈之后波函数保持单值，基于这个条件同样可以定出魔角序列。而对于相互作用的TBG，一般就没有严格可解的情况了，需要借助一些多体数值技术，相应的理论和数值方法将在第三章详细介绍。TBG相互作用多体模型的建立，例如Kang-Vafek U(4)模型 [17]，二维瓦尼尔(Wannier)局域化模型 [18]等。对于整数填充TBG，当TBG与衬底六角硼氮(hBN)对齐，可以证明基态几乎是一个斯莱特(Slater)行列式，这意味着基于斯莱特行列式的哈特里-福克(Hartree-Fock,HFA)近似变分方法能有效地模拟TBG多体基态 [19, 20, 21]，并且HFA的变分参数能直接和其中费米子的单粒子关联函数联系。微扰层面，基于赝自旋和能谷的演生SU(4)对称性考虑相应的SU(4)演生序量的涨落，利用费曼图微扰方法可以解释TBG在有限温度下的非费米液体行为 [22]。非微扰层面，小尺寸(动量空间)下，对于一般的填充特别是学界关注的分数填充，可以进行ED研究 [23]。对于考虑更大动量空间尺寸的情况，如果关注整数

填充以及TBG的2个能谷, 可以进行无符号问题的动量空间行列式量子蒙特卡洛模拟 [24, 25, 26, 27], 这种方法在计算谱函数和TBG有限温性质中是重要的。在TBG的磁有序相, 可以当成磁杂质与巡游电子相互作用系统, 可以应用DMFT求解TBG的磁性质 [28], 重费米子性质 [29]。对于分数填充的情况, DQMC很可能有符号问题, 对于某些填充, 小尺寸模拟也无法胜任, 因此理论学家也发展了TBG有效模型的密度矩阵重整化群(DMRG)算法, 在一个窄圆柱上模拟TBG 2d系统 [30, 31, 32], 能算到很大尺寸甚至外推到热力学极限, 基于 $x - k_y$ 基底下的dmrg [33] 以及Li-Haldane猜想 [34], 还可以基于纠缠熵探测TBG中的量子反常霍尔相-向列金属相之间的相变, 以及通过纠缠谱探测拓扑边界态, 这种方法能揭示关联和拓扑之间的深刻关系。从计算材料学角度, 学者们也发展了针对大尺度摩尔结构的密度泛函第一性原理算法 [35]。

虽然上述在魔角附近的理论已经有了相当多的进展, 但适用于一般转角的TBG统一模型仍然没有在学界达成共识。TBG的标准模型BM模型, 通常只适用于小转角以及中等转角(考虑能谷间耦合), 而并不能直接适用于非公度转角, 例如30° 转角的准晶TBG [36, 37] 就是一个完全没有平移对称性的系统, 不能直接应用BM理论, 而是应该利用其中的12重旋转对称性构造有限动量空间紧束缚模型, 显然这两种理论不能简单地统一到同一个框架下。注意到上述理论大多从动量空间出发, 因为实空间一般会形成尺度极大的摩尔超胞, 里面有上万个原子, 直接求解这样的大系统是不现实的也没有必要, 实际研究者只关注动量空间费米面附近平带的性质, 因此小转角和公度情况下从动量空间出发是合理的。然而我们注意到, 无论转角如何, 以及每层如何应变, TBG都可以看成是非扭转双层石墨烯的一种实空间变形, 非扭转双层石墨烯在半填充下的有效理论是狄拉克理论, 因此我们可以考虑TBG相当于其中狄拉克电子在扭转形变下所诱导的等效弯曲时空, 相应的性质归结为狄拉克电子对弯曲时空的响应。等效弯曲时空在凝聚态领域的应用其实已经比较常见, 例如对称保护拓扑相(SPT) [38, 39], 量子反常 [40, 41] 等。而且也出现了一些从实空间角度以及弯曲时空角度考虑摩尔系统的探索性工作, 例如基于变形的弯曲时空QFT在TBG [42], 扭转正方晶格 [43], 扭转拓扑绝缘体薄膜 [44] 中的应用, 不过基本限制在小形变梯度近似下进行微扰处理。还有学者基于这种实空间变形理论提出了所谓摩尔引力 [45], 全息平带理论 [46], 分形子弹性对偶 [47], TBG的弦论对应 [48] 等。然而这些理论工作虽然漂亮, 却大多缺少和实际实验现象的联系, 也

不便于预言一些TBG中的新现象。因此一般转角下的TBG理论，包括实空间理论尚不明确。最近实验上在FQH流体中观测到了手征引力子模式[49]，而TBG也能实现FQHE的格点版本分数陈绝缘体相(FCI)[50]，这也令人开始思考摩尔，FQH和弯曲时空三者之间的有趣联系。

基于上述需要和背景，受到广义的相对论以及弯曲时空量子场论等高能领域启示，本章试图从实空间理论出发，希望建立一般转角下的TBG有效理论，统一处理公度转角和非公度转角的情况，并且不限制在小形变的情况。首先本章推导一般转角下每个空间点的形变场，相应形变会诱导出时空标架，由此可以得到度规(弯曲时空)和自旋联络，自旋联络其实就充当其中的有效规范场的作用。基于卡当-黎曼理论可以写出有效狄拉克电子在这个弯曲时空的方程，这个方程的数值解在第一魔角附近的平带带宽和BM模型的结果接近。同时当转角为30º时，这套理论也能给出相应的准能带，可以猜想在存在相互作用时能复现实验上的纳米角分辨光电子能谱的结果(nanoARPES)[51]。上述两个例子其实只考虑了弯曲的空间，有理由推测上述理论也对时间依赖的TBG系统成立，因此我们考虑预言所谓转动双层石墨烯(RBG)的性质，即两层石墨烯之间相对匀速扭动的情况，这种系统对于之前的理论来说很难模拟，但弯曲时空理论就能很自然地推广过去，并能和相应的输运实验联系加以验证[52]。

## 2.2 TBG标架几何理论与一般转角下的形变场

连续介质中的形变可能有多种来源，例如刚性扭转，滑移，弛豫等。为简单起见本章只考虑刚性扭转角度$\phi$而忽略额外形变，刚性扭转的形变场为$\mathbf{u} = \phi \times \mathbf{r}$，很多小转角的TBG的工作也是基于这个形变量，即连续介质近似，然而对于一般的转角以及一般的接近热力学极限尺寸的TBG样品，这个形变场显然是不能真实描述的，因为那时形变量将无穷大，能量上不稳定。所以本章的第一步是导出一般转角下的TBG形变场分布。

首先我们选择图 2-1所示的坐标系统，确定形变场首先要选择一个参照，即相对什么对象的形变，自然我们把AA堆垛的双层石墨烯作为参照，因为AA堆叠石墨烯的有效理论是已经熟知的，我们需要在此基础上考虑一般刚性扭转后的理论，可以想象，当发生扭转，两层间距离最近的原子会相互连接形成一个弱的价键，按照这种规则连接的TBG即狄拉克电子相对非扭转双层石墨烯感受到的弯曲时空。因

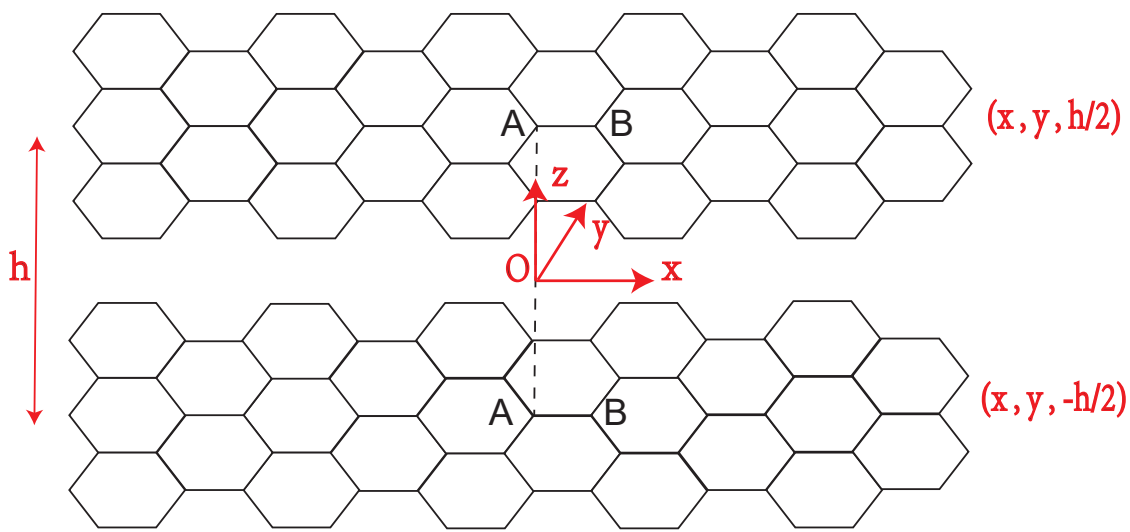

图 2-1: 本章所用的TBG系统坐标约定。转轴位于AA堆叠处，两层石墨烯分别位于(x,y,h/2), (x,y,-h/2)所在平面，如图显示双层石墨烯的零转角状态。顶层和底层的A,B子格分别正对。

此一般转角的形变场需要按照如下定义：特定原子的形变场定义为形变后，这个原子与形变前晶格中最临近的原子连线所代表的矢量，如图 2-2。显然这种定义在小转角，小半径下会回到$\mathbf{u} = \phi \times \mathbf{r}$。具体找法如下，首先画出发生转动的一层中所有共价键的中垂线，即对偶晶格，六角蜂巢晶格的对偶晶格是三角晶格。图 2-2 (b)中的蓝色实线表示发生转动的晶格，黄色实线表示对偶晶格，同时我们画出扭转之前的晶格，即 2-2 (b)中绿色虚线所示晶格，对于每一个小三角形，它的中心就是蓝色格点，如果存在某个绿色格点位于其中，则把它们连接，就能得到每个绿色格点上的形变场。注意到对于某些转角和某些位置，绿色虚线格点可能落在黄色实线上，这意味着距离这个点有多个"最临近原子"，因此这些点会具有多值的形变场，绿色格点位于黄色实线上会有2个形变，如果位于黄色格点上则会有6个形变场，这其实是一种多值奇异性，后面会看到这种奇异性对狄拉克电子在TBG的性质会有重要影响。

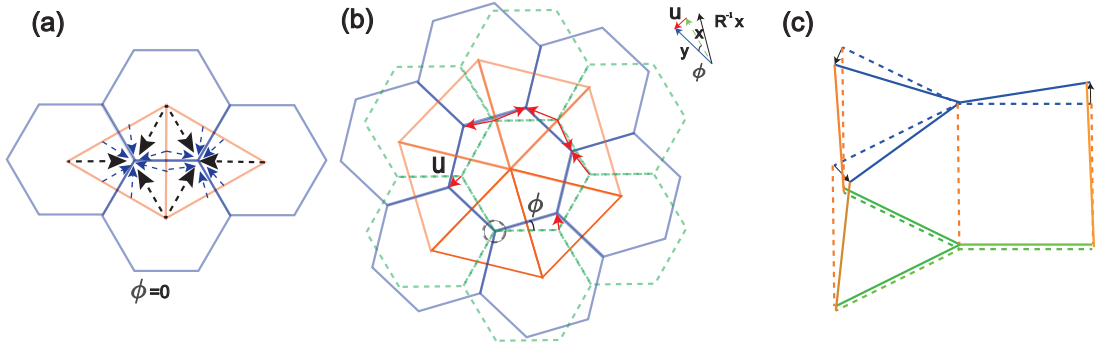

图 2-2: 一般转角下的形变场分布示意图。(a)表示零转角下的形变场分布。(b)表示转角$\phi$下的形变场，绿色虚线表示底层，蓝色实线表示顶层，黄色实线表示顶层的所有价键的中垂线集合，即对偶三角晶格。从俯视图看底层绿色格点与最临近的黄色三角形中心连接表示的矢量即顶层在这一点的形变矢量场。当绿色格点位于黄色对偶晶格上，在改点会出现多值的形变场，$\mathbf{x},\mathbf{y},\mathbf{u}$分别表示原始格点位置，形变后的位置，形变矢量。$R^{-1}\mathbf{r}$表示原始格点位置在顶层对应的斜直角坐标系下的矢量，四者关系见图(b)缩小图。(c)显示了小转角小半径下的形变场，绿黄蓝实线整体组成刚性扭转诱导的等效弯曲空间，顶层形变场符合$\mathbf{u} = \phi \times \mathbf{r}$。

基于形变场是矢量场的事实，我们可以写出形变场在不同坐标系下的变换规则。

$$\begin{pmatrix} u_x(\mathbf{r}) \\ u_y(\mathbf{r}) \end{pmatrix} = R \begin{pmatrix} u_x(R^{-1}\mathbf{r}) \\ u_y(R^{-1}\mathbf{r}) \end{pmatrix}, \ R = \begin{pmatrix} \cos\phi & -\sin\phi \\ \sin\phi & \cos\phi \end{pmatrix}, \tag{2-1a}$$

$$R^{-1}\mathbf{r} = \begin{pmatrix} x\cos\phi + y\sin\phi \\ -x\sin\phi + y\cos\phi \end{pmatrix}. \tag{2-1b}$$

其中$\mathbf{u}(\mathbf{r})$是x轴沿着底层A-B子格共价键对应坐标系下的形变场，即 2-1所示的坐标系下的形变场，而$\mathbf{u}(R^{-1}\mathbf{r}) = \mathbf{u}(\mathbf{r}')$为x轴沿着顶层(即发生扭转的那一层)A-B子格共价键对应坐标系下的形变场。其中后者的表达式是相对比较容易确认的，并且有如下的解析表达式。

$$u_x(\mathbf{r'}) = \left[ \left( \frac{3}{2}(p+q) - x_1 \right) \vartheta \left( \frac{3}{2}(p+q) + \frac{1}{2} - x_1 \right) \right.$$

$$\vartheta \left( (x_1 - \sqrt{3}x_2) - (3p-1) \right) \vartheta \left( (x_1 + \sqrt{3}x_2) - (3q-1) \right) +$$

$$\left( \frac{3}{2}(p+q) + 1 - x_1 \right) \vartheta \left( x_1 - \left( \frac{3}{2}(p+q) + \frac{1}{2} \right) \right)$$
(2-2a)

$$\left. \vartheta \left( (3p+2) - (x_1 - \sqrt{3}x_2) \right) \vartheta \left( (3q+2) - (x_1 + \sqrt{3}x_2) \right) \right]$$

$$\times \left( \frac{z}{h} + \frac{1}{2} \right) \vartheta \left( z + \frac{h}{2} \right) \vartheta \left( \frac{h}{2} - z \right).$$

$$u_y(\mathbf{r'}) = \left[ \left( \frac{\sqrt{3}}{2}(q-p) - x_2 \right) \times \vartheta \left( (x_1 - \sqrt{3}x_2) - (3p-1) \right) \right.$$

$$\vartheta \left( (x_1 + \sqrt{3}x_2) - (3q-1) \right) \vartheta \left( (3p+2) - (x_1 - \sqrt{3}x_2) \right)$$
(2-2b)

$$\left. \vartheta \left( (3q+2) - (x_1 + \sqrt{3}x_2) \right) \right] \left( \frac{z}{h} + \frac{1}{2} \right) \vartheta \left( z + \frac{h}{2} \right) \vartheta \left( \frac{h}{2} - z \right).$$

$$p = \text{floor} \left( \frac{x_1 - \sqrt{3}x_2 + 1}{3} \right), \quad q = \text{floor} \left( \frac{x_1 + \sqrt{3}x_2 + 1}{3} \right),$$
(2-2c)

$$x_1 = x\cos\phi + y\sin\phi, \quad x_2 = -x\sin\phi + y\cos\phi.$$

其中$\vartheta(x)$为亥维赛德(Heaviside)阶跃函数，$floor(x)$为向下取整函数，它们的出现意味着形变场在某些转角和某些位置存在多值奇异性。以及表达式 2-2，我们可以画出实空间的形变场分布 2-3，可以看出零转角和第一魔角的形变场分布差异主要集中在对偶的三角晶格或者说形变畴壁(domain wall)附近，可以明显看到在穿过畴壁时，形变场发生突变，因为定义形变的最近邻原子发生了变化。有了上述普适的形变场分布，为了描述相应的弯曲时空几何，我们需要导出相应的标架以及度规。标架相当于不同坐标系之间的雅可比变换，可以认为是弯曲时空的一种普适序参量，其表达式如下[38, 40]。

$$y^i(\mathbf{r}) = x^i + u^i(\mathbf{r}),$$

$$e_a^\mu = \frac{\partial y^\mu}{\partial x^a} = \delta_a^\mu + \partial_a u^\mu, \quad \xi_\mu^a = (e_a^\mu)^{-1}.$$
(2-3)

方程 2-3中的英文字母为平直时空坐标分量，希腊字母则代表形变诱导的弯曲时空的坐标分量，注意这里我们并没有要求形变很小以及形变的导数很小，甚至对

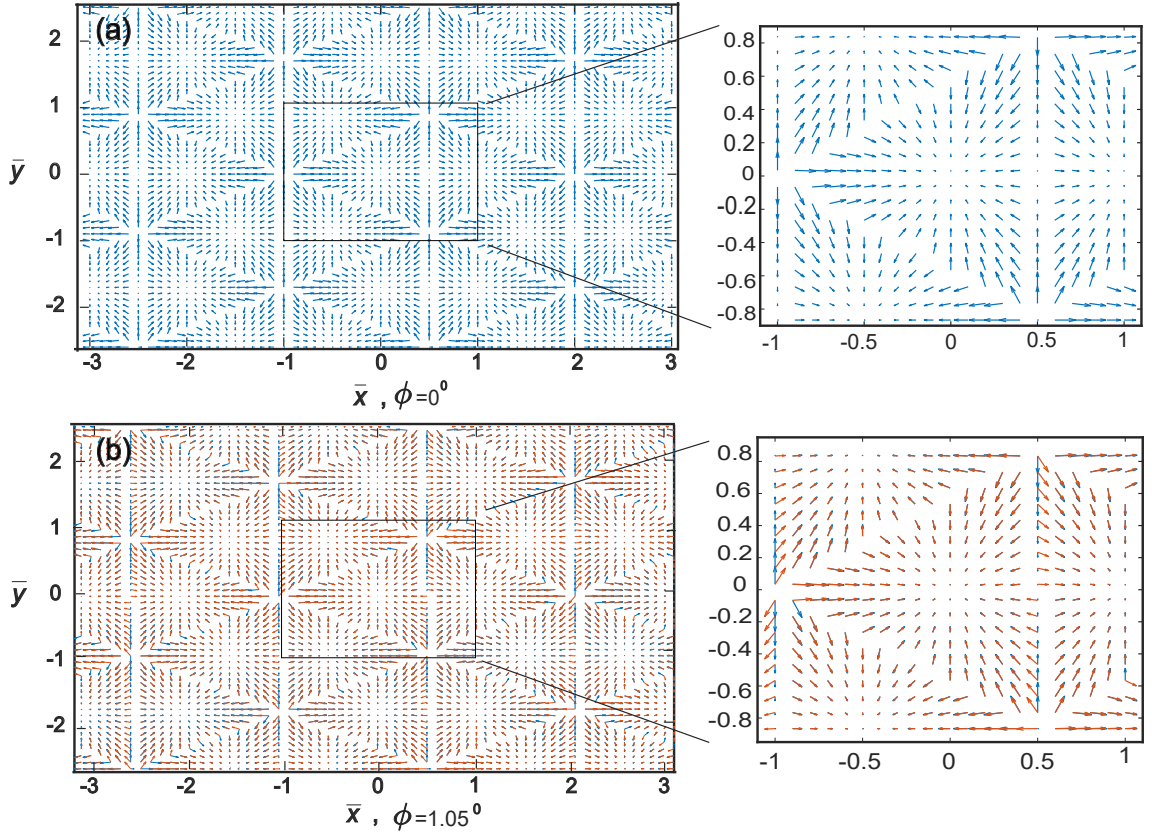

图 2-3: (a),(b)分别为零转角和TBG第一魔角1.05°下的形变场矢量分布示意图。蓝色箭头表示底层形变场而红色箭头表示顶层形变场，会发现在大部分区域2层之间的形变场差别不大，但在对偶三角晶格即形变场畴壁附近会出现层间和层内的形变场突变，显示出多值奇异性。

于某些转角和位置，标架还具有一定多值奇异性的。TBG中的有效狄拉克电子除了会感受到弯曲时空度规扰动，其实还会感受到等效规范场的作用，类似于应变石墨烯感受到的赝磁场[2, 20, 53]，即自旋联络[54, 55],相当于广义相对论中使用的克里斯托夫联络的规范变换，变换的对称群元素就是标架。接下来我们考虑引入度规，将TBG镶嵌在三维欧式空间，背景度规为平直，度规约定如下。

$$ds^2 = \eta_{ab}dx^a dx^b = -dt^2 + dx^2 + dy^2 + dz^2.  \tag{2-4}$$

即平直空间的背景度规是闵可夫斯基(闵氏)度规，$\eta_{ab} = diag(-1, 1, 1, 1)$。对于固定转角的TBG我们只需要考虑空间部分，因为我们之前假定只考虑刚性旋转，不考虑额外的弛豫，应变，所以这样刚性扭转诱导的弯曲时空是只有曲率而没有挠

率(torsion)的，也就是只有旋错(disclination)而没有位错(dislocation) [38]，此时的有效规范场或者说自旋联络也是紧致的，即规范群元 $g = exp(-i\theta_a T^a), \theta_a \in R$。尽管应变效应实验上真实存在，例如TBG的AA堆垛区会有面外排斥和面内收缩应变，AB堆垛区则相反，会有层间靠近面内扩张。图像上简单理解就是电荷中心集中在AA区域，静电相互作用使得发生上述面内面外应变以降低总能量 [18]，同时这种应变也使得费米面附近的平带更容易和更高的能带分开，增加平带与高能带之间的能隙 [20, 18, 35, 56]，简明起见本工作不考虑除了刚性扭转外的其他形变。无挠率的自旋联络表达式如下 [54]。

$$(\omega_a)_{bc} = \frac{1}{2}\eta_{bd}\xi_\mu^d(\partial_c\partial_a - \partial_a\partial_c)u^\mu + \frac{1}{2}\eta_{cd}\xi_\mu^d(\partial_a\partial_b - \partial_b\partial_a)u^\mu$$
$$+ \frac{1}{2}\eta_{ad}\xi_\mu^d(\partial_c\partial_b - \partial_b\partial_c)u^\mu + \frac{1}{2}(\partial_a\eta_{bc} + \partial_c\eta_{ba} - \partial_b\eta_{ca}). \tag{2-5}$$

其中的重复指标表示求和，首先我们注意到最后一项对于平常的欧式空间笛卡尔坐标系应该是消失的，如果我们一开始采用的是曲线坐标系，例如柱坐标球坐标等，则可能会有非零贡献，因此在我们的问题中自旋联络约化为如下形式。

$$(\omega_a)_{bc} = \frac{1}{2}\eta_{bd}\xi_\mu^d(\partial_c\partial_a - \partial_a\partial_c)u^\mu + \frac{1}{2}\eta_{cd}\xi_\mu^d(\partial_a\partial_b - \partial_b\partial_a)u^\mu$$
$$+ \frac{1}{2}\eta_{ad}\xi_\mu^d(\partial_c\partial_b - \partial_b\partial_c)u^\mu. \tag{2-6}$$

这里注意到方程 2-6中出现了$(\partial_c\partial_a - \partial_a\partial_c)u^\mu$这样的位移场混合二阶导数的对易子，根据微积分中已经熟知的知识，对于连续以及一阶空间导数连续的形变场，其混合二阶导数的对易子应该是0。但正如之前提到的那样，形变场$u^\mu$存在一定的多值奇异性并具有一些畴壁突变结构，正是这种奇异性导致了某些转角和位置出现了非零的自旋联络。经过符号计算软件mathematica的处理，形变场二阶导数对易子大致具有如下形式$(\partial_x\partial_y - \partial_y\partial_x)\mathbf{u} \approx \sum_{\langle 1,2,3 \rangle} \mathbf{A}_{123}\delta(\text{edge}_1)\text{floor}'(\text{edge}_1)\vartheta(\text{edge}_2)\vartheta(\text{edge}_3)$, $\delta(x)$，$\mathbf{A}$是特定的二维矢量函数，求和对图 2-2中所有可能的三角形畴壁进行。将自旋联络分量和赝自旋生成元缩并得到总的非阿贝尔自旋联络。

$$\omega_\mu = \frac{i}{2}(\omega_\mu)_{ab}\Sigma^{ab}, \quad \Sigma^{ab} = \frac{i}{4}[\gamma^a, \gamma^b], \quad \omega_a = e_a^\mu\omega_\mu,$$
$$\gamma^0 = \begin{pmatrix} 0 & I \\ I & 0 \end{pmatrix}, \quad \gamma^i = \begin{pmatrix} 0 & \sigma^i \\ -\sigma^i & 0 \end{pmatrix}, \quad i = 1, 2, 3. \tag{2-7}$$

其中有$[\gamma^a, \gamma^b] = \gamma^a\gamma^b - \gamma^b\gamma^a$。自旋联络对应的协变导数定义为$\partial_a \rightarrow D_a = \partial_a + \omega_a$。现在我们已经找到了TBG对应弯曲时空理论所用的黎曼度规(标架)以及自旋联络，下一步就是要写出这个狄拉克电子有效模型并进行求解。弯曲时空的狄拉克费米子的作用量以及哈密顿量是已经熟知的，即黎曼-卡当作用量，它已经在应变石墨烯[57],Kekulé扭曲下的蜂巢晶格Kitaev模型[58]，手征超流体和超导体等系统中广泛应用[59, 60]。为了简明起见，我们只考虑单能谷半填充的情况的自旋$\frac{1}{2}$狄拉克费米子的黎曼卡当作用量，以及哈密顿量密度。

$$S = i \int d^{3+1}x |\xi| (\bar{\psi}\gamma^\mu D_\mu \psi + im\bar{\psi}\psi),$$

(2-8a)

$$H = -i|\xi|[v_f(\bar{\psi}\gamma^j\partial_j\psi + \bar{\psi}\gamma^j\omega_j\psi) + im\bar{\psi}\psi],$$

(2-8b)

$$g_{\mu\nu} = \frac{\partial x^a}{\partial y^\mu}\eta_{ab}\frac{\partial x^b}{\partial y^\nu} = \xi^a_\mu \eta_{ab}\xi^b_\nu.$$

(2-8c)

其中$|\xi| = |\det(\xi^a_\mu)| = \sqrt{|\det(g_{\mu\nu})|}$为标架场的行列式，用来保证作用量积分测度是协变的，$D_\mu\psi = \partial_\mu\psi + \omega_\mu\psi$，$\mu = 0, 1, 2, 3, j = 1, 2, 3$，赝自旋的旋量约定为$\psi = (\psi_{\uparrow R}, \psi_{\downarrow R}, \psi_{\uparrow L}, \psi_{\downarrow L})^T$。$\uparrow, \downarrow$表示有效狄拉克费米子赝自旋，$L, R$表示手征性，而$\bar{\psi} \equiv \psi^\dagger\gamma^0, \gamma^\mu = e^\mu_a\gamma^a$。需要声明一点的是，本章和之前一些工作中采用的2+1d狄拉克模型不同[46]，后者将层作为一个内部自由度，而本章把层当成z方向的空间自由度，采用3+1d模型，只是z方向是开边界，表达式 2-8a可以认为是TBG中狄拉克电子的一个流体力学(hydrodynamics)模型。我们假定TBG中层间方向是弹道的散射，因此 2-8b中的z方向的裸(重整化前)费米速度$v_f$我们认为和石墨烯单层平面内一致。当层间隧穿弹道近似失效的情况将在后文简要讨论。层间隧穿在本文中如用2项控制，即$|\xi|(-iv_f)\bar{\psi}\gamma^z\partial_z\psi$ 和$-|\xi|\bar{\psi}w\psi$。可以看到测度因子$|\xi|$一般是空间依赖的并且在AA,AB堆垛区不一定为0，因此这个模型是超出手征极限模型的[15]。虽然我们的模型只考虑了单个能谷附近的狄拉克有效理论，但我们的实空间狄拉克模型 (2-8b)其实考虑了晶格尺度度规扰动的影响，对应动量空间的大动量转移，其实能一定程度反映能谷间散射这种大动量转移的相互作用。而且形变场 (2-2c)的奇异性其实也会体现在标架和度规 (2-8c)上，相当于狄拉克电子在

一个无限大三角畴壁阵列中的散射，畴壁是一系列三角形$\delta$函数组成的线缺陷，自然地形成$\delta$型势阱，很好地从图像上解释了魔角附近狄拉克电子的局域化。

接下来我们的任务就是如何求解哈密顿量 2-8b的本征值问题并将结果和标准BM模型进行对比，虽然在一般转角的弯曲时空下，平移对称性是没有的，但我们仍然可以考虑将有效狄拉克旋量具有布洛赫定理的形式即$\psi_{\mathbf{k}}(\mathbf{r}) = \exp(i\mathbf{k} \cdot \mathbf{r})u_{\mathbf{k}}(\mathbf{r})$，这样方便引入对动量的依赖，求解所谓准能带，我们在实空间有限尺寸求解上述弯曲时空狄拉克方程，采用相位扭转边界条件(twisted boundary condition) [61, 62]，边界上的相位积累来自自旋联络本身，求解域为如下的有限矩形$x \in [0, L_x], y \in [0, L_y]$，转轴在$(0,0)$位置，长度单位默认为$a_0 = 1.42$ Å即单层晶格常数。

$$\psi_{\mathbf{k}}(\mathbf{r} + \mathbf{L}_r) = \exp\left(-\int_{\mathbf{r}}^{\mathbf{r}+\mathbf{L}_r} ds\ \omega_r\right)\psi_{\mathbf{k}}(\mathbf{r}). \quad r = x, y \tag{2-9}$$

因为这个模型一般没有精确解，因此这里我们仍然采用数值差分求解偏微分方程的办法 [63]。一些参数约定如下，裸费米速度$v_f = 10^6 m/s$，层间距$h = 0.5 \times 1.42 = 0.71$ Å。有效质量$w = 0.11\ eV$。然而因为形变场本身存在奇异性，这在数值实现上是不允许的，因此我们需要唯象地加上一些展宽来近似形变场中的奇异函数，本章采用的方案是洛伦兹函数。

$$\delta(x) \approx \frac{\Gamma}{\pi}\frac{1}{x^2 + \Gamma^2}, \qquad \vartheta(x) \approx \frac{1}{\pi}\arctan\left(\frac{x}{\Gamma}\right) + \frac{1}{2},$$
$$\text{floor}'(x) = \sum_{k=-\infty}^{\infty}\delta(x-k) \approx \sum_{k=-N}^{N}\frac{\Gamma}{\pi}\frac{1}{(x-k)^2 + \Gamma^2}, \tag{2-10}$$

其中$\Gamma$是唯象引入的洛伦兹展宽，具有长度量纲，代表形变场畴壁的法向展宽，它在物理上其实也有比较明确的对应。就是一些固有量子涨落，热涨落，声子或晶格振动导致的形变场畴壁的展宽，将原本的奇异性$\delta$函数变化为洛伦兹函数。实际计算中为了最大限度保留形变场的奇异性特征，我们用符号运算软件mathematica计算了形变场混合二阶偏导数对易子$(\partial_c\partial_a - \partial_a\partial_c)u^\mu$的解析表达式，然后再把其中的奇异性函数按照方程 2-10进行替换。如果先对形变场替换成洛伦兹函数在计算其二阶混合偏导数的对易，则会得出平庸的结果。

在上述准备工作完成后，我们可以计算第一魔角$\phi \approx 1.05^o$下的TBG准能带，即MATBG准能带 2-4。其中对$\Gamma = 10^{-5}$和$\Gamma = 10^{-3}$两种形变场畴壁展宽进行计算，得到的结果大致一致，和BM模型给出的最低平带具有5meV的带宽定性上符合，能

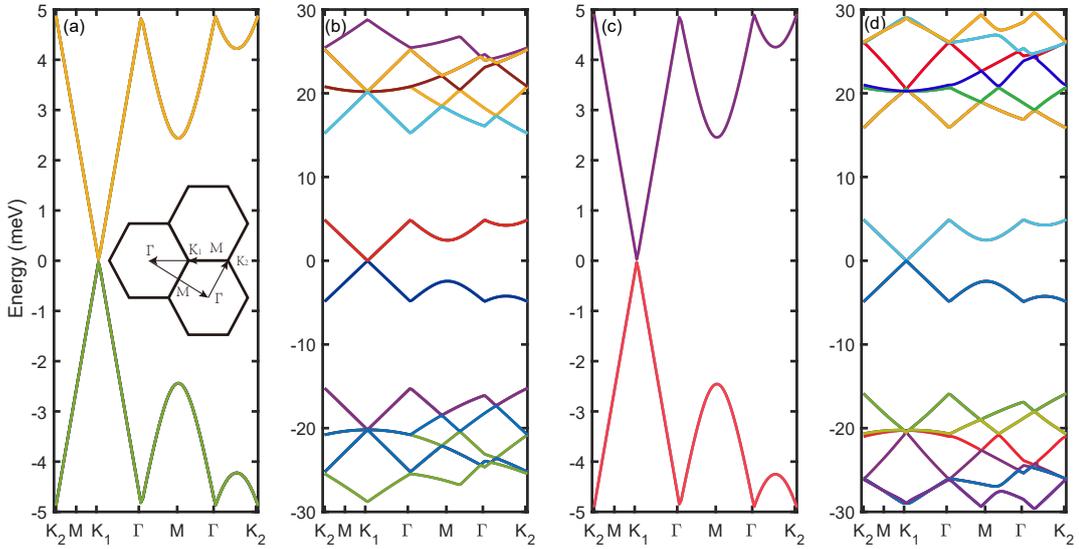

图 2-4: $1.05^o$MATBG下的准能带。其中只显示考虑半填充费米面($\epsilon = 0$)附近的若干能带，图(a),(b)的展宽$\Gamma = 10^{-5}$,而图(c),(d)的$\Gamma = 10^{-3}$，可以看到整体差别不大。(a),(b)分别显示了费米面附近的4条带以及24条带，其中最靠近费米面的4条带带宽约为5meV，和BM模型给出的MATBG最低平带带宽接近。最低的4条带中每条带2重简并，平带与更高能带的能隙大于10meV,并具有粒子空穴对称性。尺寸参数$L_x = L_y = 15$,实空间网格密度$N_x = N_y = 15$。缩小图中表示能带所对应的高对称路径，和文章 [15]相一致。

带中有一些斜率不连续点，这是TBG中费米子色散各向异性导致的 [64]。图 2-5显示出特定动量点$\frac{2\pi}{3}(\frac{1}{\sqrt{3}}, 1)$处最靠近半填充费米面的4条带，发现在$0.1^o$, $0.26^o$, $1.05^o$等转角下出现近零能态，说明这一理论并不只局限在第一魔转角。

然而，引入弯曲时空的代价是引入非厄米性(non hermiticity,NH)，即哈密顿量 2-8b一般是非厄米的，近期也有一些工作论证了弯曲时空与非厄米性的一些对应关系，例如 [65, 66, 67, 68]。例如从哈密顿量 2-8b动能项的差分形式可以比较明显看

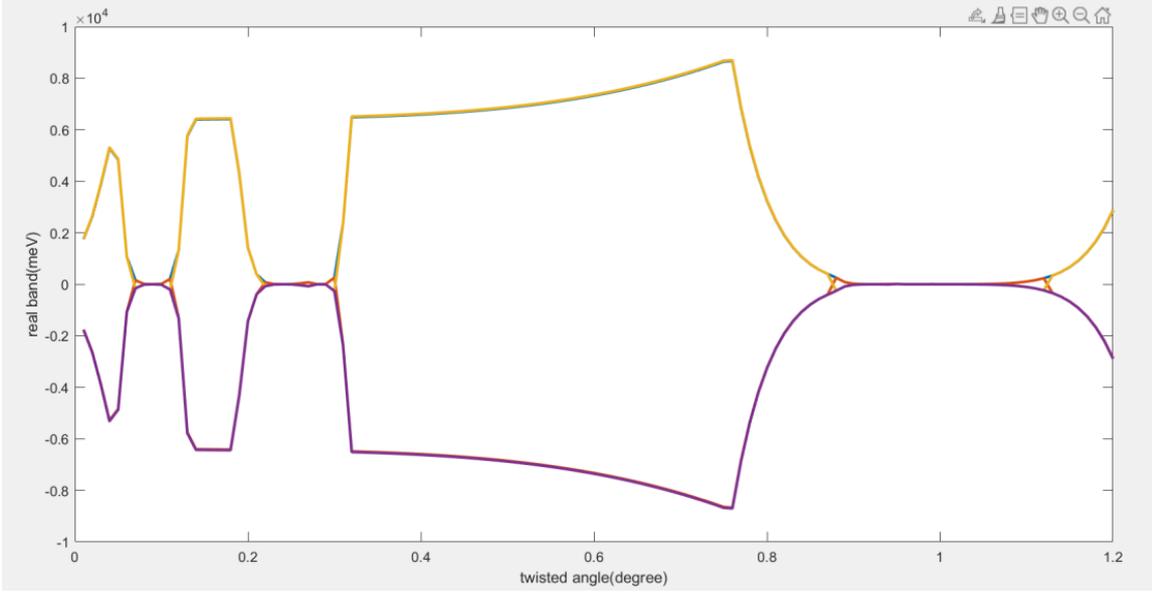

图 2-5: 动量点 $\frac{2\pi}{3}(\frac{1}{\sqrt{3}}, 1)$ 处最靠近费米面的4个准能带本征值随着转角的变化，除了第一魔角1.05$^o$，还复现了一些高阶魔角，例如0.26$^o$，0.1$^o$等 [1, 14]。(本图为补充数据，没有正式出现在发表论文中。)

出。

$$-i|\xi|\partial_x(j) \to -i|\xi_j| \left[\frac{\delta_{j,j+1} - \delta_{j,j-1}}{2\Delta x}\right],$$
$$-i|\xi|\partial_x(j+1) \to -i|\xi_{j+1}| \left[\frac{\delta_{j+1,j+2} - \delta_{j+1,j}}{2\Delta x}\right],$$

(2-11)

这个表达式中，标架以及积分测度一般有空间依赖，即 $|\xi_j| \neq |\xi_{j+1}|$，因此此时动量算符不再厄米。从整体或者积分的观点看，考虑动量算符的厄米共轭 $-i\int_{-\infty}^{+\infty} dx \psi^\dagger \partial_x \psi = -i(\psi^\dagger \psi|_{-\infty}^{+\infty} - \int_{-\infty}^{+\infty} dx (\partial_x \psi^\dagger)\psi)$，其中沿着镶嵌的背景平直欧式空间的x方向积分，可以看见动量算符厄米性要求第一项分部积分 $-i(\psi^\dagger \psi|_{-\infty}^{+\infty})$ 为0。但这个条件对于非公度以及弯曲时空的情况下是不一定成立的。作为比较，我们计算了厄米化模型相应的准能带，即在方程 2-8a 2-8b 中作如下替换 $S \to \frac{1}{2}(S + S^\dagger)$，$H \to \frac{1}{2}(H + H^\dagger)$ [69]，如图 2-6。可以发现靠近费米面附近的零能平带消失了，而且不同能带之间的能隙相较于非厄米模型从meV量级增大到eV量级。本章中除了图2-7，其余的准能带图均默认只画出实部。

出现上述现象的原因，我们猜测和阿蒂亚-辛格指标定理(Atiyah-Singer index theorem)有关 [70]，靠近费米面附近的平带由零能模算符 $-iD_\mu$ 控制，而并非它的

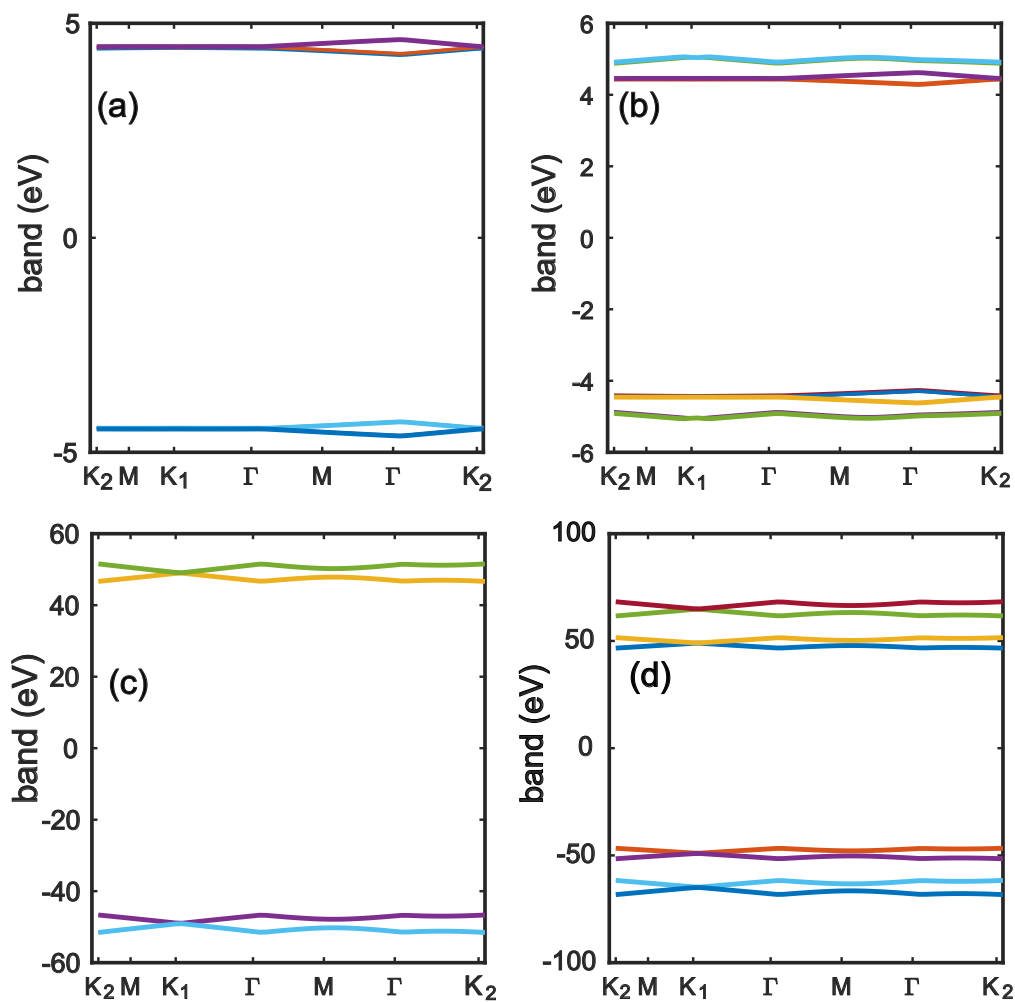

图 2-6: 厄米化之后的模型的准能带，可以发现相比非厄米能带 2-4，靠近费米面附近的零能平带消失了。其中所有能带图的模拟尺寸都是 $L_x = L_y = 7$，差分网格密度 $N_x = N_y = 7$，图(a)形变畴壁展宽 $\Gamma = 10^{-3}$，显示最靠近费米面的8条带，(b)中 $\Gamma = 10^{-3}$，显示最近12条带，(c)中 $\Gamma = 10^{-5}$，显示最近4条带。(d)中 $\Gamma = 10^{-5}$，显示最近8条带。

厄米化组合 $\frac{1}{2}(-iD_\mu + i\overline{D}_\mu)$。同时前人工作指出TBG的最低平带具有脆弱拓扑障碍 [8, 9]，或称为瓦尼尔障碍，最近工作中也将其视为等效的拓扑重费米子表示 [71]。即孤立的TBG 2带模型不能在保持所有局域对称性($C_{2z}T$对称性)情况下通过有限长程的紧束缚模型实现，由于狄拉克有效理论是相应紧束缚模型在狄拉克

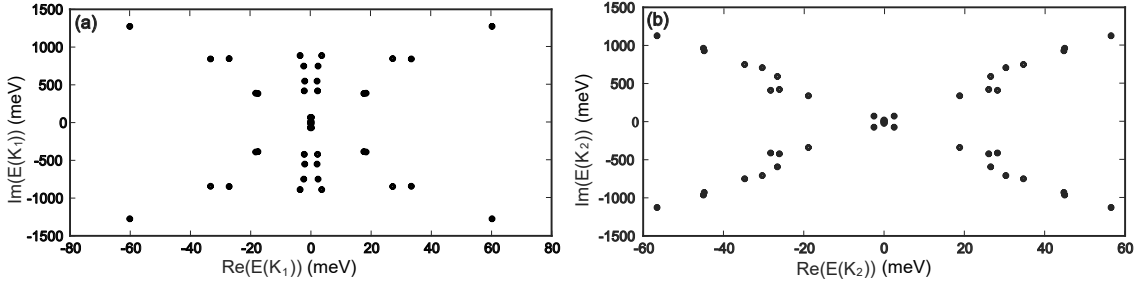

图 2-7: 在$K_1$和$K_2$两个迷你谷的复平面能谱，可以看到能量为$a + ib, a - ib,$ $-a + ib, -a - ib$ 这4个互为共轭的态在谱中同时存在。说明$PT(C_{2z}T)$对称性出现破缺。这可能是继承自类BM紧束缚模型的平带脆弱拓扑障碍，即不可能在保证保持所有局域对称性(包括$C_{2z}T$)的情况下构造有限力程的紧束缚模型。

点附近的线性化，自然也会保留这个特征使得$C_{2z}T$对称性(即PT对称性)破坏。因此我们本章接下来的讨论都集中在非厄米的情况，考虑分别为在$K_1$和$K_2$两个迷你能谷(mini-valley)的复平面能谱 2-7，可以看到能量为$a + ib, a - ib, -a + ib, -a - ib$这4个互为共轭的态在谱中同时存在，最低能带的实部和虚部量级接近，因为实部相同的能态具有相反的虚部，因此不会导致平带准粒子的增益或耗散，即平带的态密度不会因为非厄米性发生展宽。模型出现复能谱意味着PT对称性破缺，而PT对称性在BM模型中具体体现为$C_{2z}T$对称性，这个对称性在BM模型中一般是保持的，这也是我们模型与BM模型的一个显著不同。

在讨论完小转角，特别是典型的$\phi = 1.05^o$第一魔转角之后，接下来我们讨论非公度的$30^o$转角的TBG来显示我们理论的对非公度情况的普适性，$30^o$TBG是典型的准晶，即只有旋转对称性而没有平移对称性，$30^o$TBG具有12重旋转对称性[36]，并不能直接纳入BM模型的框架下。基于我们的几何理论，我们考虑解释实验上$30^o$准晶TBG(QCTBG)的镜面狄拉克锥导致的镜面处能隙打开行为[51]，在动量空间将$0^o$转角层的狄拉克锥$K_{0^o}$，作关于$30^o$转角层价键为镜面的镜像，得到相应的镜像狄拉克锥$K_R$。

NanoAPRES实验发现$K_{0^o}$与$K_R$连线与$30^o$转角层价键交点$M_{30^o}$出会发现能隙打开，能隙大约280meV。文章[51]给出的机制是$K_{0^o}$与$K_R$之间声子媒介的拉满散射导致$M_{30^o}$处能隙的打开，因此我们在模拟的时候选择比较大的形变场畴壁展宽$\Gamma = 0.5$表示比较强的声子作用。图 2-8显示了$30^o$转角下的准能带，其中缩略图

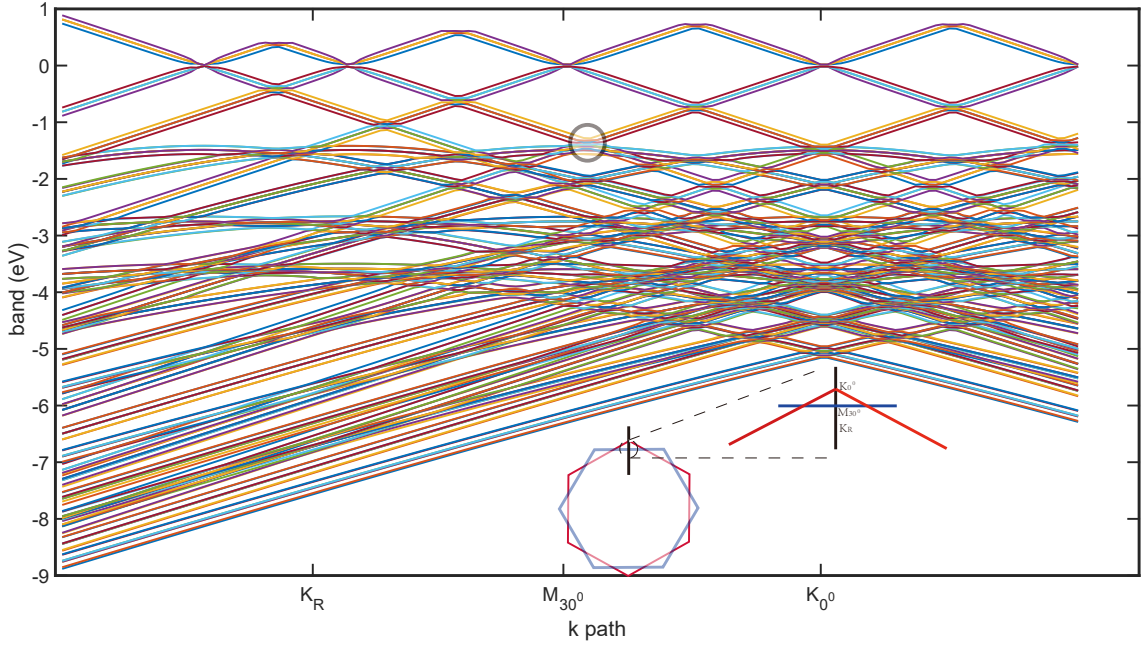

图 2-8: 30° QCTBG沿着垂直于镜面直线的准能带。在镜面点$M_{30°}$大约$-1.8eV$位置有较多能带交叉，能带束宽度约为280meV，预期增大电声耦合就可能在这个位置打开能隙，从而可能复现nanoARPES的实验结果。

中红色六边形表示$0°$转角石墨烯层，蓝色表示30°层，相应$K_{0°}$，$K_R$，$M_{30°}$的位置已经标记，黑圈表示$M_{30°}$点附近可能存在能隙打开的位置，黑圈位置有许多准能带在这里交叉，能带束的宽度正好和nanoARPES实验给出的280meV相近，结果部分定性符合实验。受限于本章讨论的仍然是单体理论，没有加上其他的相互作用项，准能带中未能打开$M_{30°}$处能隙，仍然需要引入其他和声子有关的相互作用，留待未来工作。

本节最后讨论一下当层间电子隧变变成扩散型而非之前的弹道型会对能带有什么影响，因为垂直于层方向相当于开边界，我们预想应该会有一个纯虚的有效费米速度。从微观的紧束缚模型出发，我们有如下结果。

$$t_\perp(c_1^\dagger c_2 + c_2^\dagger c_1) = t_\perp(c_1^\dagger c_{1+dz} + c_2^\dagger c_{2-dz}) \approx$$
$$t_\perp(c_1^\dagger c_1 + c_2^\dagger c_2 + hc_1^\dagger \partial_z c_1 - hc_2^\dagger \partial_z c_2)$$
$$= t_\perp(c_1^\dagger c_1 + c_2^\dagger c_2) - i[(it_\perp h)c_1^\dagger \partial_z c_1 + (-it_\perp h)c_2^\dagger \partial_z c_2]. \tag{2-12}$$

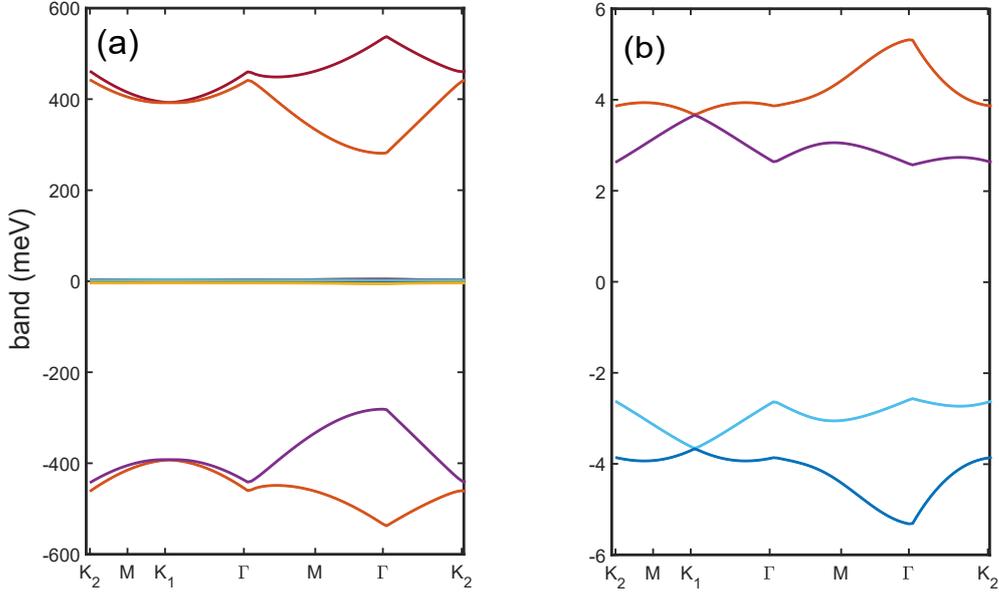

图 2-9: 如图为采用扩散型层间虚费米速度的MATBG的准能带，$\Gamma = 10^{-3}$。(a)显示最靠近费米面$\epsilon = 0$的12条带，(b)显示最靠近费米面的8条带。高对称路径的选取仍然和 2-4相同，相应的平带带宽约为3meV，甚至比 2-4中的平带带宽更小，但费米面处的狄拉克锥消失了。

层间隧穿变成扩散型要求TBG中狄拉克费米子的平均自由程小于层间距，然后我们假定差分近似有效$\partial_z \psi \approx \frac{1}{h}(\psi(2) - \psi(1))$，其中1,2分别标记底层和顶层。可以看到z方向的费米速度表示为$v_\perp = \pm it_\perp h$，形式和z方向的衰逝波类似，符号表示层间2个相反方向的隧穿并发生衰减。z方向的费米速度可以根据Koster-Slater参数化来估计 [18]，即$\left| \frac{v_\perp}{v_f} \right| = \left| \frac{t_\perp h}{t_{//} a} \right| = \left| \frac{V^0_{pp\sigma} h}{V^0_{pp\pi} a} \right| = \frac{0.48 \times 3.35}{2.7 \times 1.42} \approx 0.419$。相应的第一魔角准能带如下图 2-9。转角$\phi = 1.05^o$，展宽$\Gamma = 10^{-3}$。高对称路径的选取仍然和 2-4相同，相应的平带带宽约为3meV，甚至比 2-4中的平带带宽更小，但费米面处的狄拉克锥消失了，移位到了大约$\pm 3.7meV$的位置。定性上和之前的弹道型模型符合。以下除非特别说明，本章均采用弹道输运型模型。

小结一下，本节发展了基于实空间形变场诱导的标架，建立了TBG的一般转角形变场理论以及弯曲时空狄拉克模型，理论上能适用于一般的转角。模型复现了

第一魔角处的最低平带，其带宽和BM模型给出的较为接近。同时我们论证了其中出现非厄米性的原因。对于非公度情况，我们求解了$30^o$ QCTBG的准能带，得出了和实验上$M_{30^o}$处能隙定性符合的能带束宽度，可以论证当加入足够的声子相互作用，能复现NanoARPES实验上的结果。同时我们讨论了扩散型隧穿对第一魔角能带的影响，得到了和TBG弹道型输运定性上一致的最低平带。本小节主要是用我们的弯曲时空理论复现一些TBG中已经熟知的结果，下一小节我们将考虑对我们理论作进一步推广，并尝试预言一些新的现象。

## 2.3 转动双层石墨烯的弗洛凯工程-弯曲时空理论研究

上一节我们讨论了如何对静态的一般转角TBG建立一般的基于形变场的弯曲时空几何理论，一定程度上复现了之前的理论和实验结果。因为我们这套理论适用于一般转角，一个很自然的推广就是能不能将这套理论应用于动态的非平衡情况，例如两层石墨烯之间转角发生时间变化的情况，即所谓的转动双层石墨烯(rotating bilayer graphene, RBG)，为了简单起见，本节只讨论2层发生匀速相对转动的情况，即$\phi = \omega_t t$。之前TBG的情况其实是对应静态的弯曲空间，当发生相对匀速转动，则会真正演生一个弯曲时空，据笔者所知，这个问题从理论上是我们组首先开始研究的。

首先我们简要回顾RBG的研究动因和背景介绍，当TBG的研究热潮重新掀起时，研究者发现小转角TBG中作相对滑移可以发生拓扑荷的泵浦[72]，即索利斯泵浦(Thouless pumping)[73]，同时这种相对滑移也作为一种弗洛凯(floquet)工程调控，而注意到目前的TBG弗洛凯工程主要集中在光场调控领域[74, 75, 76]。然而对于RBG以及相应的弗洛凯调控工程的研究特别是理论方面研究仍然较少[77, 78, 79]。这里需要注意和澄清的是，本节提到的RBG并非大多数弗洛凯调控采用的方式，大多数弗洛凯调控是用交变电磁场对格点系统进行驱动，使得能带随时间周期性演化[74]，而RBG其实是一种周期性机械驱动对应的弗洛凯调控，在稳态下RBG系统每旋转$120^o$就会回到原来的状态，对应一个驱动周期$T = \frac{2\pi}{3\omega_0}$，同时这样的周期性转动驱动会导致摩尔原胞的大小，层间耦合以及连接状态也会整体的进行周期演化，我们预期RBG这样的弗洛凯调控能相比单纯基于周期电磁场驱动的较传统弗洛凯调控能产生更丰富的非平衡物态。下面考虑TBG理论到RBG的推广，首先RBG中因为时间依赖性，会产生非零的自旋联络时间分量$\omega_t$，而连续时间周期

演化也可能会产生非平庸的几何相位,同样也会产生等效的规范场,具体来说绝热演化产生贝里相位(Berry phase),非绝热演化则产生AA(Araronov-Ananda)相位[80],时间方向的自旋联络规范场和AA相位对应的规范场协同作用,我们预期其中会产生新的拓扑相。显然如果限制在传统的BM模型范式,或者没有建立一个适用于一般转角的模型,RBG是无法进行理论预测的。然而上一节提及的弯曲时空理论可以很自然地推广到RBG,可以考虑时间依赖哈密顿量如下。

$$H = -i|\xi|[-\overline{\psi}\gamma^0\omega_t\psi + v_f(\overline{\psi}\gamma^j\partial_j\psi + \overline{\psi}\gamma^j\omega_j\psi) + im\overline{\psi}\psi]. \tag{2-13}$$

其中$\phi = \omega_0 t$,相应的标架,度规和自旋联络都要推广到3+1d,其中生成元$\Sigma^{0j} = \frac{i}{4}[\gamma^0, \gamma^j], (j = 1 \cdots 3)$的出现也会导致哈密顿量 2-13不再厄米,因此 2-13实际是一个非厄米弗洛凯连续场论模型。因为对于RBG,准能带也是随时间变化的,一般只有时间方向有周期性而空间方向没有周期性,因此动量空间的图像以及基于陈数等的动量空间拓扑不变量已经不能很好很方便地描述RBG,所以我们应该考虑用一个实空间的拓扑指示器对RBG进行描述,很幸运的是,我们找到了这样的一个实空间拓扑指标,称为博特(Bott)指数[81, 82],而且这个指标能和实空间的拓扑电荷泵浦发生联系[83],为输运实验验证这种理论提供了可能。

首先考虑时间依赖情况下的自旋联络 2-6非零分量,除了上一节讨论的之外,还有如下的非零分量。注意含时情况下有些混合二阶偏导数对易子之间并非独立。

$$\phi = \omega_0 t, \partial_t = \omega_0 \partial_\phi,$$
$$\partial_\phi = x\partial_y - y\partial_x = x_2\partial_{x_1} - x_1\partial_{x_2}, \tag{2-14}$$
$$x_1 = x\cos\phi + y\sin\phi, \ x_2 = -x\sin\phi + y\cos\phi.$$

仿照方程 2-1b考虑将混合二阶偏导数对易子改写在$x_1, x_2$基底下,可以得到:

$$(\partial_\phi\partial_x - \partial_x\partial_\phi)\begin{pmatrix} u^x \\ u^y \end{pmatrix} = \left[-A_\phi C_\phi(\partial_{x_1}\partial_{x_2} - \partial_{x_2}\partial_{x_1}) - \frac{1}{2}E_\phi\partial_{x_1} - \frac{1}{2}D_\phi\partial_{x_2}\right]\begin{pmatrix} u'^x \\ u'^y \end{pmatrix},$$

$$\tag{2-15a}$$

$$(\partial_\phi \partial_y - \partial_y \partial_\phi) \begin{pmatrix} u^x \\ u^y \end{pmatrix} = \left[ -B_\phi C_\phi (\partial_{x_1} \partial_{x_2} - \partial_{x_2} \partial_{x_1}) + \frac{1}{2} D_\phi \partial_{x_1} - \frac{1}{2} E_\phi \partial_{x_2} \right] \begin{pmatrix} u'^x \\ u'^y \end{pmatrix},$$

(2-15b)

$$
\begin{aligned}
A_\phi &= x_1 \cos\phi - x_2 \sin\phi, \quad B_\phi = x_1 \sin\phi + x_2 \cos\phi, \\
C_\phi &= \begin{pmatrix} \cos\phi & -\sin\phi \\ \sin\phi & \cos\phi \end{pmatrix}, \quad D_\phi = \begin{pmatrix} 1 + \cos 2\phi & -\sin 2\phi \\ \sin 2\phi & 1 + \cos 2\phi \end{pmatrix}, \\
E_\phi &= \begin{pmatrix} \sin 2\phi & -(1 - \cos 2\phi) \\ 1 - \cos 2\phi & \sin 2\phi \end{pmatrix}.
\end{aligned}
$$

(2-16)

其中方程 2-15a 右边的表达式可以解析求解，完成了上述准备我们就可以求解哈密顿量 2-13 的瞬时本征值问题了，驱动周期为 $T = \frac{2\pi}{3\omega_0}$。现在假定我们已经求解了一个周期内每个瞬时的本征值问题，那么下一步就是构造相应的博特指数，它是一种常用于刻画准晶拓扑的实空间指示器，由投影坐标算符表示如下。

$$
\begin{aligned}
P(t) &= \sum_{\text{Re}(E_n(t)) < \mu} |\psi_n(t)\rangle \langle \tilde{\psi}_n(t)|, \\
U_r &= I - P + P \exp\left( i2\pi \frac{r}{L_r} \right) P, \quad r = x, y.
\end{aligned}
$$

(2-17a)

$$
\begin{aligned}
I_{Bott}^r(t) &= \frac{1}{2\pi} \int_0^t \text{Im}[\text{Tr}\left( \ln\left( U_r(t' + dt') U_{-r}(t') \right) \right)] \\
&= \frac{1}{2\pi} \int_0^t dt' \partial_{t'} [\text{Im}\left( \ln\left( \det\left( U_r(t') \right) \right) \right)]. \quad r = x, y.
\end{aligned}
$$

(2-17b)

$$I_{Bott} = I_{Bott}^x + I_{Bott}^y.$$

(2-17c)

需要注意的是，由于哈密顿量 2-13 一般非厄米，因此博特指数中的投影算符一般要用双正交基构造，即 $\{|\psi_n\rangle\}$ 为哈密顿量 2-13 的右本征态，相应矩阵记为 $V$，则 $\{|\tilde{\psi}_n\rangle\}$ 对应矩阵 $(V^{-1})^\dagger$ [84, 85]，为了简明起见我们考虑 RBG 在转动过程中费米面始终位于 $\epsilon = 0$，只考虑实部在费米面以下的子空间。另外为了更清楚地看

到Bott指数和拓扑电荷泵浦之间的联系，可以考虑从电极化的现代理论出发[83]。

$$\langle r(t)\rangle = \langle\psi(t)|\hat{r}|\psi(t)\rangle = \frac{L_r}{2\pi}\mathrm{Arg}\left(U_r(t)\right), \quad r = x, y. \tag{2-18}$$

其中幅角完整写出来是如下运算的组合$\mathrm{Arg} = \mathrm{ImTr ln}$。因此极化电流泵浦的时间积分可以与博特指数具有如下联系。

$$j_r(t)\Delta t = \mathrm{Arg}\left(U_r(t+\Delta t)U_{-r}(t)\right),$$
$$I_{Bott}^r(T) = \int_0^T dt j_r(t). \quad r = x, y. \tag{2-19}$$

下面我们给出RBG Bott指数的结果，考虑它的时间依赖和驱动频率依赖。可以看到Bott指数是在低频驱动下是量子化跳变的，从方程 2-19可以看出低频下一个驱动周期内泵浦出有限尺寸区域$L_x \times L_y$的电荷也是量子化的，并且存在一些时刻有整数单位电荷进出该区域。改变驱动频率，则一个驱动周期内泵浦的电荷数也会发生改变，即在频域存在拓扑相变，这种拓扑相可以视为自旋联络和AA联络共同控制的，为除了光场调控，滑移运动之外的TBG弗洛凯调控提供了新的途径。值得一提的是，当驱动频率比较高时，拓扑荷泵浦会逐渐偏离量子化，这是非绝热性导致的，类似的现象也在波导阵列的光子学拓扑泵浦系统中被观测到[86]。

本节的最后，我们尝试从电荷输运角度进一步理解博特指数的物理图像，以及考虑是否有可能的实验方案对这样的拓扑泵浦输运进行验证。首先Bott指数可以理解成霍尔电导或陈数的一个实空间有限尺寸版本，类比电磁学中的法拉第定律，赝磁通变化产生的电压为$U = \frac{d\Phi}{dt} \approx \omega_0 A$，其中A与驱动频率$\omega_0$无关，驱动频率$\omega_0$则反映了赝磁通的变化率。而微分霍尔电导定义为$\sigma_H = C\frac{e^2}{h} = \frac{dI}{dU} \approx \frac{dI^{Bott}}{d\omega_0}$，因此Bott指数的平台跳变可以理解为有限实空间区域内陈数或霍尔电导的改变，这与Streda定理的论证是类似的[87]。在实际实验中，考虑图 2-12所示示意图，摩尔材料的底层通常是放在晶格适配的衬底上，然后顶层进行转角堆垛。实验中可以用一个扫描隧道显微镜(STM)或原子力显微镜(AFM)中的针尖固定其中的顶层，针尖与2d材料的接触面设计成$L_x \times L_y$的矩形，然后匀速转动底层衬底来实现一个RBG系统。当驱动达到稳态，根据电流的连续性有$\partial_t Q + I_x + I_y + I_z = 0$，$\partial_t Q = 0$，$I_x + I_y = I_{Bott} = -I_z = -I_{measure}$，因此在z方向测得的电流积分等于x,y两个方向在一个周期内泵浦的电荷总数(差一个符号)，即测量了针尖对应有限区域内的总Bott指数。这种方案与最近兴起的一项

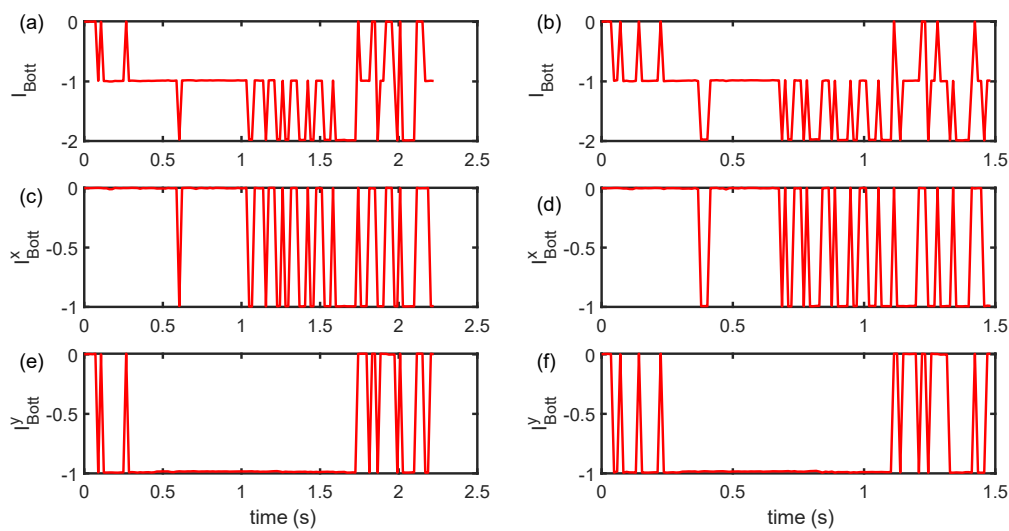

图 2-10: 特定驱动频率下的RBG Bott指数随着时间的依赖关系。时间上限为一个驱动周期$T = \frac{2\pi}{3\omega_0}$，参数选择如下，展宽$\Gamma = 0.5$，尺寸$L_x = 15$，$L_y = 15\sqrt{3}$，网格密度$N_x = N_y = 15$。(a),(c),(e)分别显示驱动频率$\omega_0 = 0.943rad/s$下的总Bott指数，Bott指数x分量,y分量。(b),(d),(f)则是驱动频率$\omega_0 = 1.414rad/s$下对应量之间的变化，时间步长均为$T/125$。

新技术，量子扭转显微镜(quantum twisting microscope,QTM) [88, 89]有一定类似之处。同时我们注意到最近实验上也提出了另一种可能实现RBG的方案，即通过在底层石墨烯上放置一个有限尺寸的单层石墨烯片，石墨烯片上再放置一层金属转子，通过AFM针尖对顶端的金属转子进行接触机械调控，可以实现转角的原位实时机械调控 [79]。

## 2.4 本章小结与展望

本章讨论了在高能物理特别是弯曲时空量子场论启示下的TBG新理论。即从实空间出发，将扭转看成一种空间分布的形变，相应会诱导出一种等效的弯曲时空，TBG中的狄拉克电子在感受到这样的弯曲时空后会相应地做出响应，例如在第一魔转角处发生局域化产生平带。这种理论可以处理任意的转角，无论公度与非公度。而且理论上也能自然地推广到其他的扭转布拉伐格子 [90]以及扭转同质多层的情况，只要非扭转理论已知并能比较简易地和弯曲时空进行耦合。对非扭转理论

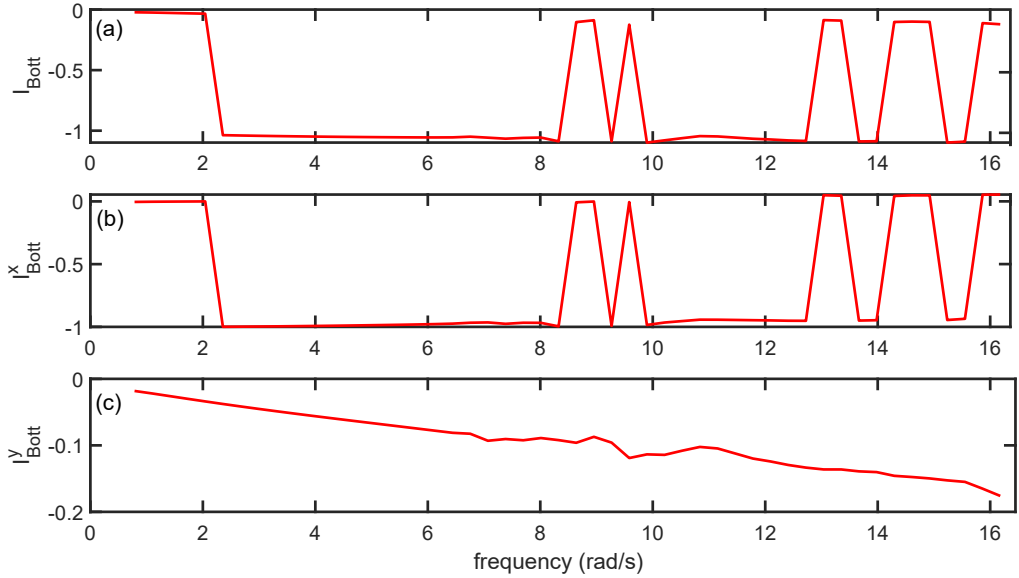

图 2-11: 一个驱动周期下的RBG Bott指数随着驱动频率的依赖关系。参数同样是 $\Gamma = 0.5$，尺寸 $L_x = 15, L_y = 15\sqrt{3}$，网格密度 $N_x = N_y = 15$，(a),(b),(c)分别对应 $I_{Bott}$，$I_{Bott}^x$，$I_{Bott}^y$。频率步长为 $0.314rad/s$，每幅图中展示了50个数据点。可以看到低频驱动下Bott指数量子化行为比较好，当驱动频率变高，Bott指数即拓扑荷泵浦逐渐偏离量子化。

是紧束缚模型的情况，我们预期弯曲时空同样也会引入自旋联络规范场，对每个紧束缚hopping参数产生空间依赖的相位调制，类似一个磁通空间调制的霍夫斯塔德(Hofstadter)模型[91]，这是有趣并值得后续研究的。扭转形变场会诱导出弯曲时空标架和度规，可以视为一种序参量，将狄拉克费米子与背景时空进行最小耦合，形变场在特定转角和位置会出现多值奇异性由此诱导规范的奇异性和非零自旋联络，形变场畴壁阵列相当于一系列 $\delta$ 线缺陷，使得电子容易局域化，自旋联络则充当赝磁场的作用，可以产生非平庸拓扑性质。基于这套理论我们复现出了第一魔转角处的最低平带，带宽和BM模型给出的定性一致。对于30° QCTBG我们也重复了其准能带，在30° 镜面处存在能带束，可以论证当引入其他相互作用，$M_{30°}$ 点可以打开能隙。说明这个理论同时适用于公度和非公度转角。此外，这套理论还能预言转动双层石墨烯即RBG的性质，这个系统是之前BM模型及其衍生模型无法

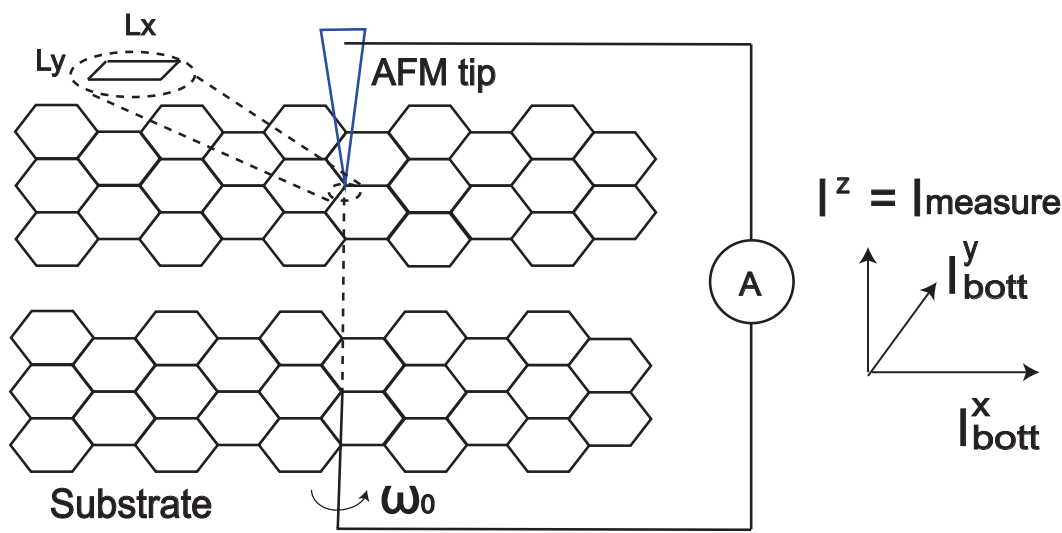

图 2-12: 如图为可能实验实现测量RBG中旋转驱动下可能得量子化电荷泵浦的方案示意图，相应的原子力显微镜针尖设计成$L_x \times L_y$尺寸矩形，转动衬底进行驱动，用电流计或电荷计记录一个驱动周期内的z方向电流积分，根据连续性方程，稳态下一个周期内z方向测得电荷与x,y方向发生泵浦的总电荷数数值上一致。

预测和模拟的，因为一般转角特别是非公度转角的情况不能直接纳入BM模型，然而本章基于形变场的弯曲时空理论能自然推广到3+1d的非平衡情况。基于实空间拓扑指示器Bott指数，可以探测有限尺寸实空间的拓扑荷泵浦，Bott指数能与电荷输运联系在一起，有望通过输运实验进行验证。Bott指数在频率空间存在跳变，说明RBG在驱动过程存在拓扑相变，为TBG中的弗洛凯调控提供了新的思路。

　　然而需要声明的是，本章建立的理论仍然是单电子有效理论，即最后都归结为求解单电子的本征值问题，暂时还没有引入多电子之间的相互作用。但在摩尔电子系统中，研究者们比较感兴趣的关联现象，例如超导，分数陈绝缘体(FCI)，分数量子霍尔效应等，都需要强的电子相互作用。如何基于弯曲时空理论解释强关联的摩尔奇异物象仍然是未来值得研究的充满挑战的方向。下一章我们将基于BM类模型研究多体相互作用摩尔系统，并尝试借助高能物理中的色单态，量子信息中的纠缠等概念对其进行描述，真正从多体波函数角度寻找摩尔系统中可能存在的强关联奇异物象。

# 第3章 扭转双层棋盘格分数陈绝缘体候选者-粒子纠缠谱计数与高能色单态规则的类比

## 3.1 研究背景和意义

前面章节我们着重讨论了高能物理对凝聚态系统描述与求解的启示，并且主要集中在单电子层面，本章主要讨论摩尔系统的多体层面，以及如何基于高能物理中的色单态以及量子信息中的粒子纠缠谱刻画其中可能存在的奇异拓扑物态。首先我们仍然简要回顾一下本章将要介绍的2个重要概念，扭转双层棋盘格(twisted bilayer checkerboard lattice, TBCB)以及分数陈绝缘体(fractional chern insulator,FCI)。

费米子的单层棋盘格模型早期主要用于研究其中的量子反常霍尔相以及向列相,和石墨烯类似也有2个子格，其晶格和3d阻挫烧绿石的2d投影以及铜氧化物铜氧面上氧原子的晶格类似，因此和阻挫磁体的自发对称破缺相以及铜氧高温超导有一定联系[92]，在单层棋盘格中也有很多工作讨论FCI等强关联拓扑物态的实现[93, 94, 95]。在其单层的BZ $(\pi, \pi)$动量点存在二次型的准粒子色散，和石墨烯的狄拉克型线性色散不同，棋盘格的二次型节点携带陈数$C = \pm 2$，符号取决于对应的能谷。最近有工作[96]在非相互作用水平研究了TBCB的性质，发现其中同样存在一个魔角序列，与TBG不同的是，TBCB魔角序列由无量纲参数$\alpha \approx \frac{w_{AB}}{t\theta^2}$给出($w_{AB}$为层间子格间耦合，t为单层紧束缚参数,$\theta$为转角)，而非类似TBG中的由$\alpha \approx \frac{w_{AB}}{t\theta}$决定[96]。而且TBCB费米面附近的平带能在更宽的参数区域内稳定存在，这对实验实现的要求也有所降低。同时类似的扭转正方晶格也被广泛研究[97, 98]，并且相应模型可以应用到FeSe的摩尔超晶格中[99]。

从TBCB魔角附近的超平带，以及在mBZ的近均匀贝里曲率(Berry curvature)分布，可以联想到分数填充的TBCB拓扑平带可能实现分数陈绝缘体(FCI)相。FCI可以视为是格点版本的分数量子霍尔效应(FQHE)，或者说是晶格平移对称性所扩充(enriched)的FQHE。最初的FQHE在强磁场下异质结量子

阱的2d电子气中由崔琦等学者实现，实验条件较为苛刻[100, 101]。因此学者们后来希望在格点系统寻找只需要弱磁场甚至零磁场的FQHE类似相即FCI，最初Bernevig等学者从理论和数值上论证了$C = 1$的拓扑平带(对应最低朗道能级，即LLL)在$\nu = 1/3$填充存在FCI相，并提出粒子纠缠谱(particle entanglement spectrum,PES)在纠缠能隙以下的态计数服从广义泡利原理作为FCI相存在的直接证据[93]。FCI相后来在MATBG体系实验实现[50]，遗憾的是仍然需要弱磁场对FCI进行稳定化而非实现零磁场FCI，近期Ashvin等理论学家也基于HFA，DMRG等数值方法在MATBG体系预言了零场FCI可能存在的相区域[102]。然而在此之后实验学者在较大转角的扭转双层$MoTe_2$[103, 104, 105]以及电位移场调控下的菱面5层石墨烯(R5G)-六角硼氮(hBN)衬底异质结构中发现了分数量子反常霍尔效应(FQAHE)[106]，其中的分数霍尔电导就是靠FCI相实现的，相比TBG中的FCI相，$tMoTe_2$的FCI相的实现仅需要较大转角，合适的电位移场(不需要加外磁场)，而相应最低平带不存在瓦尼尔拓扑障碍，这些特性大大简化了其中FCI等奇异相的实验和理论研究难度[107, 108, 109]，而菱面5层石墨烯-hBN结构以及相应多层体系甚至本身不需要转角就能实现FCI相以及FQAHE(但需要hBN衬底引入摩尔势)，这更令学者们开始思考转角摩尔超晶格到底是否是实现FCI以及FQAHE的必要条件[110, 111, 112]，最近学者甚至在$\nu = 3/4$填充R6G-hBN观测到在拓扑量子计算中更有价值的非阿贝尔FQAHE候选者[113]。而FQAHE作为零场FCI中的现象，也很可能并非只能通过FCI实现，未来有望在例如分数化的拓扑绝缘体等体系中实现[114]。

从单粒子的水平看，有利于FCI相出现的条件大致有3个。1. 具有孤立的理想平带，带宽很窄，远小于相互作用强度以及单粒子能隙。2. 贝里曲率在布里渊区(或mBZ)足够均匀，用贝里曲率的标准差$\sigma[\eta] = [\frac{1}{2\pi} \int d^2k (F(\mathbf{k}) - \bar{F})^2]^{1/2}$度量。3. 迹条件(trace condition)偏差，即量子几何条件，用量子度规的迹与贝里曲率绝对值的差在mBZ上的积分$T[\eta] = \frac{1}{2\pi} \int d^2k (Tr(g) - |F|)$度量，一般对于量子度规g和贝里曲率F，可以证明有不等式$tr(g) \geqslant |F|$，当上述等号成立，对应的布洛赫函数(周期部分)是关于$z = k_x + ik_y$的全纯函数，并可以验证不等式取等条件等价于复变函数中的柯西黎曼条件[102]，详见附录5。近期也有学者将上述三个条件统一为所谓的"涡旋可附加性"条件[115, 116]，相关理论正在发展当中并向$C > 2$的高陈数等复杂情况推广。转角以及摩尔条纹在其中的作用主要是减小电子动能使得费米面附近

能带尽可能平坦，而在手征极限下层间相同子格的耦合消失即$w_{AA} = 0$，使得平带波函数是关于$z = k_x + ik_y$的全纯函数从而满足迹条件偏差为0，因此对转角和层间耦合进行调控是可能实现FCI相的。而从多体水平来看，首先和FQHE类似需要足够强的电子间相互作用，通常相互作用远大于单粒子平带带宽，小于单粒子平带与更高能带之间的带隙。同时要求足够短使得粒子之间相互作用近似在实空间是$\delta$函数型的接触势，这样使得粒子在动量空间近似是均匀分布，能抑制电荷密度波(CDW)序等FCI的竞争相。一般可以通过调节栅极间距，电场以及衬底的有效介电常数等来调控摩尔系统中的屏蔽库仑作用。从比较初步的角度看，出现FCI相的必要条件是多体能谱中出现能量远低于其他态的低能态，它们之间有明显的能隙，并且当沿着某个方向绝热插入磁通时，这些低能态只会相互转换而不会跑出低能态组成的简并子空间(简并基态流形)。因此多体的精确对角化(ED) 能谱和磁通泵浦下的谱流动是数值研究模型是否存在FCI相的主要方式之一。然而上述多体方法有可能存在有限尺寸效应，即谱能隙可能在有限尺寸下出现(即使是摩尔系统这种在动量空间的ED)，但热力学极限下能隙仍然有关闭的可能。以及长宽比效应，即当其中一个方向的尺寸远小于或大于另一个方向时，也有可能出现虚假的能隙，当两个方向的尺寸同时接近无穷(但保持不兼容长宽比)达到热力学极限，所谓低能简并流形和高能非普适流形之间的能隙还是有可能关闭，即实际成为一维的热力学极限而非FCI对应的二维热力学极限。最重要的是，单凭多体能谱和磁通泵浦的能隙无法排除公度CDW的可能性，公度CDW同样也会产生类似的多体谱和磁通泵浦 [93]，而且CDW与FCI互为竞争相。稍微简易一点的区分CDW和FCI的方法是看基态平均粒子数在动量空间的分布，因为CDW是一种电荷密度平移对称破缺产生的序，因此CDW的粒子数分布通常集中在个别动量点，而FCI的粒子数分布则在BZ内近似均匀。然而相比上述特征，最能反映FCI拓扑性质的是其基态流形粒子纠缠谱(PES)的能隙和相应能隙下的态计数，也称为准空穴计数 [93, 117, 118, 119]，当准空穴计数满足所谓的广义泡利原理(对于C=1的$\nu$=1/3情况，与Laughlin态的准空穴计数一致。)，就能断定基态是FCI而排除拓扑平庸公度CDW的可能性，这一点在本章后面部分将重点讨论。

本章基于工作 [120]，将主要讨论TBCB的平带，拓扑性质和量子几何，以及在数值层面讨论在什么参数下可能实现FCI相，高能物理中的色单态类比，量子信息的粒子纠缠谱(PES)能带给我们在这一问题上带来什么启示。为避免歧

义，本章采用小写k指代单粒子动量，大写K代表多体动量或多粒子总动量，黑体$\mathbf{k} = (k_x, k_y)$代表单粒子2d波矢量，希腊字母$\kappa = w_{AA}/w_{AB}$代表手征比。

## 3.2 单粒子视角下的TBCB-拓扑平带与量子几何

在特定系统中寻找潜在FCI相的第一步是寻找合适的参数使得在单粒子水平满足前言所提到的理想拓扑陈带的三大条件[102]，狭义上来说上述条件只对$C = 1$的FCI直接适用，如何推广到高陈数FCI将在后文详细讨论。一个很自然的想法就是考虑手征极限下的魔角TBCB(MATBCB)，考虑如下TBCB的类BM模型[96]。

$$
\begin{aligned}
H = \sum_{k, q_{i,b}, q_{i,t}} & h(k, q_{i,b}, q_{i,t}) + \\
& \sum_{k, (q_{j,t}, q_{i,b})} \begin{pmatrix} c^\dagger_{k,q_{j,t}} & c^\dagger_{k,q_{i,b}} \end{pmatrix} \begin{pmatrix} 0 & T_{q_{j,t}, q_{i,b}} \\ T^\dagger_{q_{j,t}, q_{i,b}} & 0 \end{pmatrix} \begin{pmatrix} c_{k,q_{j,t}} \\ c_{k,q_{i,b}} \end{pmatrix},
\end{aligned} \tag{3-1}
$$

$$
\begin{aligned}
h(k, q_b, q_t) = & \\
& \begin{pmatrix} c^\dagger_{k,q_t} & c^\dagger_{k,q_b} \end{pmatrix} \begin{pmatrix} H_0^{\phi/2}(k - q_t) & T_{q_t, q_b} \\ T^\dagger_{q_t, q_b} & H_0^{-\phi/2}(k - q_b) \end{pmatrix} \begin{pmatrix} c_{k,q_t} \\ c_{k,q_b} \end{pmatrix},
\end{aligned} \tag{3-2}
$$

$$
\begin{aligned}
H_0^\phi(k) = & 2t' \cos(k_x - k_y)(\cos\phi\sigma_y + \sin\phi\sigma_x) + \\
& 4t \cos(\frac{k_x + \pi}{2}) \cos(\frac{k_y + \pi}{2})(\cos\phi\sigma_x - \sin\phi\sigma_y) + \Delta\sigma_z,
\end{aligned} \tag{3-3}
$$

$$
\begin{aligned}
T_{q_t, q_b} = & (w_{AA}I + w_{AB}\sigma_x)[\delta(q_t^x = q_b^x + \frac{k_\phi}{\sqrt{2}})\delta(q_t^y = q_b^y + \frac{k_\phi}{\sqrt{2}}) \\
& + \delta(q_t^x = q_b^x - \frac{k_\phi}{\sqrt{2}})\delta(q_t^y = q_b^y - \frac{k_\phi}{\sqrt{2}})] + \\
& (w_{AA}I - w_{AB}\sigma_x)[\delta(q_t^x = q_b^x + \frac{k_\phi}{\sqrt{2}})\delta(q_t^y = q_b^y - \frac{k_\phi}{\sqrt{2}}) \\
& + \delta(q_t^x = q_b^x - \frac{k_\phi}{\sqrt{2}})\delta(q_t^y = q_b^y + \frac{k_\phi}{\sqrt{2}})],
\end{aligned} \tag{3-4}
$$

其中方程 3-3中$H_0$为单层棋盘格的紧束缚哈密顿量，t和t'分别为AB子格之间和相同子格之间的紧束缚参数，所有的动量点是相对单层BZ的$(\pi, \pi)$点定义的，

当 $t = 2t'$，可以证明在 $k_x = k_y = 0$ 附近展开到二阶，能得到二次型的色散关系，泡利矩阵 $\{\sigma_i\}, i = x, y, z$ 定义在A,B子格空间，$\phi$ 为两层之间的相对转角，$\Delta$ 为子格势，可以通过衬底实现，用来打开二次型节点处的能隙，本节单粒子情况暂定 $\Delta = 0$。$T_{q_t, q_b}$ 则表示顶层和底层2个二次型动量节点之间的隧穿矩阵，$q_t, q_b$ 分别为顶层和底层的动量，$w_{AA}, w_{AB}$ 分别是同子格和子格之间的层间耦合，和TBG情况类似，$w_{AA} = 0$ 表示手征极限 [15]。而 $w_{AA} \neq 0$ 的情况会使得最低平带一定程度展宽，不利于FCI形成。TBCB的类BM模型就是单层项以及所有最近邻不同层二次型节点耦合哈密顿量的总和 3-1，与TBG中有3个主要隧穿方向类似，TBCB中有4个主导层间隧穿方向 [90]。数值计算中，动量 **k** 可以在第一mBZ中连续取点，而二次型节点对应的倒格点我们一般选足够多的格点，即一个有限尺寸的大倒格点阵，大倒格点其实对应着数值上选取的动量截断或短波紫外截断(分辨率)，也决定了半填充处下方能带的数量，对于某些问题，考虑足够大的动量截断以及平带下方所有填充带的Hartree-Fock修正 [111, 121] 或重整化修正 [110] 是重要的，例如R5G-hBN异质结系统的FCI问题，在无相互作用情况下拓扑平带不会打开能隙，只有在Hartree-Fock或重整化修正下相应拓扑平带与其他能带的带隙才会被打开，本文将不展开讨论。手征极限下的第一魔转角TBCB单粒子能带如图 3-1。

根据文献 [96] 的结果，选择参数 $t = 1000meV, t' = 500meV, \phi = 1.608^o, w_{AB} = 2.05meV$ 时，靠近费米面 $\epsilon = 0$ 的平带带宽最小，大概为 $1.5 \times 10^{-3} meV$，并且在图 3-1中的红点和蓝点出现二次型的能带接触节点(类比TBG的狄拉克点)。MATBCB单粒子能带 3-1和文献 [96] 给出的结果类似。很自然地，接下来我们需要先研究最低的2条平带是否在单体水平拓扑非平庸。一种简单的方案是看2条带的绕数(winding number)，即平带布洛赫函数周期部分沿着某一方向(例如 $k_x$)的威尔逊回路(Wilson loop)随着另一个方向动量 $k_y$ 的缠绕关系。从电极化的现代理论看，绕数也等同于平带波函数瓦尼尔中心(Wannier center)随着 $k_y$ 变化或者泵浦时在实空间单胞中的移动情况。绕数通常是整数，对应瓦尼尔中心在 $k_y$ 泵浦一个周期会移动整数个单胞 [30, 31]。

因为2条带之间有交点，一般我们需要考虑如下SU(2)版本的威尔逊回路。

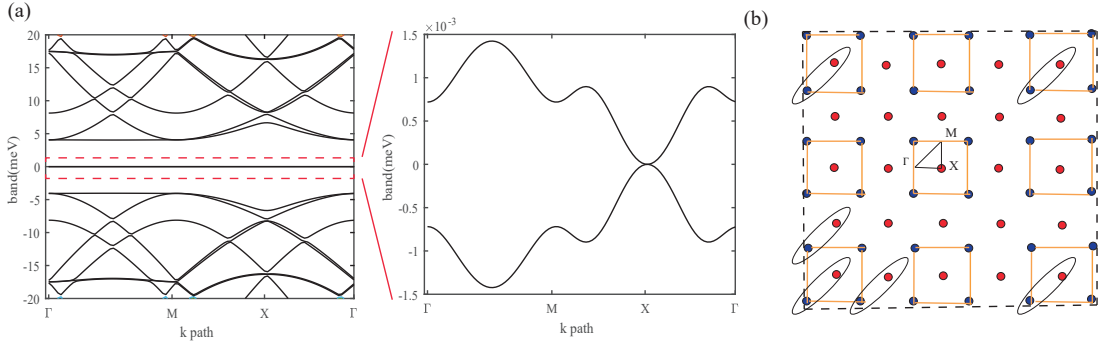

图 3-1: (a)为第一魔角1.608°下的TBCB高对称点单粒子能带，通过类BM模型对角化得到。沿着高对称路径画出能带可以看到费米面附近有2条平带，带宽约为$1.5 \times 10^{-3} meV$，在X点具有二次型的色散能带接触。红色虚线方框为2条平带位置，放大图中可以看到2条带的细致带宽，以及在手征极限下具有粒子-空穴对称性。图(b)显示出不同层的二次型节点位置，红格点和蓝格点分别表示底层和顶层的二次型节点，任意两个在动量空间相邻的红蓝节点直接都存在形式如 3-4的隧穿作用项，黄色实线框为其中一种第一mBZ的取法。

$$[O(k_x, k_y)]_{\alpha\beta} = \langle u_\alpha(k_x, k_y) | u_\beta(k_x + dk_x, k_y) \rangle, \quad dk_x = \frac{2\pi}{N_x}, \quad \alpha, \beta = 1, 2$$

$$W(k_y) = \left[ \prod_{n_x=1}^{N_x-1} O \left( k_x = \frac{2\pi}{N_x}(n_x - \frac{N_x+1}{2}), k_y \right) \right] O'(k_y),$$

$$[O'(k_y)]_{\alpha\beta} = \sum_{G_x, G_y, \tau, \mu} \langle u_\alpha(k_x = \frac{\pi(N_x-1)}{N_x}, k_y) | G_x, G_y, \tau, \mu \rangle \tag{3-5}$$

$$\langle G_x - 2\pi, G_y, \tau, \mu | u_\beta(k_x = -\frac{\pi(N_x-1)}{N_x}, k_y) \rangle.$$

其中$[O(k_x, k_y)]_{\alpha\beta}$是固定$k_y$下沿着$k_x$方向相邻2个动量的布洛赫函数周期部分的内积。根据$\langle u_k | u_{k+dk} \rangle \approx \langle u_k | (I - idk(i\partial_k)) | u_k \rangle \approx \exp(-idk(i\langle u_k | \partial_k | u_k \rangle))$，而$i\langle u_k | \partial_k | u_k \rangle$正是实空间坐标坐标算符(或者极化)的期望，因此可以看到在连续极限下，威尔逊回路 3-5取对数正是贝里联络或者极化算符沿着$k_x$方向的积分。求解威尔逊回路中有几点需要注意，首先数值上因为动量差分的有限性，重叠矩阵$[O(k_x, k_y)]_{\alpha\beta}$数值上将不会是精确的幺正矩阵，后面其实会看到这个矩阵的模部分对应动量空间的一些二阶非线性效应即量子度规的贡献，而幺正部分对应贝里曲

率，此时就需要我们对重叠矩阵做奇异值分解(SVD)或极分解，分离出重叠矩阵的幺正部分，具体如下。

$$O(\mathbf{k}) = U(\mathbf{k})S(\mathbf{k})V^{\dagger}(\mathbf{k}), \quad \tilde{O}(\mathbf{k}) = U(\mathbf{k})V^{\dagger}(\mathbf{k}), \tag{3-6}$$

$$O(\mathbf{k}) = P(\mathbf{k})R(\mathbf{k}), \quad R(\mathbf{k}) = \sqrt{O^{\dagger}(\mathbf{k})O(\mathbf{k})},$$
$$P(\mathbf{k}) = \tilde{O}(\mathbf{k}) = O(\mathbf{k})R^{-1}(\mathbf{k}) = O^{1/2}(\mathbf{k})[O^{\dagger}(\mathbf{k})]^{-1/2}. \tag{3-7}$$

在上述操作之后我们就得到了重叠矩阵的幺正部分$\tilde{O}$，这个步骤特别在求解极化算符的本征问题以及构造混合$n - k_y$最大局域化瓦尼尔轨道问题中是重要的(本文暂不详细讨论此问题)，因为需要对角化的算符形式为$\exp(-iP)$，只有保证这个算符是幺正算符，其本征态才能正交归一[30, 31]。其次就是威尔逊回路是一个闭环，然而之前提到TBCB动量空间我们其实选用的是开边界条件，也就是最终我们的动量会来到第一-mBZ的边界附近，这时候为了保证闭环就需要做紧致化处理，即把动量拉回到第一mBZ，同时为了保证这个操作后的波函数和上一个k点波函数的重叠仍然是一个一阶小量，相应的波函数分量$G_x$也要相应拉回补偿$2\pi$即平移一个倒格点，如方程 3-5所示。如果不做这一拉回操作，实际我们计算的仍然是一个开的威尔逊线，通常会依赖于布洛赫函数的规范而不是拓扑不变量，上述操作也称为插入镶嵌矩阵[96, 122]。

从绕数图 3-2 (a)我们可以看出MATBCB的2条平带具有非平庸陈数$C = \pm2$，并且绕数关于$k_y$的斜率较为均匀，因此我们推测贝里曲率在mBZ的分布应该是比较均匀的，有利于FCI的实现。虽然直观上来看，$C = 2$以及更高陈数的拓扑平带能直接通过堆叠(stacking) $C = 1$的拓扑平带直接构造，然而在这个堆叠过程中并不能保证能谷的位置保持在例如K点，有可能堆叠过程中能谷位置变化到$\Gamma$点[123]，而我们本章提到的TBCB模型并不会遭遇这样的困难，这也是这个模型相比在$C = 1$系统实现$C = 2$ FCI的优势，因为TBCB中$C = 2$拓扑平带是自然实现的而不涉及堆叠过程。然而需要注意，之前的FCI三判据[102]其实狭义上只能直接应用在具有最低朗道能级(LLL)对应的拓扑平带，即陈数$C = \pm1$的情况，而对于$C > 1$的情况，将没有直接的朗道能级对应，这源于在单粒子水平，粒子相当于多携带了一个$SU(C)$的内部对称性（类比高能中的色自由度），会带来更多的

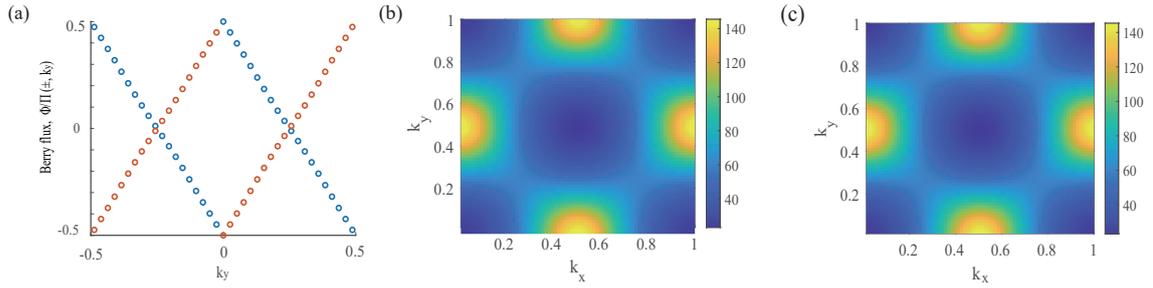

图 3-2: (a)MATBCB 2条费米面附近平带对应SU(2)威尔逊回路本征值随着$k_y$的变化，幅角约定为$[-\pi,\pi]$，可以看到绕数的本征值总是等大相反，当$k_y$遍历第一mBZ，绕数绕过±2圈，因此最低的2个平带具有陈数$C=\pm2$。(b)显示了基于方程 3-12得到的贝里曲率分布，实际第一mBZ的k点密度为$74\times74$(其他的强度图均采用该尺寸)。(c)显示了相应的量子度规 3-12的迹的分布。(b),(c) 两者的分布是几乎定量相同的,后面会看到定量上两者整个mBZ的累计偏差在$10^{-7}$量级。

约束 [124, 125, 119, 126, 127]。最近有理论工作指出，对于理想的$C>1$的拓扑平带波函数，可以在保证全纯性的条件下分解为$C=1$的波函数 [128, 129]，当贝里曲率具有特定动量平移对称性，这一分解能够正交进行，即分解得到的$C=1$态能被独立填充。这一理论为继续应用量子几何指示器三判据提供可能，后文将详细介绍如何将这套理论应用于TBCB系统，具体涉及贝里曲率的分布。

下面简单介绍量子几何张量(QGT)以及TBCB中贝里曲率的数值算法。类比之前黎曼几何中距离的定义，我们同样可以定义特定参数空间中2个态之间的"距离"。区别是波函数由于具有相位，因此相应的具有复结构，需要借鉴高能和微分几何中的复几何 [130, 131]，考虑如下无穷小波函数距离。

$$ds^2 = |\langle u_{\mathbf{k}}|u_{\mathbf{k}}\rangle|^2 - |\langle u_{\mathbf{k}}|u_{\mathbf{k}+d\mathbf{k}}\rangle|^2 = I - \langle u_{\mathbf{k}+d\mathbf{k}}|u_{\mathbf{k}}\rangle\langle u_{\mathbf{k}}|u_{\mathbf{k}+d\mathbf{k}}\rangle \tag{3-8}$$

假定所有k点上的布洛赫函数都正交归一，利用如下恒等式

$\partial_{k^\mu}[\langle u_\mathbf{k}|u_\mathbf{k}\rangle] = 0, \quad \partial_{k^\mu}\partial_{k^\nu}[\langle u_\mathbf{k}|u_\mathbf{k}\rangle] = 0,$ 可以推出:

$$
\begin{aligned}
\langle u_\mathbf{k}|u_{\mathbf{k}+d\mathbf{k}}\rangle &\approx I + dk^\mu\langle u_\mathbf{k}|\partial_{k^\mu}u_\mathbf{k}\rangle + \frac{1}{2}\langle u_\mathbf{k}|(dk^\mu\partial_{k^\mu})^2 u_\mathbf{k}\rangle, \\
\langle u_{\mathbf{k}+d\mathbf{k}}|u_\mathbf{k}\rangle &\approx I + dk^\mu\langle\partial_{k^\mu}u_\mathbf{k}|u_\mathbf{k}\rangle + \frac{1}{2}\langle(dk^\mu\partial_{k^\mu})^2 u_\mathbf{k}|u_\mathbf{k}\rangle,
\end{aligned}
\tag{3-9}
$$

$$\langle u_\mathbf{k}|\partial_{k^\mu}\partial_{k^\nu}u_\mathbf{k}\rangle + \langle\partial_{k^\mu}\partial_{k^\nu}u_\mathbf{k}|u_\mathbf{k}\rangle = -\langle\partial_{k^\mu}u_\mathbf{k}|\partial_{k^\nu}u_\mathbf{k}\rangle - \langle\partial_{k^\nu}u_\mathbf{k}|\partial_{k^\mu}u_\mathbf{k}\rangle.$$

由此我们可以立刻得到如下量子几何张量, 重复指标表示遍历求和:

$$
\begin{aligned}
ds^2 &\approx \eta_{\mu\nu}dk^\mu dk^\nu, \\
\eta_{\mu\nu}(\mathbf{k}) &= \langle\partial_{k^\mu}u_\mathbf{k}|(I - |u_\mathbf{k}\rangle\langle u_\mathbf{k}|)|\partial_{k^\nu}u_\mathbf{k}\rangle.
\end{aligned}
\tag{3-10}
$$

上述表达式 3-10可以直接推广到多带的情况, 即非阿贝尔的QGT, 距离的定义和相应表达式修改如下, 导出过程完全类似。

$$
\begin{aligned}
ds^2 &= |\sum_n\langle u_\mathbf{k}^a|u_\mathbf{k}^n\rangle\langle u_\mathbf{k}^n|u_\mathbf{k}^b\rangle - \sum_n\langle u_{\mathbf{k}+d\mathbf{k}}^a|u_\mathbf{k}^n\rangle\langle u_\mathbf{k}^n|u_{\mathbf{k}+d\mathbf{k}}^b\rangle|, \\
\eta_{\mu\nu}^{ab} &= \langle\partial_{k^\mu}u_\mathbf{k}^a|(I - \sum_n|u_\mathbf{k}^n\rangle\langle u_\mathbf{k}^n|)|\partial_{k^\nu}u_\mathbf{k}^b\rangle = \langle\partial_{k^\mu}u_\mathbf{k}^a|(I - P(\mathbf{k}))|\partial_{k^\nu}u_\mathbf{k}^b\rangle
\end{aligned}
\tag{3-11}
$$

其中a,b,n等为能带指标, 我们假定所有动量上不同能带之间都已经正交归一化$\sum_n\langle u_\mathbf{k}^a|u_\mathbf{k}^n\rangle\langle u_\mathbf{k}^n|u_\mathbf{k}^b\rangle = \delta^{ab}$, 关于n的求和是对我们所关注的能带子空间进行, 本章的例子是TBCB的2条最低平带, $P(\mathbf{k})$是多带子空间对应的投影算符, 我们可以看到, 定义态之间距离的QGT并不简单地是内积的海森矩阵, 而还要减去我们感兴趣的子空间的投影算符, 物理上理解, 这是由于量子态是希尔伯特空间中的射线, 同一条射线上的态是等价的, 因此我们需要减去这个冗余性[131]。量子几何张量是一个复函数, 而黎曼度规是实的, 因此还需要加以组合才能定义通常意义的黎曼度规[102]。

$$g_{\mu\nu}^{ab} = \frac{1}{2}(\eta_{\mu\nu}^{ab} + \eta_{\nu\mu}^{ab}), \quad F^{ab} = i\epsilon^{\mu\nu}\eta_{\mu\nu}^{ab}. \tag{3-12}$$

QGT的关于动量分量的对称组合或者说实部对应量子度规,也称为富比尼-斯塔迪度规(Fubini-Study metric), 一定程度可以度量波函数在参数空间中变化时保真度(fidelity)的改变, 反对称组合或虚部则是熟知的贝里曲率, 因此QGT相比贝里曲

率反映更多的信息。QGT还可以采用如下表达式 3-13，避免了有些情况下差分可能带来的误差，只用计算内积，可以证明在一阶近似下和表达式 3-11等价：

$$\eta_{\mu\nu}^{ab}(\mathbf{k}) \approx \frac{\langle u_{\mathbf{k}+dk_\mu}^a|u_{\mathbf{k}+dk_\nu}^b\rangle - \sum_n\langle u_{\mathbf{k}+dk_\mu}^a|u_{\mathbf{k}}^n\rangle\langle u_{\mathbf{k}}^n|u_{\mathbf{k}+dk_\nu}^b\rangle}{dk_\mu dk_\nu}. \tag{3-13}$$

然而数值上求解QGT需要注意的是在靠近mBZ边界附近，可能会有边界规范问题，以及贝里曲率分布的非线性特性，此时即使采用方程 3-5类似的镶嵌矩阵方法，数值上也较难避免QGT分量在边界发生突变，因此实际做法是按照威尔逊小方格(Wilson plaquette)积分映射为格点规范(LGT)问题，这种算法在2d mBZ（环面torus $T^2$，闭流形）上不需要固定规范[132]，具体如下。

$$
\begin{aligned}
&[U_r(\mathbf{k})]_{ab} = \langle u_a(\mathbf{k})|u_b(\mathbf{k}+d\mathbf{k}_r)\rangle, r = x, y.\\
&W(\mathbf{k}) = [U_y(\mathbf{k})]^{-1}[U_x(\mathbf{k}+dk_y)]^{-1}U_y(\mathbf{k}+dk_x)U_x(\mathbf{k})\\
&= \exp(-g_{k_x k_x}dk_x^2 - g_{k_y k_y}dk_y^2)\exp(iFdk_xdk_y),\\
&W = USV^\dagger, \quad \exp(iFdk_xdk_y) = UV^\dagger, \quad S = \exp(-Tr(g)dk_x^2),\\
&C = \frac{1}{2\pi}\int_{mBZ}d^2k Tr(F) = \frac{1}{2\pi i}\int_{mBZ}Tr(\ln(W(\mathbf{k})))\\
&= \frac{1}{2\pi i}\int_{mBZ}\ln(\det(W(\mathbf{k}))),\\
&F = \frac{\mathbf{Im}(\ln(W))}{dk_xdk_y}, \quad Tr(g) = -\frac{\mathbf{Re}(\ln(W))}{dk_x^2}.
\end{aligned}
\tag{3-14}
$$

利用SVD，我们可以提取出威尔逊小方格积分中的幺正的部分以及模部分，它们分别对应贝里曲率和量子度规的迹[133]，同时此处假定差分步长 $dk_x = dk_y$。可以看到这个表达式3-14能很好地描述贝里曲率以及量子度规迹分布的可能的非线性特征。在TBCB的最低两条平带的例子中 $W(\mathbf{k})$是$2\times2$矩阵，由于2条平带具有相反的陈数$C = \pm2$，因此总陈数为0，需要对每个动量上的$W(\mathbf{k})$区分本征值的符号再积分，才能分别得到$C = \pm2$的结果。图 3-2 (b),(c) 画出在第一mBZ的贝里曲率分布以及量子度规迹的分布，结果和工作[71]给出的类似。它们在定量上是几乎相同的。综上，基于QGT能得到MATBCB的FCI三判据定量如下：

1. 平带带宽 $W = 1.5 \times 10^{-3}$ meV。

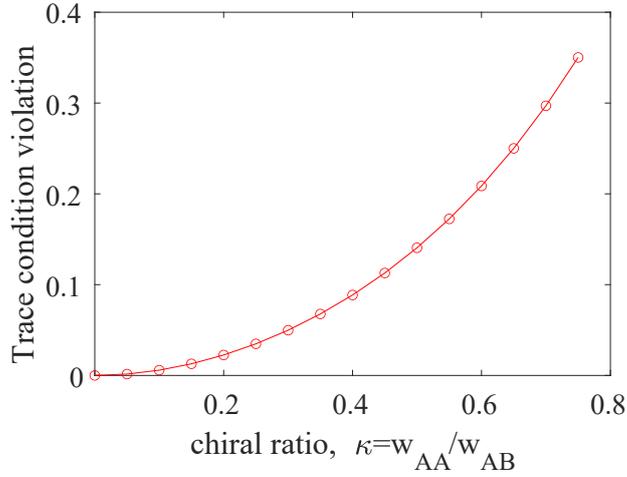

图 3-3: 如图为迹条件偏差$T[\eta]$关于手征比$\kappa$的变化，中间两条平带的单粒子能隙选取为$\Delta = 1meV$。可以看出$T[\eta]$随着$\kappa$单调递增，但在大范围的$\kappa$参数下，$T[\eta]$仍然保持较小水平，说明在单粒子水平FCI相仍然对于$\kappa$非理想性具有一定稳定性。

2. 贝里曲率标准差$\sigma[\eta] = 6.0579$。

3. 迹条件偏离$T[\eta] = 1.6357 \times 10^{-7}$。

可以看出TBCB的QGT指示器数值相较于TBG[102]具有更窄的单粒子带宽，更小的迹条件偏离，以及稍大的贝里曲率标准差。同时这个$C = 2$平带的贝里曲率具有$F(\mathbf{k}) = F(\mathbf{k} + \mathbf{Q})$,$\mathbf{Q} = (G_1/2, G_2/2)$的特征，这个条件的满足使得$C = 2$的平带可以正交分解成两个$C = 1$的平带[128, 129]，意味着两个$C = 1$的平带可以独立地被填充，$C = 1$的FCI量子几何三大判据在MATBCB系统仍然适用。基于量子几何，从单粒子水平看TBCB平带比较适合产生FCI相，从贝里曲率分布 3-2(b) 可以看到在$(0, \pi), (\pi, 0)$存在峰值，这些动量点可以构建所谓的拓扑重费米子表示，说明拓扑重费米子也有可能是FCI相的潜在竞争者[71]，相关问题将留待后续研究。同时我们可以画出迹条件偏离这个和"涡旋可附加性"[115, 116]最相关的量随手征比$\kappa$的变化如图3-3，可以看出在大范围的$\kappa$参数下，$T[\eta]$仍然保持较小水平，说明在单粒子水平FCI相仍然对于$\kappa$非理想性具有一定稳定性。

小结本节，我们从单体层面分析TBCB平带是否适合产生FCI相，论证了非相互作用的MATBCB的拓扑平带以及非平庸陈数$C = \pm 2$，定量给出了平带的量子几

何判据,更为定量的多体层面的研究将留到下节,多体版本的量子几何张量也正在发展当中 [134]。

## 3.3 FCI的多体拓扑简并基态候选者-动量空间精确对角化,磁通泵浦

上节我们从单体层面初步分析了TBCB是否是一个实现$C = 2$的FCI相的理想体系,从量子几何三大判据大概可以定性认为是可以存在FCI的。但FCI作为FQHE的格点版本,也是一种拓扑序,即强纠缠物态,从单体视角即单粒子波函数作为mBZ环面(torus)上复矢量丛的分类 [130],即陈数,量子几何等给出的信息还是相当有限。因此怎么分析FCI的多体能谱和多体波函数是一个始终无法回避的问题。另外相互作用也在其中和拓扑(纠缠)有着错综复杂的关系,关联作用也需要从多体层面进行考虑。

首先考虑相互作用的TBCB类BM模型 [96, 19],具体如下:

$$
H = \sum_{\mathbf{k},n} \epsilon_{\mathbf{k},n} C_{\mathbf{k},n}^\dagger C_{\mathbf{k},n} + \frac{1}{2A} \sum_{i,j,k,l,\mathbf{k},\mathbf{k}',\mathbf{q}} V_{i,j,m,l}(\mathbf{k}, \mathbf{k}', \mathbf{q}) \tag{3-15a}
$$
$$
(C_{\mathbf{k}+\mathbf{q},i}^\dagger C_{\mathbf{k},j} - \bar{\rho}_{\mathbf{q}=0}\delta_{i,j})(C_{\mathbf{k}'-\mathbf{q},m}^\dagger C_{\mathbf{k}',l} - \bar{\rho}_{\mathbf{q}=0}\delta_{m,l}),
$$

$$
V_{i,j,m,l}(\mathbf{k}, \mathbf{k}', \mathbf{q}) = \sum_{\mathbf{Q}} V(\mathbf{q} + \mathbf{Q}) O_{i,j}(\mathbf{k}, \mathbf{q}, \mathbf{Q}) O_{m,l}(\mathbf{k}', -\mathbf{q}, -\mathbf{Q}), \tag{3-15b}
$$

$$
O_{i,j}(\mathbf{k}, \mathbf{q}, \mathbf{Q}) = \sum_{\mathbf{G},\tau,\mu} u_{\mathbf{k}+\mathbf{q}+\mathbf{Q},i}^*(\mathbf{G}, \tau, \mu) u_{\mathbf{k},j}(\mathbf{G}, \tau, \mu) =
$$
$$
\sum_{\mathbf{G},\tau,\mu} u_{\mathbf{k}+\mathbf{q},i}^*(\mathbf{G} - \mathbf{Q}, \tau, \mu) u_{\mathbf{k},j}(\mathbf{G}, \tau, \mu). \tag{3-15c}
$$

$$
V(\mathbf{q}) = \frac{2\pi e^2}{\epsilon_r \epsilon_0} \frac{\tanh(|\mathbf{q}|d/2)}{|\mathbf{q}|}. \tag{3-15d}
$$

其中$\epsilon_{\mathbf{k},n}$是动量$\mathbf{k}$和能带带$n$标记的单粒子TBCB平带能量,即动能,通常在魔转角附近远远小于相互作用,简单起见我们只考虑单能谷情况,并且手动加一个

子格在位能$\Delta\sigma_z$在单粒子能带 3-1X点打开能隙,本文中$\Delta = 1meV$,实际可以通过加衬底实现。只考虑导带(2带模型的ED计算量太大),因此导带具有确定的陈数$C = +2$。$C_{\mathbf{k},n}$为平带费米子算符,具体形式仍然可以仿照方程 4-15通过Jordan-Wigner变换用泡利算符弦表示。$A = N_x N_y$为归一化面积,$N_x \times N_y$为相应第一mBZ的k点尺寸,实际动量离散化为$k_x = \frac{2\pi}{N_x}(0, 1 \cdots, N_x - 1)$,$k_y = \frac{2\pi}{N_y}(0, 1 \cdots, N_y - 1)$。$V_{i,j,m,l}(\mathbf{k}, \mathbf{k}', \mathbf{q})$表示裸相互作用在平带基底下的投影,投影操作实际是一个近似,没有考虑遥远的已填充能带对平带上相互作用的修正作用,更精确的方法是考虑已填充带的HFA [102, 21],或者考虑微扰重整化下的平带参数跑动 [110, 135],更精确的数值计算留待未来工作。$V(\mathbf{q})$表示裸相互作用势,通常采用屏蔽库仑势,本文采用双栅极的屏蔽库仑势,具体可以利用电磁学中的电像法导出 [136],简单起见本文不分辨层间和层内,原则上可以进行分辨 [31]。$O_{i,j}(\mathbf{k}, \mathbf{q}, \mathbf{Q})$是平带波函数的投影形状因子,对层,子格等自由度的求和也统一写在对$\mathbf{G}$的求和中,其解析形式其实有很丰富的U(4)代数结构 [19, 96],数值上由于每个k点的平带布洛赫函数具有相位不确定性,在ED模拟中需要固定规范即固定相邻k点布洛赫态的相位差,来保证布洛赫函数以及形状因子在第一mBZ的光滑性,由于平带拓扑非平庸,相位不光滑性是固有存在的,我们选择把它放在mBZ的边界上 [30]。规范固定为$O(\mathbf{k}, \mathbf{q}, \mathbf{Q}) \to |O| \exp[i \frac{2\pi C}{G_1 G_2}(q_x + Q_x)(q_y + Q_y)]$,其中$\mathbf{Q}$是二次型节点组成的大倒格点 3-1对应的动量转移,对于$\mathbf{k} + \mathbf{q}$超出第一mBZ的情况,需要拉回第一mBZ,同时在二次型节点对应的倒格点上作动量补偿,类似方程 3-5。这种规范固定实际是规范$O(\mathbf{k}, \mathbf{q}, \mathbf{Q}) \to |O| \exp[i \int_{\mathbf{k}}^{\mathbf{k}+\mathbf{q}+\mathbf{Q}} F dk_x dk_y]$一种线性化近似,即动量空间两点之间单粒子布洛赫函数的相位差由相应贝里曲率在对应张开的矩形上的积分决定,动量空间的单粒子相干性对实现FCI候选者有关键作用。

做完上述准备之后我们就可以考虑对相互作用哈密顿量 3-15a进行精确对角化(ED),由于前面提到的指数墙问题,即使对于相互作用TBCB的类BM模型,在动量空间也不能直接算很大尺寸从而渐进到热力学极限,即使直接储存态矢量也相当困难,因此首要的任务是减小矩阵维数。利用对称性可以有效降低矩阵维数,因为对称算符与哈密顿量对易,它们有共同本征矢量,对称算符的本征值也可以用来标记哈密顿量的不同参数域(sector)。这个过程相当于把整个哈密顿矩阵块对角化,每个块的维数通常会远小于总维度。常用

的守恒量一般是总粒子数(由填充数决定),总动量。考虑每个离散动量点有单个轨道,可以占据或不占据,因此哈密顿量总维度$2^N$,粒子数为$N_e$的块则相应有$C_N^{N_e} = \frac{N!}{N_e!(N-N_e)!}$维,$N = N_x N_y, \nu = N_e/N$,越接近半填充这个子空间越大。在粒子数固定的对角块可以进一步用总动量分块,虽然对于实空间ED需要先构造平移不变基底再对哈密顿量投影,但动量空间则不用,可以直接用总动量标记,类似自旋链ED用总磁化标记[137],总动量同样约束在第一mBZ,$K_i \rightarrow mod(K_i, N_i), i = x, y$,mod表示求余。

对于一般情况下的投影相互作用$V_{i,j,m,l}(\mathbf{k}, \mathbf{k'}, \mathbf{q})$,求和指标较为复杂,本文数值上采用ncon张量缩并求解器辅助完成所有相互作用张量的求解[138]。其中每个张量可以看成一个线路元件,用负整数数标记的是外腿,而正整数标记的是需要求和(缩并)的内腿,ncon只要输入张量表,对应张量腿,以及缩并顺序就能求解复杂的求和,例如求解投影相互作用张量的过程如图 3-4,能带指标$b_1 \cdots$实际都是一维,可以省略。

对于$C = 2$的理想拓扑平带,它可以视为2个$C = 1$的朗道能级的堆垛(QH bilayer stacking),进一步可以证明从Halperin (mmn)试探波函数出发,在$\nu = \frac{1}{2sC+1} = \frac{1}{4s+1}, s > 0, s \in Z$,得到平移对称不变的FCI相[128],因此TBCB最低导带在$\nu = 1/5$我们预期应该会出现FCI相,根据经验一般需要较强的相互作用才更有利于形成FCI相,因此我们采用双栅极间距$d = 3000$(单位为摩尔晶格常数),相对介电常数$\epsilon_r \approx 1$,来减小屏蔽作用。对于整数填充态,能严格证明其基态是斯莱特行列式[19, 96],然而这样的分数填充态基态一般是斯莱特行列式的叠加,存在符号问题,无法用行列式蒙特卡洛(DQMC)模拟[27],只能采用ED或DMRG,如图 3-5给出$4 \times 5$k点密度下的多体能谱,初步推断拓扑基态简并子空间的简并度(ground state degeneracy, GSD)$GSD = 10$,简并子空间与第一激发态之间的能隙大约为12meV。可以看到$K = 5, 6 \cdots 9$以及$K = 15, 16 \cdots 19$分别有5个动量上靠近的低能近简并态,它们之间大致可以通过矢量$\mathbf{G}_1/2$联系,为高陈数能带的分解提供依据[128]。但仅凭这些我们还无法断定它们就是拓扑简并基态,仍然需要看它们在绝热演化下是否稳定,是否会跑出假定存在的"简并子空间"。同时数值上2个方向的尺寸必须接近,即长宽比适配,否则并不能渐进趋向2d热力学极限,从而可能产生虚假的多体能隙[93]。

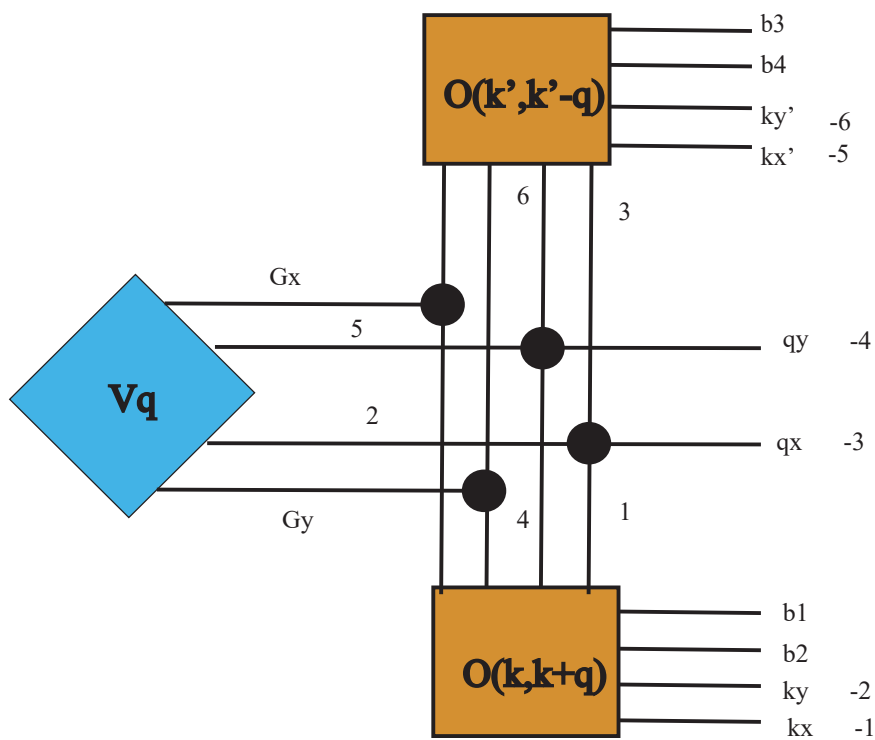

图 3-4: 如图为投影相互作用的张量图表示, 负整数数标记外腿, 即最终张量保留的指标, 正整数标记的是需要求和(缩并)的内腿, 缩并次序默认1, 2···, 可以通过先缩并维度较高的内腿来减小计算代价, 即减小中间步骤中需要储存的张量维度。张量O表示平带波函数的重叠即投影形状因子, $V_q$ 为裸势能即屏蔽库仑势, 每个黑点表示高阶克罗内克张量, 要求连接到黑点的所有腿取相同指标。

磁通泵浦或者说多体谱流动其实就是考虑某个方向, 例如 $k_y$ 方向将所有mBZ采样的动量点做连续平移, 相当于 $k_y$ 方向卷成周期边界条件(PBC)形成一个圆柱, 沿着圆柱轴方向绝热加入磁通, 同时也等价于之前提到的相位扭转边界条件[73],求解数个周期绝热磁通泵浦下的电荷变化同样可以得出分数霍尔电导。但这里我们主要目的是看磁通绝热改变边界条件下拓扑基态简并流形是否稳定, 并且如何相互缠绕(braiding)转化, 简并基态相互转化的过程在拓扑量子计算的门操作中具有潜在应用[139]。实际数值计算中逐个磁通进行分块矩阵ED, 最后再把所有块的最低本征值拼在一起, 平带波函数和投影相互作用等也会随着磁通改变你, 需要注意的是, 当 $k_y$ 截线随着磁通泵浦超出第一mBZ, 不能将 $k_y$ 再拉回第一mBZ, 正确做法是继续在第二mBZ等k点求解, 否则会因为边界规范问题导致谱流动出现不连续。

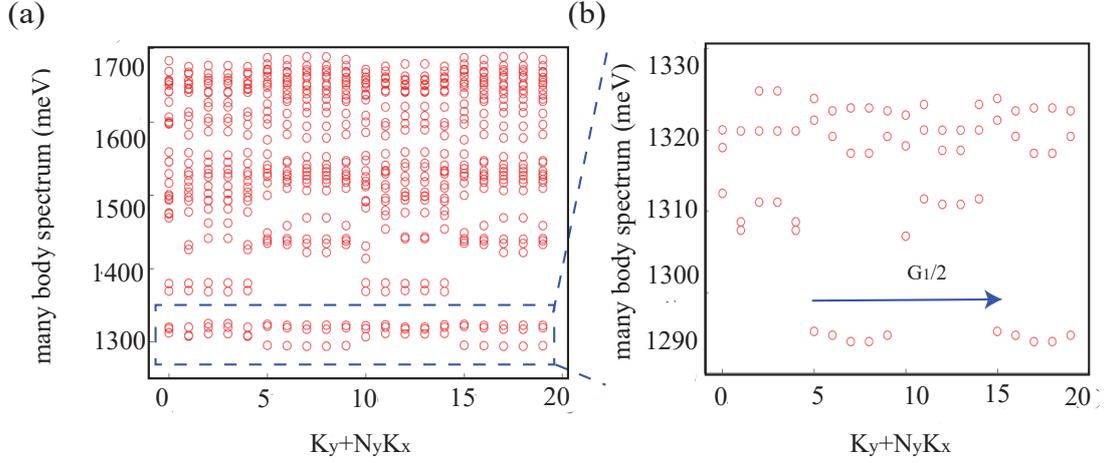

图 3-5: (a)为$4 \times 5$ k点第一mBZ离散化下MATBCB投影平带相互作用模型的ED多体能谱,二次型大倒格点尺寸$7 \times 7$,填充数$\nu = 1/5$,可以看到$GSD = 10$,简并子空间与激发态之间能隙大约为12meV。图(b)为(a)相应蓝色虚线框内低能态的放大细致图,可以看到$K = 5, 6 \cdots 9$以及$K = 15, 16 \cdots 19$的态近似通过$\mathbf{G}_1/2$联系。

图 3-6显示了$4 \times 5$ 第一mBZ k点密度下的能谱流动,在1294.5meV左右发现拓扑简并基态的可能信号,有10重近似简并度,其中每条谱流都几乎有2重简并,它们随着磁通插入相互转化而不会跑出子空间,这个简并流形的宽度约为1.5meV,简并态与最低激发态的最小能隙大约为4-5meV。简并度除了差一个因子2,和文献[128]给出的简并度解析解$\frac{(N_e+C-1)!}{N_e!(C-1)!}$一致($N_e = 4, C = 2$),这是因为TBCB具有时间反演对称性,即时间反演后映射为同一能谷而文献[128]考虑的单-双层扭转石墨烯在时间反演下会映射为另一个能谷,因此两者简并度差2倍。图中磁通范围到$\frac{2\pi}{N_y}$,因为磁通插入导致$k_y$发生轮换,然而相互作用项$V(\mathbf{q})$只和相对动量转移有关,因此更大的磁通下的谱流动只是$\Phi \in [0, \frac{2\pi}{N_y}]$内谱流的简单平移(当动量点足够多,k空间的开边界效应可以忽略)。此外,我们还可以求解$\nu = 1/5$填充MATBCB的多体陈数[140],通过沿着torus的两个方向绝热插入磁通$C_{many} = \frac{1}{2\pi} \int_0^{2\pi} d\theta_x \int_0^{2\pi} d\theta_y F(\theta_x, \theta_y)$,可以得到总多体陈数$|C_{tot}| = 4$,多体陈数$|C_{many}| = 2/5$先相当于$C_{tot}$平均在10重简并基态上,其直接与多体电导相关。具体算法是在二维的磁通参数空间$\theta_x, \theta_y$考虑一个等效的磁通晶格,按照每个磁通对所有动量截线作整体平移,最后可以依据公式 (3-14)在磁通晶格上的应用得到多体陈数,过程和单体陈数计算类似。可以验证

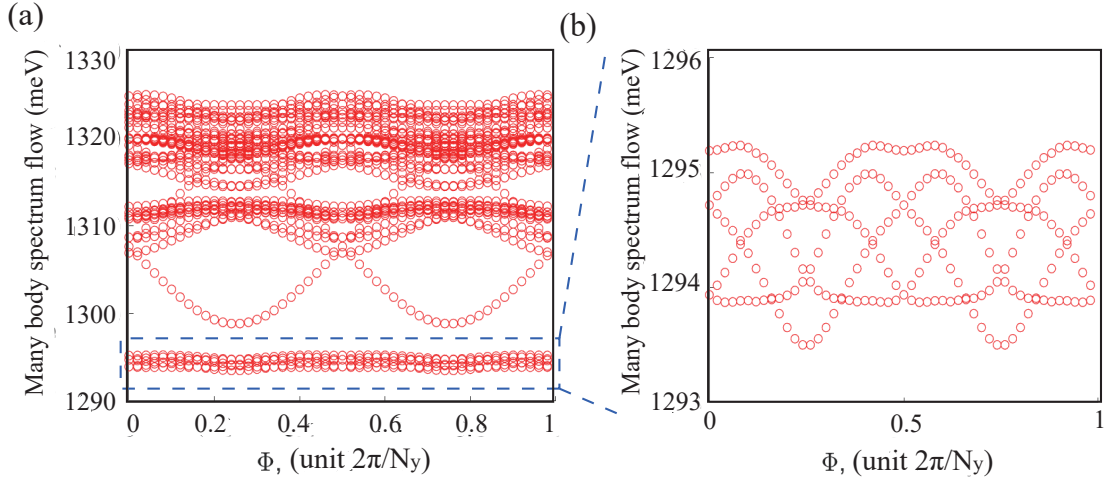

图 3-6: TBCB投影平带相互作用模型绝热插入磁通时多体能谱的谱流动。低能子空间具有10重简并,其中每条谱具有几乎严格的二重简并,呈现出来的5条谱之间略有劈裂并相互转化,说明这个简并度在绝热微扰(强度即变化率远小于谱能隙)下是稳定的。

多体陈数,单体陈数与填充数之间满足$|C_{many}| = |C_{single}|\nu$ [141]。分数的多体陈数排除了对称破缺陈绝缘体相例如量子反常霍尔晶体等的可能性,因为后者的多体陈数是整数。

同时我们还可以考虑上述$C = 2$ FCI相拓扑简并基态候选者在单粒子动量空间的分布,如图 3-7所示,10重简并基态分别用图 3-5相应的总动量标记,可以发现它们的单粒子动量分布基本一致。和$C = 1$ FCI [93]近似均匀的单粒子动量分布,FCI与CDW处于单纯竞争关系不同的是,$C = 2$ FCI单粒子动量分布会稍微集中在贝里曲率大的动量 3-2,说明$C = 2$ 的拓扑平带中,CDW和FCI很可能不再是像$C = 1$ FCI中是单纯的竞争关系,而是存在一定的协作和共存。文献 [142]也提到了类似的现象,其中单层棋盘格的FCI相可能和所谓近晶(smetic)序一定程度共存。从分数多体陈数可以看出,虽然CDW和FCI一定程度有共存可能,但主导的仍然是FCI相。

然而正如前面提到,上述结果还不足以断定FCI相的存在,公度CDW在多体能谱和磁通泵浦方面也会呈现类似的行为,FCI相作为强关联拓扑态,其本质还是需要用多体拓扑进行判断。因此下一节我们还要从多体纠缠的角度,即从量子信息层

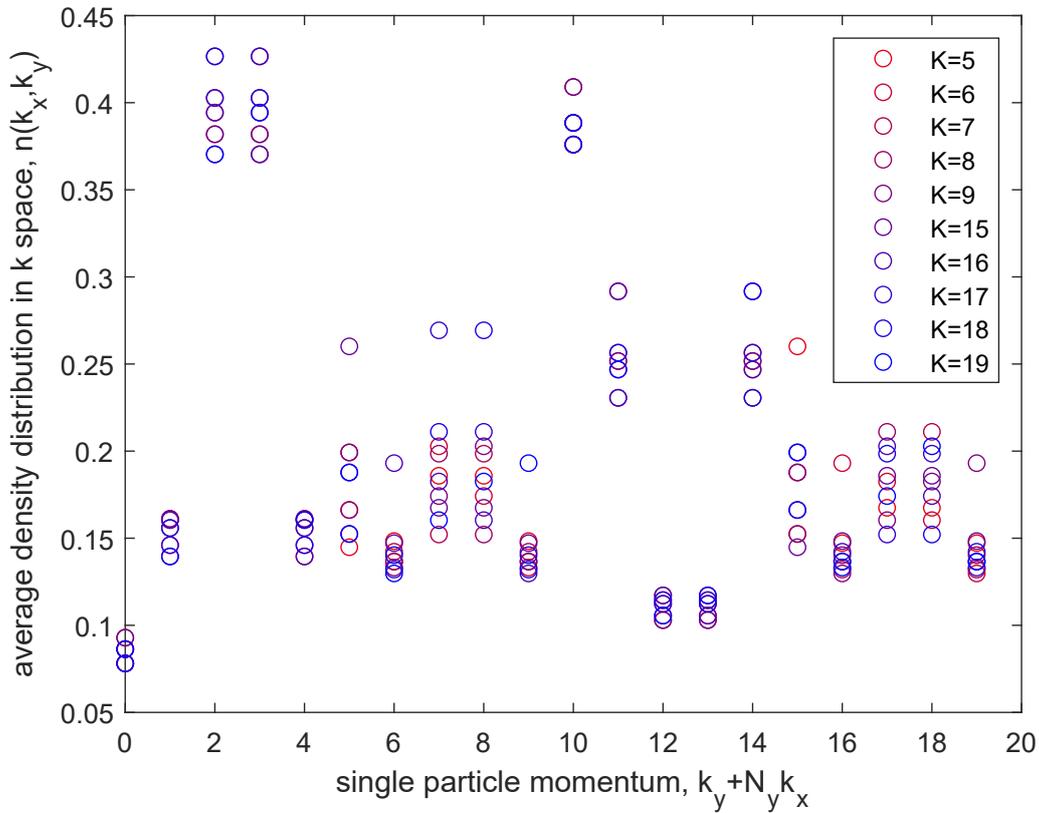

图 3-7: 如图为FCI拓扑简并基态候选者在单粒子动量空间的平均粒子数分布，对于不同的总动量，在单粒子动量上的细致分布是近似的。可以看到和$C = 1$ FCI相近似均匀分布的情况不太一样的是，$C = 2$ FCI候选者的基态，粒子在动量空间倾向于分布在贝里曲率 3-2较大的位置。说明$C = 2$ 的拓扑平带中，CDW和FCI很可能不再是像$C = 1$ FCI中是单纯的竞争关系，而是存在一定的协作和共存。从分数多体陈数可以看出，虽然CDW和FCI一定程度有共存可能，但主导的仍然是FCI相。

面看这个FCI相候选者的拓扑特征，即考虑粒子纠缠谱(PES)。

## 3.4 FCI相的关键证据-粒子纠缠谱态计数与广义泡利原理

我们先简要介绍一下PES的概念和算法。纠缠谱的概念最早由霍尔丹(haldane)等学者提出 [34]，对于纯态与波函数二分化的施密特(schmidt)分解有关，而对于更一般的混态与约化密度矩阵的本征值有关。物理上根据Li-Haldane猜想 [34]，某个抽

象空间的块内(bulk)纠缠谱和相应空间"边界"(edge)上的能谱或激发在低能下具有对应关系。常用的纠缠谱有2种,即实空间的纠缠谱和粒子数空间(Fock空间)的粒子纠缠谱 [143]。实空间纠缠谱对应边界的能谱,即拓扑物态中感兴趣的边界激发,常用于DMRG模拟 [30, 102]。遗憾的是本章研究动量空间模型,动量空间的纠缠对应k空间开边界上的激发似乎没有一个很好的物理理解,可能和无序响应有关,相关问题超出了本文的研究范围。综上所述这里我们考虑PES,福克空间的"边界激发"其实就对应准空穴激发,FQHE的典型模型态$\nu = 1/3$拉夫林(Laughlin)态,莫尔-里德(Moore-Read, MR)态准空穴激发的计数符合广义泡利原理 [93, 143],对于实际的$C = 1$ FCI系统作为实际格点体系比FQHE液体中的拉夫林态复杂一些,但我们预期低能的PES仍然和$\nu = 1/3$拉夫林态以及广义泡利原理计数一致,其他更高的非普适谱和描述准空穴的谱应该要通过一个纠缠能隙(PES gap)分开,即FCI相比拉夫林态多出来的只是PES能隙上的非普适高纠缠态。

一般波函数(纯态)以及密度矩阵(混态)的PES算法如下。首先我们预设系统有一个二分化(bipartition),记为A,B。粒子数分别为$N_A, N_B, N_A + N_B = N_e$。对于偶数粒子系统,方便起见我们选择$N_A = N_B = N_e/2$,奇数粒子系统选择$N_A = (N_e - 1)/2, N_B = (N_e + 1)/2$。对于纯态可以考虑如下施密特分解 [34]。

$$|\psi\rangle = \sum_i \exp(-\xi_i/2)||\psi_i^A\rangle|\psi_i^B\rangle. \tag{3-16}$$

其中简略写法$|\psi_i^A\rangle|\psi_i^B\rangle$表示A, B两个子系统的多体波函数的张量积,$\{\xi_i\}$即纠缠谱,要求不同纠缠本征值的基底相互正交$\langle\psi_i^A|\psi_j^A\rangle = \langle\psi_i^B|\psi_j^B\rangle = \delta_{ij}$,这可以用SVD实现,并且不要求A,B两个子系统具有相同维数。为了数值上实现SVD,需要先把$|\psi\rangle$改写成矩阵的形式,行列分别标记A,B。一般$|\psi\rangle$是一系列斯莱特行列式的叠加,对于每一项斯莱特行列式,我们需要形式化把它写成一次量子化的形式 [144] 3-17,根据线性代数中熟知的行列式拉普拉斯展开定理 [145]对斯莱特行列式展开,即$N_e$阶行列式等于所有可能的$N_A$阶子式和$N_B$阶代数余子式乘积的求和,代数余子式符号$sign(A, B)$由所选取的前$N_A$列的逆序符号决定。斯莱特行列式的系数和代数余子式符号填入$|\psi\rangle$对应行列,然后对这个矩阵做SVD即得到相应纯态的PES。$C(N_e, N_A) = (N_e)!/[N_A!(N_e - N_A)!]$表示归一化。对于玻色子系统,需要二分化一个积和式,据我们所知,就没有类似费米子的这种简单处理了 [124]。

$$|\psi_0\rangle = \begin{pmatrix} a_1 \\ a_2 \\ \cdots \\ a_n \end{pmatrix} = a_1 \begin{pmatrix} 1 \\ 0 \\ \cdots \\ 0 \end{pmatrix} + \cdots + a_n \begin{pmatrix} 0 \\ 0 \\ \cdots \\ 1 \end{pmatrix} = a_1\phi_1 + \cdots + a_n\phi_n$$

$$= \frac{1}{\sqrt{C(N_e, N_A)}}[a_1(\sum_{A,B}(-1)^{sign_1(A,B)}\phi_1^A\phi_1^B) + \cdots + a_n(\sum_{A,B}(-1)^{sign_n(A,B)}\phi_n^A\phi_n^B)].$$

$$(3\text{-}17)$$

对于拓扑多重简并基态的情况，我们需要处理如下的混态密度矩阵。

$$\rho = \frac{1}{D}\sum_{i=1}^{D}|\psi_i\rangle\langle\psi_i|. \tag{3-18}$$

其中D为低能态个数，此时不同纯态$|\psi_i\rangle$可能有不同的二分化，因此不能再用简单的SVD直接处理。但我们总可以考虑如下的$C(N, N_A) \times C(N, N_B)$的大矩阵($N = N_xN_y$，$C(N, N_A)$表示相应组合数)，即具有确定子系统粒子数的所有基底，当特定A,B子系统构型的并集和总粒子构型一致，A,B就实现了$N_e$个粒子的一种二分化。对$|\psi_i\rangle$逐项SVD然后外积得到约化密度矩阵，最后把不同纯态的密度矩阵相加，删去所有全零行列，就能得到最终的混态约化密度矩阵，它的本征值也对应了混态的PES，即：

$$|\psi_i\rangle = U_iS_iV_i^\dagger,$$

$$\rho_A = Tr_B[\rho] = \frac{1}{D}\sum_{i=1}^{D}Tr_B[|\psi_i\rangle\langle\psi_i|] = \frac{1}{D}\sum_{i=1}^{D}U_iS_i^2U_i^\dagger.$$

$$(3\text{-}19)$$

基于上述理论，可以求出MATBCB的低能态PES，混态 3-19用 3-5中最低的10个态构造。可以看到大概在$\xi = 6.5$的位置有一个明显的PES能隙，其下方的准空穴态计数对于每个子空间A的K区域(sector)计数分别为4($K^A = 0\cdots4$),5($K^A = 5\cdots19$)，总计95个态，符合自旋$\frac{1}{2}$粒子保持自旋单态的配分构型数，即$(20*19)/(2*2) = 95$，详细原因将在后文阐述。

关于如何从物理上理解PES，特别是图 3-8在纠缠能隙以下的准空穴态计数，我们还需要简单回顾一下广义泡利原理[93]。对于MATBCB动量空间的2d $N_x\times N_y$系统，总轨道数$N_x\times N_y$，我们将它们折叠成一个一维轨道例如$K_y+N_yK_x$，

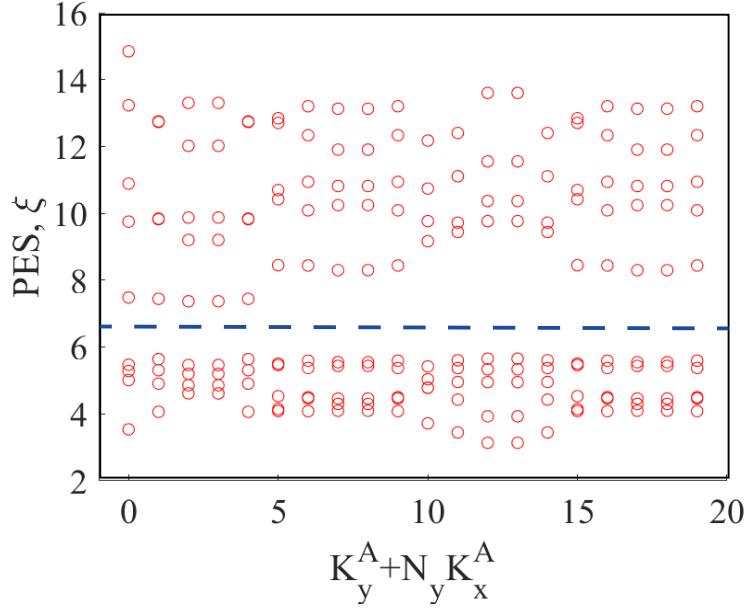

图 3-8: 如图为MATBCB最低能量的10个多体基态构造的密度矩阵对应的粒子纠缠谱(PES)。PES能隙大概在$\xi = 6.5$的位置，能隙以下有95个态，$K^A = 0 \cdots 4$有4个态，其余K sector有5个态，符合Halperin自旋单态广义泡利原理的准空穴计数。

根据周期边界条件，我们需要把这个一维轨道首尾相接形成一维环。$(k,r)_{C=1}$ 广义泡利原理$(k \leqslant r)$指的是对于陈数为1的电子，在这个折叠一维轨道中填充时，相邻的r个轨道最多填充k个电子。通常$\nu = k/r$，即k和r分别是分数填充的最简分子分母。对于$C = 1, \nu = 1/3$的情况，即满足$(1,3)_{C=1}$计数原理的构型数具有一个简单的严格解$N_x N_y \frac{(N_x N_y - 2N_A - 1)!}{(N_x N_y - 3N_A)!(N_A!)}$，详见 [93]，一般我们认为准空穴态在所有的K区域是均匀分布的，即每个K区域的计数为$\frac{(N_x N_y - 2N_A)!}{(N_x N_y - 3N_A)!(N_A!)}$，但需要注意有时候这个每个K区域的计数并不一定是整数，这是由于一些公度效应导致的，这样的准空穴构型在折叠轨道空间通常具有一个更小的周期，使得不需要平移$N_x N_y$次就能回到自身，因此当$N_e$跟其中一个方向的尺寸$N_x$或$N_y$具有非平庸最大公约数时，不同K区域的准空穴通常会进行重排并不再相等，对于$N_e$与$N_x, N_y$均互质的情况，准空穴态在所有的K区域等量分布 [93]。对于更一般的$(k,r)_{C=1}$情况，具体计数与一个多分量特殊函数有关，称为Jack多项式 [146]，具有较为复杂的形式。然而小尺寸情况下数值上对所有满足$(k,r)_{C=1}$的构型进行穷举是容易的，需要注意的是周期边界条件导致首尾的构型也必须满足广义泡利原理，这体现了准空穴计数对底流形拓扑的依

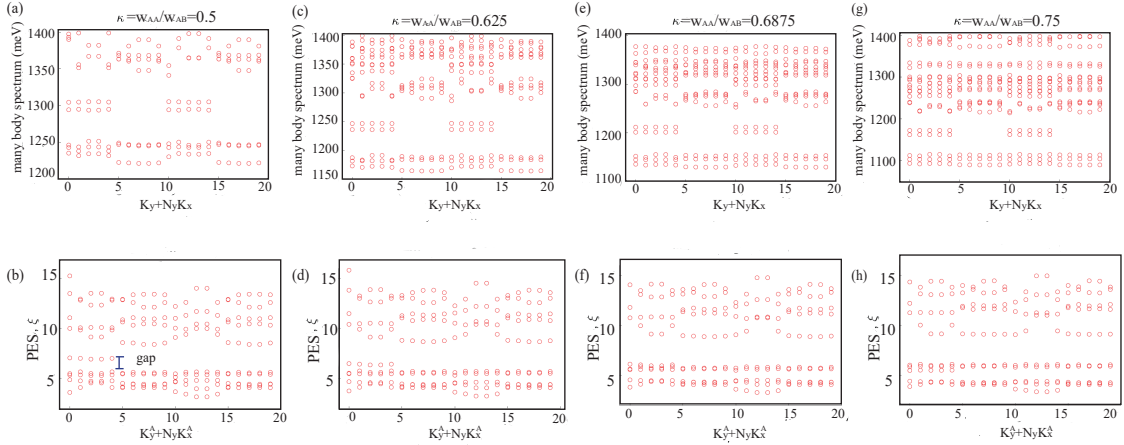

图 3-9: 如图(a),(c),(e),(g)分别是$\kappa = w_{AA}/w_{AB} = 0.5, 0.625, 0.6875, 0.75$时MATBCB的多体能谱。(b),(d),(f),(h)分别是对应手征比下的PES。从多体能谱上看随着$\kappa$增大，10重简并基态与高能态之间的能隙逐渐减小，相应的PES中纠缠能隙上方的5个态也逐渐向纠缠低能态靠近，相应的PES能隙由图(b)中的蓝色线段表示。最终这5个态会落在低能的准空穴态上。这个相变可以理解为硬核条件的破坏，当$\kappa$增大发生相变之后，具有不同赝自旋的粒子可以占据同一轨道，因此PES能隙以下的态会多出$\frac{20}{22} = 5$个。

赖。作为校正，附录 5将给出工作 [93]所计算的$C = 1$霍尔丹-哈伯德模型最低平带的PES结果，可以验证其计数符合$(1,3)_{C=1}$广义泡利原理，但需要用比原文更强的相互作用才能实现。

现在让我们回到对MATBCB中准空穴计数 3-8的理解，以及高能物理思想对这一现象的启示，虽然$C = 2$的系统可以看成2个$C = 1$子系统的堆垛。对于单粒子波函数，也已有工作证明$C > 1$的理想陈带可以全纯分解成数个$C = 1$的理想陈带(理想指满足量子几何条件)，通过将mBZ等分为C份(副本)然后对角化在不同副本之间转移的边界条件，并要求变换矩阵是动量k的全纯函数 [128]。但如果我们应用简单的$(1,5)_{C=1}$广义泡利原理，得到的准空穴应该是110，比实际通过PES看到的多了15个，因此我们应该考虑在二分化时施加更多约束。$C > 1$与$C = 1$的主要区别是准粒子相当于额外携带了一个内部的$SU(C)$自由度 [124, 147](这里C表示陈数绝对值，以下同理)，因此我们有理由要求二分化仍然需要保持这个SU(C)对称性，这个对称性类比高能术语我们也称为色(color)自

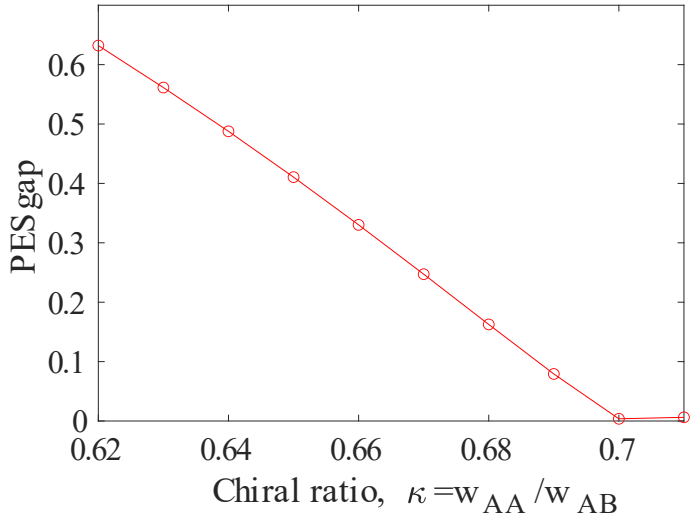

图 3-10: 如图为PES能隙随着手征比$\kappa$的变化。PES 能隙定义为$|\min(\xi(n = 5, K^A = 0 \cdots 4)) - \max(\xi(n = 4, K^A = 0 \cdots 4), \xi(n = 5, K^A = 5 \cdots 19))|$即原PES能隙上方5个态最小值与PES能隙下方0个态的最大值之差的绝对值。PES能隙随着$\kappa$减小，在$\kappa_C \approx 0.7050$时接近0，意味着可能发生从FCI相到软核FCI相(或拓扑重费米子半金属相)的相变。

由度，当对系统做二分化的时候，每个子系统要求不显示出色极化，即每个子系统都要尽可能呈现色单态,也称为Halperin color singlet。基于这一假设我们再看TBCB的$C = 2$的情况，SU(2)相当于自旋$s = 1/2$的情况，子系统A需要分成自旋单态，对于$\nu = 1/5, N_x = 4, N_y = 5, N_A = 2$，A正好分配2个赝自旋相反的粒子，自旋不同的粒子并不需要受广义泡利原理约束，仅有硬核要求即2个电子不能占据同一轨道，因此退化成排列组合问题，首先$N_x \times N_y = 20$个轨道中填2个自旋相反粒子有20*19种排列，而每次抽到特定自旋粒子的概率为1/2，因此准空穴态为$(20*19)/(2*2) = 95$，符合 3-8的PES能隙下的态计数。对于粒子数更多的情况，仍然是尽可能多的把数个自旋单态分配到A子系统，然后相同自旋之间的粒子满足广义泡利原理，而自旋不同的粒子不需要，只需要满足硬核条件。对于$C > 2$的情况，可以图像化为排布C种不同颜色的球，同样颜色的满足原理而不同颜色的不需要[124]，C种颜色的球需要尽可能数量一致来保证整体接近色单态,对于子系统A粒子数$N_A$并不能被整除C的情况，就会有比较复杂的计数，但基本原则还是让子系统

的色极化最小化，留待未来工作。

同时我们还可以结合实验需要，考虑量子几何意义上的非理想情况，例如非手征极限下，FCI相是否仍然能稳定存在，以及是否存在FCI相到其他相的相变。类似文献 [31, 30]，我们可以考虑相变参数为手征比 $\kappa = w_{AA}/w_{AB}$，即同子格层间耦合与不同子格层间耦合的比，$\kappa = 0$ 正好对应手征极限。这里采用的判据是PES中低能准空穴态与高能态之间的能隙，对于 $4 \times 5$ 动量空间，$\nu = 1/5$ 填充，不同手征比下的多体能谱以及PES的情况如图 3-9，在图 3-10 中我们画出PES能隙随手征比的减小趋势，可以得到相变点大致位置 $\kappa_C \approx 0.7050$。当 $\kappa$ 增大，原本在手征极限下PES准空穴态以上的5个态会逐渐靠近准空穴态，当 $\kappa = \kappa_c$ 这5个态和准空穴态的能隙变为0，使得准空穴计数不再严格服从自旋单态广义泡利原理。这可以理解为硬核条件的破坏，即不同自旋的粒子可以填充同一个轨道，相比原来PES能隙下的95个态，会多出 $(4 * 5)/(2 * 2) = 5$ 个态。根据文献 [148] 的TBG系统DQMC模拟，手征比增大后摩尔系统可能会出现拓扑重费米子半金属相，和这里相变后两种赝自旋的粒子占据同一轨道形成赝自旋束缚态的图像很相似，因此我们这里推测 $\kappa_C > 0.7050$ 的相可能是一个软核FCI，也可能是一个拓扑重费米子半金属，$\kappa_C > 0.7050$ 的关联相的确认有待以后进一步研究。

综合上述结果，我们认为TBCB体系很有可能在 $\nu = 1/5, \phi = 1.608^o, w_{AB} = 2.05meV$ 实现FCI相，基态具有10重简并，PES揭示的准空穴计数也服从自旋单态规则和广义泡利原理，但由于简并子空间和低激发态之间能隙较小，只占子空间宽度的2-3倍左右，即10meV的量级，因此要在MATBCB实现 $C = 2$ 的FCI相，很可能需要像TBG那样加一个弱磁场稳定其中的FCI相 [50]，相应有外加磁场的霍夫斯塔德模型研究留待未来工作。限于篇幅，更多填充数下的结果详见附录 5。同时我们注意到最近有研究组理论上提出了 $C = -2$ 的FCI也可能在 $tMoTe_2$ 中通过有效斯格明子(skyrmion)晶格的协助得以实现 [149]。

## 3.5 本章小结与展望

本章采用量子几何，动量空间精确对角化的方法研究了MATBCB的单粒子平带及其拓扑性质，量子几何条件是否适合产生FCI相，以及通过研究多体能谱，绝热磁通泵浦，PES等方面，确定了 $C = 2$ FCI相候选者的基态简并度，发现基态简并具有稳定的10重，对应的PES准空穴计数服从Halperin自旋单态广义泡利原理的一个

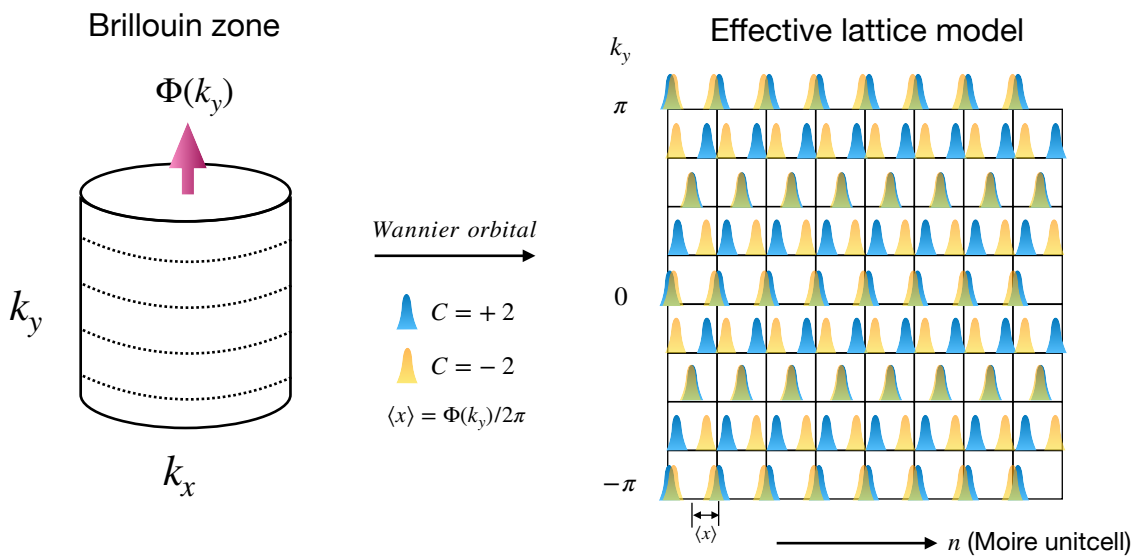

图 3-11: 如图为杂化基底 $n_x - k_y$ 下最大局域化瓦尼尔函数随着动量 $k_y$ 的变化，当 $k_y$ 变化 $2\pi$，即沿着相应圆柱轴线绝热插入 $2\pi$ 磁通，瓦尼尔中心沿着 x 方向移动 $\pm 2$ 个单胞，和之前的拓扑平带绕数一致。(该示意图为笔者合作者 Dr. Zhu 绘制)

特例。但由于简并基态与第一激发态能隙并不大，我们预期需要加一个外磁场来稳定其中的 $C = 2$ FCI 相，此时需要处理一个霍夫斯塔德模型，这是一个有趣的开放问题。除此之外，本章还有许多开放问题，例如 $\nu = k/r, (k > 2)$ 时在 TBCB 是否可能存在非阿贝尔 FCI 相，它的 PES 计数是否符合 Halperin 单态规则和广义泡利原理，需要怎么推广计数原理，特别是在不同 K 区域是否和模型态以及 $C = 1$ FCI 是否有对应等 (折叠问题) [124]，此时内部的赝自旋内部对称性可能和空间平移对称性纠缠在一起，它们的复合对称性一般是保持的，但当平移对称性破缺时，内部赝自旋对称性也可能随之破缺，此时基于 Halperin 赝自旋态的 PES 广义泡利原理也很可能不再成立，有可能要用另一套基于赝势的理论 [150] 进行解释。另一个开放问题是完整的 CDW-FCI 相图的研究，这一相图可以由转角，$w_{AA}/w_{AB}$，有效介电常数等变量共同调控。以及 FCI 冷原子光晶格量子模拟方案的实现等 [151, 152, 153]。这些问题都值得进一步研究。

从实验观点看，目前 FQAHE 的实验现象仍然有大量理论不能完全解释的地方，它们都可以作为开放问题。例如为什么需要外加电位移场来实现 FQAHE，以及实验出现了非常反直觉的有限温度现象，即当温度下降，R5G-hBN 系统中

的FQAHE会向整数量子反常霍尔效应(IQAHE)过渡。首先适当强度的电位移场有
利于在RnG-hBN这类系统形成具有较为理想的量子几何的重整化拓扑平带[110]，
其次对RnG施加hBN摩尔势之后，RnG中的电子倾向于局域化，重整化费米速度
减小，以相对论的观点看在随电子运动的坐标系感受到一个缓慢运动的电场，
相应地会感生出一个较大的有效磁场，这个磁场可能就是产生FQAHE的必要条
件之一。而关于FQAHE降低温度向IQAHE的过渡的理论解释，最近有工作表明
可能与其中的杂质无序协助效应有关[154, 155]。杂质对应的局域化轨道能量比拓扑
平带低，局域化轨道上的电子将不会对多体电导产生贡献，当温度下降，拓扑
带中的电子会向下被杂质局域化，平带上的电子减少不足以继续维持FQAHE，
因此过渡到IQAHE。然而当温度升高，杂质能级上的电子可以补偿到拓扑平带，
准空穴的构型数量对热力学熵有显著贡献，通过相应的自由能判据就可以定
义FQAHE-IQAHE过渡发生的特征温标[154]，从有限温度PES能隙的演变也可以看
出这一过渡，相应简并基态的密度矩阵需要推广到有限温度玻尔兹曼权重。相关问
题在高陈数以及非阿贝尔FCI中的推广值得进一步研究。

另一个比较具体的开放问题是能否探测TBCB中(C=2)FCI候选者的边界态，
它是否具有类似手征共形场论(chiral cft)的特征?[102]，前面提到根据Li-Haldane猜
想，探测边界态能谱需要考虑系统块体态的实空间纠缠谱，因此需要至少把一
个方向变换到实空间，然后在实空间考虑二分化和施密特分解，遗憾的是因为
涉及实空间通常屏蔽库伦作用力程很长，ED计算尺寸十分有限，有限尺寸效应
将会很大，因此ED失效。更实际的方法是采用无穷密度矩阵重整化群(infinite
dmrg, idmrg)，外推到无穷大尺寸的热力学极限，或者采用有限dmrg进行有限尺寸标
度[31]，同时idmrg采用矩阵乘积态拟设(MPS)能够很自然地求解每个价键上的纠缠
谱[30]。然而因为TBCB这类系统和TBG一样，存在脆弱拓扑障碍[62, 9]，即不可能
在同时保持所有局域对称性的情况下构造相应的2d最大局域化wannier轨道[18]，实
际dmrg在2d实空间进行大尺寸模拟也并不现实，因为y方向需要卷成圆柱形，尺寸
受到限制，仍然会存在一个关于圆柱周长的指数墙问题，通常要求关联长度小于
圆柱周长的一半dmrg才是可靠的。幸运的是，通过引入$n_x - k_y$混合基底最大局域
化瓦尼尔函数(hWS)可以一箭双雕解决上述问题，即x方向变换为实空间y方向保持
动量空间，可以分别对每个$k_y$构造1d最大局域化瓦尼尔函数[156]，并且可以证明对
于1d，最低局域化瓦尼尔函数就是投影坐标算符(电极化算符)的本征值，从而绕开

了2d的脆弱拓扑障碍。同时y方向保持动量好量子数，有利于dmrg中每个K区域实现更大的MPS价键维数(bond dimension)，即可以对称分辨地更精确描述纠缠，同时又能够比较准确描述长波低能物理[33]。$C = 2$最大局域化hWS的瓦尼尔中心随着$k_y$的变化示意图如下 3-11。

然而引入上述hWS的代价是x方向一般对于真实的屏蔽库伦相互作用，需要考虑比较长程的截断，就需要考虑比之前纯动量空间多很多的相互作用构型，相应的矩阵乘积算符(MPO)将会具有很大的维数[31, 30]。需要基于控制论以及MPO正则化SVD截断进行压缩[30, 157]。对于$n_x - k_y$基底下的dmrg模拟，已经有一些张量网库能直接实现例如TenPy[33]。然而遗憾的是MPO的压缩模块尚未集成其中，限于笔者水平，尚不能将这种方法实现在TBCB的dmrg模拟。然而对于MPO的高效生成，可以考虑类比高能物理中求解相互作用关联函数的路径积分方法，可以构造类似的MPO生成函数$G = \prod_i (I + \lambda_i O_i)$，类似关联函数通过对路径积分求泛函微分实现，基于MPO编码的相互作用也可以通过生成函数对连续参数$\lambda_i$的微分来系统生成。微分可以用机器学习中的自动微分技术实现。[158, 159, 160]。即基于生成函数的TBCB等摩尔系统的dmrg模拟也很有潜在价值的开放问题。

另外，量子模拟平台特别是里德伯(Rydberg)量子模拟近年来迅速发展，其中可以通过里德伯阻塞对系统局域希尔伯特空间进行约束，即阻塞半径内只能有一个激发里德伯原子，从而实现凝聚态中的奇异物态，最近也有工作在里德伯量子模拟平台提出实现玻色型FCI相的方案[161, 162]。因为里德伯阻塞和广义泡利原理存在一定的类似之处，区别在于里德伯阻塞一般发生在实空间，而广义泡利原理一般要求在动量空间的相邻r个轨道最多只能占据k个粒子，因此可以合理地猜测，如果能在动量空间构建一个有效的里德伯阵列，并依据我们预期的广义泡利原理设计相应里德伯阻塞，就有可能实现费米型FCI的里德伯量子模拟，这也是本章在量子模拟层面的重要开放问题。

# 第4章 磁无序近藤半金属$CeFe_2Al_{10}$传统有效模型与高能第一性原理研究

。

## 4.1 研究背景和意义

前两章中我们基于高能理论从演生论的角度理解凝聚态系统的低能有效理论，一个自然的问题是高能理论能否从还原论的方面理解凝聚态物理，而非从演生论角度，本章将尝试部分回答这一问题，即能否从高能物理第一性原理看凝聚态。传统的凝聚态基于密度泛函理论(DFT)的第一性原理一般只能处理弱相互作用体系，对于相互作用绝缘体，唯象上也可以用采用加哈伯德相互作用的方法处理，即$DFT+U$方法。但对于强关联金属类或半金属类材料，加U仍然不能很好地描述，因为巡游电子对关联的贡献不能再简单地由单个U描述，而是有一定频率分布的动态平均场热库(bath)。本章研究近藤半金属$CeFe_2Al_{10}$的反常磁激发，从非弹性中子散射实验(INS)并没有测出磁有序，故不能用通常的描述磁性绝缘体的自旋交换模型进行仿真，即不能简单地用自旋波图像理解，当我们对每个磁杂质格点加上磁无序约束考虑修正自旋波理论(modified spin wave theory, MSWT)，仅能解释自旋激发谱即INS的一些定性特征。唯象模型上为考虑到杂质和巡游电子的相互作用，需要进一步引入安德森杂化模型，第一性原理中实际求解类似的杂质-巡游电子相互作用问题则需要结合密度泛函和动力学平均场，即$DFT+DMFT$。通过Kohn-Sham格林函数的局域投影和杂质格林函数达到自洽来求解强关联巡游系统。但这一方法仍然存在不少困难，首先是需要一些先验参数，例如Hubbard相互作用以及Hund相互作用不能事先知晓，DFT交换关联势其实已经求出了Green函数的一部分自能，而剩余部分的自能目前还没有一个统一的一般形式，也需要唯象地给出。另一个困难就是传统DMFT求解杂质Green函数需要基于连续时间量子蒙特卡洛方法(CTQMC)求解，而大多数关联金属材料的实验都是在接近零温的极低温

进行的，CTQMC计算低温系统要付出很大的计算代价，除非采用最新发展的自然轨道重整化群方法 [163]。因此如何改进算法使得对强关联金属类化合物的模拟更为普适，具有更强的可移植性，并依赖于更少的先验参数等，这些改进仍然是一个开放问题。基于如上背景和研究需求，本章研究一种针对普适晶体关联材料的算法，即高能中的格点规范在凝聚态系统的可行性研究，在将晶体中传递库仑相互作用的电磁场当成准经典场处理后，大大简化为经典场的路径积分蒙特卡洛抽样问题，此外本章我们还提出一种抽样中固定经典电磁场规范的可行简化算法，避免像高能规范场解析理论那样引入较为复杂的鬼场(ghost field)。

本章我们需要面对的问题是解释并模拟近藤半金属$CeFe_2Al_{10}$的磁无序以及磁激发，首先我们可以从一些凝聚态有效模型图像上理解该系统，即从Hartree-Fock近似，修正自旋波理论，自旋激子图像，安德森杂质模型等引入。以及后续为了更精确地模拟磁激发，引出第一性原理方案即密度泛函+动力学平均场的必要性。但这一系统会体现凝聚态的第一性原理仍然具有一定唯象性，即有时需要为了接近实验结果进行调参，而高能第一性原理则具有更多的普适性，原则上只需要晶体结构，晶格常数，原子位置等基本信息就能计算可观测量，能够推广到一般的晶体材料的类似磁激发的系列线性响应的模拟。

## 4.2 近藤屏蔽与安德森杂化模型

本节从近藤屏蔽和安德森杂化模型图像上理解巡游电子与磁性杂质相互作用产生磁无序的现象。近藤磁序屏蔽现象广泛存在于稀土金属间化合物中，例如近几十年INS，电热输运，光学响应实验研究较多的$CeT_2Al_{10},(T = Fe, Ru, Os)$，这类化合物在低温下是否出现自发磁有序和过渡金属元素$T$有关，根据D.T.Adroja等实验学者的INS实验研究 [164, 165, 166, 167, 168]揭示了当$T = Ru, Os$，低温下存在长程磁序即自旋波激发，而$T = Fe$时呈磁无序，激发谱不能用自旋波解释。同时从输运实验 [167, 169]，光电导实验 [168]上看，$CeT_2Al_{10},(T = Fe, Ru, Os)$这类金属间化合物属于近藤半金属，具有很窄的杂化能隙，传统的DFT+DMFT可能难以精确计算这一能标。本章的主要问题是如何理解与模拟$CeT_2Al_{10}$材料在低温下仍然呈现磁无序。从巡游电子角度看这种现象是发生了所谓近藤屏蔽，即巡游电子与杂质电子形成了自旋单态，整体对外不显示磁有序，这是单纯的自旋交换模型无法解释的。为初步

简单解释这一现象,可以从微观格点模型出发,即所谓的安德森杂化模型 [170]:

$$H = \sum_{\langle ij \rangle, \sigma} (t_{ij} c_{i\sigma}^\dagger c_{j\sigma} + h.c) + U \sum_i d_{i\uparrow}^\dagger d_{i\uparrow} d_{i\downarrow}^\dagger d_{i\downarrow} +$$
$$\sum_{i,\sigma} (V c_{i\sigma}^\dagger d_{i\sigma} + h.c) + \sum_{i,\sigma} (E_d d_{i\sigma}^\dagger d_{i\sigma}). \tag{4-1}$$

其中$t_{ij}$是格点$i, j$之间巡游电子的跃迁(hopping),$U$是哈伯德相互作用,$V$是杂质电子与巡游电子的杂化强度,$c_{i\sigma}, d_{i\sigma}$分别是自旋分辨的巡游电子和杂质电子的二次量子化算符。可以把巡游杂质模型粗浅地看成2个轨道,则杂化强度其实就是2个轨道之间的跃迁强度。$E_d$是杂质电子能级,一般我们认为它在没有修正情况下没有色散即杂质平带。类似于导出自旋交换模型,如果认为U很大,把V当成微扰,基于施里弗-沃尔夫(Schrieffer-Wolf)变换,在二阶微扰下我们可以得出杂质自旋和巡游电子自旋的交换作用,即近藤(Kondo)相互作用,当巡游电子包围杂质电子自旋,且它们自旋相反,则形成近藤屏蔽(近藤自旋单态),即杂质磁有序被屏蔽。此外如果我们进一步考虑近藤相互作用的二阶过程,即把巡游电子积掉只剩下杂质自旋,可以得到等效的杂质之间的长程相互作用,即RKKY相互作用,这种相互作用可以随着距离变号,也是近藤或稀土化合物中可能出现螺旋序等复杂磁序的原因之一。令近藤相互作用为$J$,巡游电子在费米面态密度为$\rho$,温度为$T$。在典型的Doniach相图中,近藤相互作用与RKKY相互作用竞争,形成反铁磁序以及费米液体磁无序相等 [171]。

为了更好地理解杂化强度如何影响磁有序态,让我们先暂不看$CeFe_2Al_{10}$这类具体材料,而从一般安德森杂化模型出发在Hartree-Fock自洽平均场的水平去判定特定杂化强度下是否存在磁有序。为简单起见我们考虑$S = \frac{1}{2}$的情况,因方程 4-1中含有算符四次项,需要用平均场约化为二次项。通过格林函数的算符运动方程法作平均场近似截断(等价于Hartree-Fock近似)和涨落耗散定理 [170],可以得到如下杂质电子的态密度关于平均场的自洽关系,杂化导致的共振宽度以及外场和哈伯德相互作用导致的杂质能级修正具体表示如下:

$$\rho_{d\sigma}(\omega) = \frac{1}{\pi} \frac{\Gamma}{(\omega - \widetilde{E}_{d\sigma})^2 + \Gamma^2},$$
$$\Gamma(\omega) = \pi \rho^0(\omega) |V|^2, \tag{4-2}$$
$$\widetilde{E}_{d\sigma} = E_d + \sigma \mu_B h + U \langle n_{d\bar{\sigma}} \rangle.$$

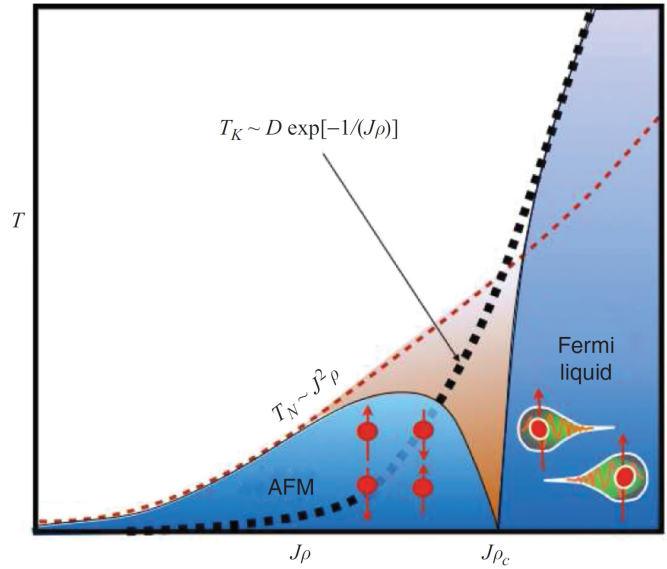

图 4-1: 如图为近藤系统的Doniach相图，节选自科尔曼《多体物理导论》[171]，图中横轴为近藤耦合与费米面态密度乘积，纵轴为温度，分为反铁磁相，临界区(非费米液体)和费米液体相(近藤屏蔽相)，可以看出当近藤耦合主导，将产生磁无序的费米液体，而RKKY耦合主导，则会产生有序相。

限于篇幅这里不给出具体推导，详见李正中《固体理论》[170]。其中$\rho^0$是平均态密度，一般可以认为是巡游电子在费米面的态密度。这里可以看到杂化强度模方正比于杂质共振态能级展宽，即由于杂化的存在使得杂质电子寿命变得有限，如果近似用费米能级处的态密度固定，则展宽$\Gamma$也是固定的，当$\Gamma \to 0$即杂化为0，杂质态密度分布$\rho_{d\sigma}$回到分立能级对应的$\delta$函数。展宽使得态密度变成了洛伦兹型。杂质能级$E_{d\sigma}$会随哈伯德相互作用和外场发生修正，$\sigma$代表两种自旋。根据平均粒子数和态密度定义，有如下自洽方程：

$$\langle n_{d\sigma} \rangle = \int_0^{E_F} \rho_{d\sigma}(\omega)d\omega = \frac{1}{\pi}\cot^{-1}(\frac{E_d + U\langle n_{d\bar{\sigma}} \rangle + \sigma\mu_B h - E_F}{\Gamma}). \tag{4-3}$$

其中$E_F$为巡游电子费米能级，具体写出来是如下自洽方程组：

$$\langle n_{d,\uparrow} \rangle = \frac{1}{\pi}\cot^{-1}(\frac{E_d + U\langle n_{d,\downarrow} \rangle + \mu_B h - E_F}{\Gamma}),$$
$$\langle n_{d,\downarrow} \rangle = \frac{1}{\pi}\cot^{-1}(\frac{E_d + U\langle n_{d,\uparrow} \rangle - \mu_B h - E_F}{\Gamma}). \tag{4-4}$$

因为这样的磁有序-无序相变是二级相变，我们需要看磁化率在哪里发散来判

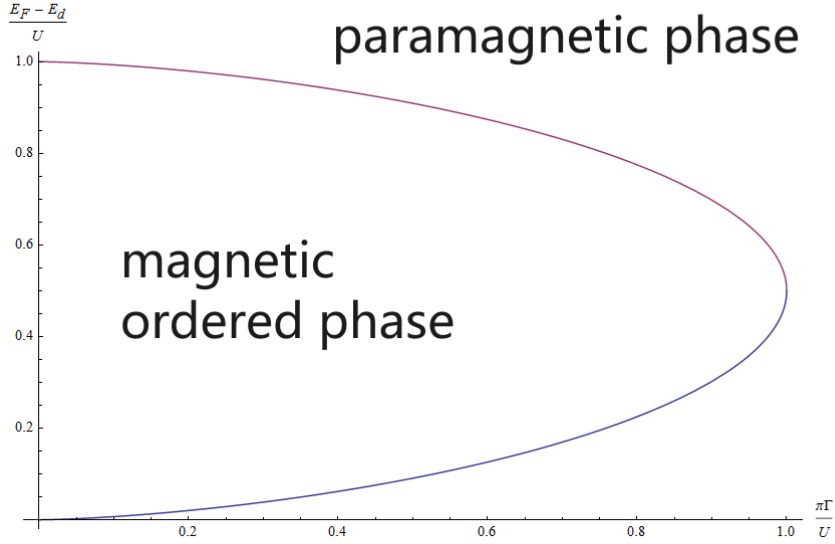

图 4-2: 如图为HFA水平下$S = \frac{1}{2}$系统安德森杂化模型的磁性相图，参数曲线 4-7包围的区域内为磁有序相，曲线外为磁无序相即顺磁相 [170]。

定相边界。假设外加弱磁场h，可以关于h展开方程 4-4，得到静态磁化率：

$$\chi_I = \lim_{h \to 0} \frac{\mu_B}{h}(\langle n_{d,\uparrow} \rangle - \langle n_{d,\downarrow} \rangle) \approx \frac{2\mu_B^2 \rho_{d\sigma}(E_F)}{1 - U\rho_{d\sigma}(E_F)}. \tag{4-5}$$

可以看到这个响应率和随机相近似(RPA)中的结果类似，$U\rho_{d\sigma}(E_F) = 1$即方程 4-5分母发散将会意味着发生相变，也是RPA失效的参数区域。假设从顺磁相出发，即$\langle n_{d,\uparrow} \rangle = \langle n_{d,\downarrow} \rangle = n_0$，联立自洽方程和磁化率发散条件，可以得到：

$$U = \frac{\pi\Gamma}{\sin^2(\pi n_0)},$$
$$n_0 = \frac{\Gamma}{U}\cot(\pi n_0) + \frac{E_F - E_d}{U}. \tag{4-6}$$

消去参数方程中的$n_0$，得到如下Hartree-Fock近似(HFA)水平的相边界：

$$\sin^{-1}(\sqrt{\frac{\pi\Gamma}{U}}) = \sqrt{\frac{\pi\Gamma}{U}(1 - \frac{\pi\Gamma}{U})} + \pi\frac{E_F - E_d}{U},$$
$$\pi - \sin^{-1}(\sqrt{\frac{\pi\Gamma}{U}}) = -\sqrt{\frac{\pi\Gamma}{U}(1 - \frac{\pi\Gamma}{U})} + \pi\frac{E_F - E_d}{U}. \tag{4-7}$$

以$\frac{\pi\Gamma}{U}$为横轴，$\frac{E_F - E_d}{U}$为纵轴可以画出HFA水平下的相边界如图 4-2：

另外我们看到当方程 4-4其他参数保持有限时$\Gamma \to \infty$，有$\langle n_{d,\uparrow} \rangle = \langle n_{d,\downarrow} \rangle = 0$。所以相边界右侧一定是磁无序相。从相图可以看出当近藤耦合$J_K \approx |V|^2/U \approx$ $\pi\Gamma/U$越大越容易形成磁无序，而$(E_F - E_d)/U$处于中等强度容易出现磁有序，类似部分填充发生劈裂的Hubbard带即化学势$\mu$正好处于Hubbard能隙中，这是符合预期的。对于$CeFe_2Al_{10}$这样含有稀土元素过渡元素的金属间化合物，磁性杂质一般是高自旋的，具有多重态。此时也有和方程 4-4类似的自洽方程，只是一般的平均场相边界就无法解析求解，需要数值迭代自洽解，但我们可以合理猜测图像上和$S = 1/2$的情况是类似的[170]。因此$CeFe_2Al_{10}$中的磁无序很可能源于强杂化。结合HFA磁性相图 4-2以及之前的有限温度Doniach相图 4-1，可以看到它们是一般的安德森杂质模型$J_K - \mu - T$三参数相图的不同二维截面，磁有序区域对应一个三维的闭区域，在这个区域外则对应磁无序相，我们预想$CeFe_2Al_{10}$在实验条件下落在了外部的磁无序相。

有了上述简单图像，下一节我们将讨论自旋关联函数和INS实验的联系，以及怎么从施加无序约束的自旋交换模型定性解释实验测出的自旋激发谱。

## 4.3 中子散射实验结果分析与修正自旋波模拟

本节简要回顾一下近ampere半金属$CeFe_2Al_{10}$的非弹性中子散射(INS)实验结果与修正自旋波理论(modified spin wave theory,MSWT)，定性解释磁激发谱的分立点状信号，谱图的特征可以用自旋激子图像简单理解。中子由于没有电荷但有磁矩，因此不会混杂电荷相关的元激发，成为探测磁性材料的磁序和磁矩相关激发的重要手段，其中弹性中子散射主要起到结构测定的作用，即中子衍射，可以显示晶体结构。而非弹性部分主要用于测量磁激发，涉及磁性元激发的能量动量转移。中子受到自旋散射的微分散射截面如下[172] [173]：

$$\frac{d^2\sigma}{dEd\Omega} = r_0^2 \frac{k'}{k} \sum_{\alpha,\beta} (\delta_{\alpha\beta} - \tilde{\kappa}_\alpha \tilde{\kappa}_\beta) \sum_{\lambda,\lambda'} p_\lambda$$

$$\sum_{l,d} \sum_{l',d'} F_d^*(\vec{\kappa}) F_{d'}(\vec{\kappa}) \exp(i\vec{\kappa} \cdot (\vec{R}_{l'd'} - \vec{R}_{ld}))$$

$$\langle\lambda|\hat{S}_{ld}^\alpha|\lambda'\rangle\langle\lambda'|\hat{S}_{ld'}^\beta|\lambda\rangle\delta(\hbar\omega + E_\lambda - E_{\lambda'}),$$

$$F_d(\vec{\kappa}) = \int d^3r \exp(i\vec{\kappa} \cdot \vec{r})\sigma_d(\vec{r}).$$

(4-8)

其中$F_d$是电子自旋在单胞中的傅里叶变换，也称为磁形状因子。$k, k'$ 分别为中子的入射和出射波矢量，$S_{ld}^\alpha$为单胞$l$格d的原子位置出自旋的$\alpha$分量。可以看出当中子出入射几何固定，自旋非极化的情况下，中子微分散射截面正比于自旋关

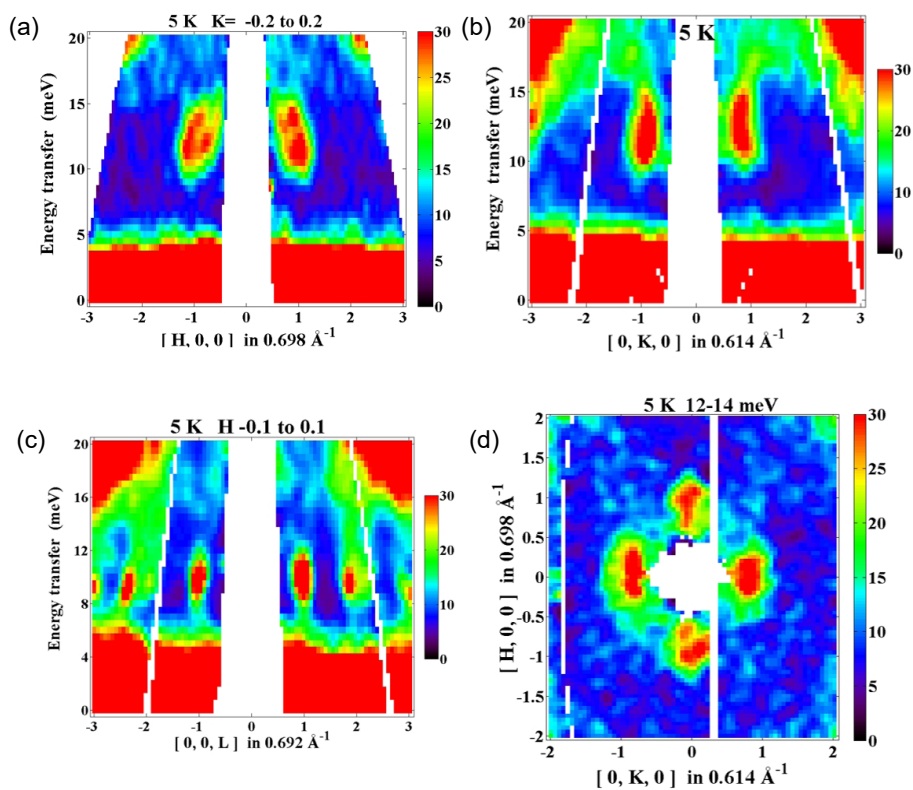

图 4-3: $CeFe_2Al_{10}$单晶的非弹性中子散射实验结果，原图节选自[167]以及D.T.Adroja教授在2017年KPS学术会议的报告[174]，图(a),(b),(c),(d)分别为关联函数$S(\mathbf{q},\omega)$四维数据的4种2d截面。该体系没有磁有序，即使在低温下只有一些点状激发，并不像自旋波那样出现系列亮线即在磁振子色散上具有谱权重。由于孪晶效应，(a)图的$S(q_x,\omega)$和(b)图的$S(q_y,\omega)$的强度分布定性上类似，而(c)图$S(q_z,\omega)$就显示出了磁激发的各向异性。(d) 表示$S(q_x,q_y,\omega)$等能面上的强度分布，从近似$C_4$分布的强度可以佐证孪晶效应的存在。实验中测$S(q_x,\omega)$, $S(q_y,\omega)$, $S(q_x,q_y)$时中子束沿着c轴，而测$S(q_z,\omega)$时中子束沿b轴。

联函数，因此以下如果没有特别说明，都默认通过计算自旋关联函数来比较实验上测出的中子散射截面，而中子散射截面的能量依赖主要体现在方程 4-8中的$\delta$函数，相应的谱权重反映了准粒子色散，这也是INS实验能间接观测自旋关联和准粒子激发能谱的原理。下面我们分析在5K低温下的$CeFe_2Al_{10}$的INS结果，相关实验由DT.Adroja实验组完成[167]。

首先需要明确的是，INS散射截面或自旋关联函数是一个四维函数$S(\vec{q},\omega)$，

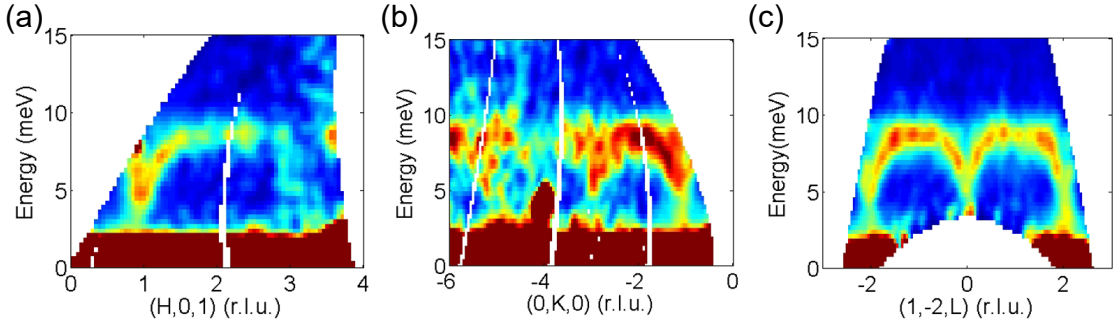

图 4-4: $CeRu_2Al_{10}$的INS实验结果，原图节选自 [167] 以及D.T.Adroja教授KPS学术会议的报告 [174]，图中显示的是典型的自旋波激发，能看出色散的大致轮廓。即在一些类似色散的线上能看到明显的谱权重。(a),(b),(c)分别表示自旋关联函数作为四维数据对应的沿着各个主晶轴的二维强度截面$S(q_x, q_y = 0, q_z = 2\pi, \omega), S(q_x = 0, q_y, q_z = 0, \omega), S(q_x = 2\pi, q_y = -4\pi, q_z, \omega)$。$q = 0$的谱权重已经扣除，因为实验关心的是动量转移非零的非弹性散射。

同时具有能动量依赖，对应中子被自旋散射时发生能动量转移，图 4-3中对应着4种截面，数据中已经扣除了声子散射等非磁部分的贡献，底下是一些本底噪声。$CeFe_2Al_{10}$只有一些分立点状激发，并不是自旋波那种线形激发，因此判断低温下$CeFe_2Al_{10}$没有磁有序而是发生了近藤屏蔽。典型的磁有序激发是类似$CeRu_2Al_{10}$的激发谱 [167]。其中的强度分布可以大致看出自旋波色散的轮廓 4-4。而$S(q_x, \omega)$，$S(q_y, \omega)$的分布定性上是类似的，说明xy平面上可能存在孪晶效应，$S(q_x, \omega)$与$S(q_z, \omega)$有较大差别，说明其中准粒子激发具有一定各向异性 [167]。

为了定性解释如上自旋关联函数，我们采用修正自旋波理论(modified spin wave theory, MSWT) [175, 176]。基本想法是仍然假定从磁有序态出发，自旋极化出现在稀土原子Ce上，不同的Ce原子之间通过RKKY相互作用耦合，在$CeFe_2Al_{10}$的单胞中有4个Ce原子，初始磁序为简单起见采用↑,↓,↓,↑，可以施加一个约束使得每个Ce原子上的平均自旋为0，相应地会引入一个拉格朗日乘子。通过Dyson-Maleev变换 [175] 将自旋算符形式化变换为玻色子算符并作平均场近似，自洽求解平均场参数之后就能按照通常的磁有序的情况求解自旋关联函数。$CeFe_2Al_{10}$四子格自旋的Dyson-Maleev(DM)变换表示如下。

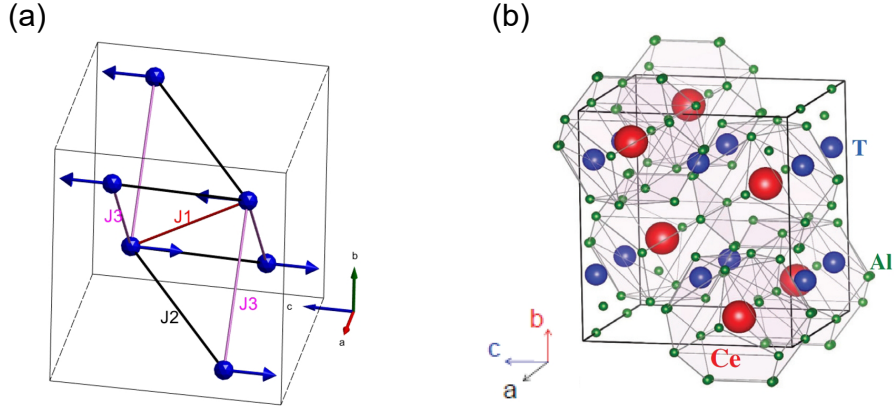

图 4-5: (a)MSWT模拟采用的初始磁序，这里假定磁序和$CeRu_2Al_{10}$相同。$J_1, J_3$为反铁磁耦合,$J_2$为铁磁耦合。磁序示意图由spinw自旋波模拟程序生成 [177],由合作者Dr. Jin提供。如图(b)为$CeT_2Al_{10}$的晶格结构，晶格常数$a = 9.0159, b = 10.2419$, $c = 9.0882$,单位埃。这类化合物是笼形化合物，原图节选自 [169]，一个单胞中有4个Ce原子用红色球表示，蓝色球表示过渡金属原子T，绿球表示Al原子，在Ce原子周围形成笼形结构。

$$S_l^- = a_l^\dagger, \quad S_l^+ = (2S - a_l^\dagger a_l)a_l, \quad S_l^z = (S - a_l^\dagger a_l),$$
$$S_p^- = -b_p^\dagger, \quad S_p^+ = -(2S - b_p^\dagger b_p)b_p, \quad S_p^z = -(S - b_p^\dagger b_p),$$
$$S_n^- = -c_n^\dagger, \quad S_n^+ = -(2S - c_n^\dagger c_n)c_n, \quad S_n^z = -(S - c_n^\dagger c_n),$$
$$S_m^- = d_m^\dagger, \quad S_m^+ = (2S - d_m^\dagger d_m)d_m, \quad S_m^z = (S - d_m^\dagger d_m).$$

$$(4-9)$$

其中$l, p, n, m$ 分别标记4个Ce原子对应的子格，这里假定平衡位置上$l, m$自旋向上，$p, n$自旋向下，值得一提的是，由于无序约束$\langle S^z \rangle = S - \langle a_i^\dagger a_i \rangle = 0$对四个子格均成立，这个约束的引入涉及到自旋的非常大改变和翻转，因此不适合用Holstein-Primakoff(HP)变换进行玻色化 [170]，因为后者适用自旋偏离平衡位置不大的情况，否则线性化会存在较大偏差，高阶展开也无法确定到哪一阶才足够准确，而HP变换的原始形式不适合作平均场近似。基于上述方案，相应的四子格自旋示意图可表示如下 4-5。

相应地海森堡哈密顿量具有如下形式:

$$H = J_1 \sum_{\langle l,p \rangle} (S_l \cdot S_p + S_n \cdot S_m) + J_2 \sum_{\langle n,p \rangle} (S_n \cdot S_p + S_l \cdot S_m) + J_3 \sum_{\langle l,n \rangle} (S_l \cdot S_n + S_p \cdot S_m)$$
(4-10)

我们将自旋算符代入DM变换 4-9以及如下的形式波戈留波夫(Bogoliubov)变换,可以得到哈密顿量的玻色子算符表示。

$$\sqrt{2}\hat{a}_{\vec{k}} = \cosh\theta_{\vec{k}}\hat{\alpha}_{\vec{k}} + \sinh\theta_{\vec{k}}\hat{\beta}_{\vec{k}}^\dagger + \cosh\phi_{\vec{k}}\hat{\gamma}_{\vec{k}} + \sinh\phi_{\vec{k}}\hat{\delta}_{\vec{k}}^\dagger$$

$$\sqrt{2}\hat{b}_{\vec{k}} = \sinh\theta_{\vec{k}}\hat{\alpha}_{\vec{k}}^\dagger + \cosh\theta_{\vec{k}}\hat{\beta}_{\vec{k}} + \sinh\phi_{\vec{k}}\hat{\gamma}_{\vec{k}}^\dagger + \cosh\phi_{\vec{k}}\hat{\delta}_{\vec{k}}$$

$$\sqrt{2}\hat{c}_{\vec{k}} = \sinh\theta_{\vec{k}}\hat{\alpha}_{\vec{k}}^\dagger + \cosh\theta_{\vec{k}}\hat{\beta}_{\vec{k}} - \sinh\phi_{\vec{k}}\hat{\gamma}_{\vec{k}}^\dagger - \cosh\phi_{\vec{k}}\hat{\delta}_{\vec{k}}$$

$$\sqrt{2}\hat{d}_{\vec{k}} = \cosh\theta_{\vec{k}}\hat{\alpha}_{\vec{k}} + \sinh\theta_{\vec{k}}\hat{\beta}_{\vec{k}}^\dagger - \cosh\phi_{\vec{k}}\hat{\gamma}_{\vec{k}} - \sinh\phi_{\vec{k}}\hat{\delta}_{\vec{k}}^\dagger$$
(4-11)

可以证明变换后的算符仍然满足玻色子的对易关系,即形式波戈留波夫变换保持玻色算符的对易关系。对其中的四次算符相做平均场近似,平均场自由能对方程 4-11中参数$\theta_k, \phi_k$变分极小。就能得到平均场基态能,自洽求解平均场参数直到收敛,就可以完全仿照磁有序的情况计算如下自旋关联函数[176]。

$$S_{\vec{q}}(\omega) = (2N)^{-1} \sum_{\vec{k}} [\cosh(\theta_{k+q} - \theta_k) + \cosh(\phi_{k+q} - \phi_k) + 2]\delta(\omega - \epsilon_{k+q} + \epsilon_k)(n_{k+q} + 1)n_k \sim$$

$$(2N)^{-1} \sum_{\vec{k}} [\cosh(\theta_{k+q} - \theta_k) + \cosh(\phi_{k+q} - \phi_k) + 2]\frac{\Gamma}{\pi}\frac{(n_{k+q} + 1)n_k}{(\omega - \epsilon_{k+q} + \epsilon_k)^2 + \Gamma^2}.$$
(4-12)

其中我们仍然需要引入唯象展宽$\Gamma$来近似描述$\delta$函数,这里假定这里温度足够低,有效自旋激发只发生准粒子带内跃迁,带间跃迁项可以忽略。方程 4-8中的磁形状因子我们采用如下的$Ce^{2+}$离子的唯象表达式[173]。

$$f_i(Q) = A_i \exp(-a_i s^2) + B_i \exp(-b_i s^2) + C_i \exp(-c_i s^2) + D_i,$$

$$F(Q) = \sum_i f_i(Q), \quad s = \frac{\sin(\theta_M)}{\lambda} = \frac{Q\sin(\theta_M)}{2\pi}.$$
(4-13)

相关的$Ce^{2+}$离子唯象磁形状因子参数如下表 4-1:

.

| $Ce^{2+}$ | $A_i$ | $a_i$ | $B_i$ | $b_i$ | $C_i$ | $c_i$ | $D_i$ |
|---|---|---|---|---|---|---|---|
| $j = 0$ | 0.2953 | 17.6846 | 0.2923 | 6.7329 | 0.4313 | 5.3527 | -0.0194 |
| $j = 2$ | 0.9809 | 18.0630 | 1.8413 | 7.7688 | 0.9905 | 2.8452 | 0.0120 |
| $j = 4$ | -0.6468 | 10.5331 | 0.4052 | 5.6243 | 0.3412 | 1.5346 | 0.0080 |
| $j = 6$ | -0.1212 | 7.9940 | -0.0639 | 4.0244 | 0.1519 | 1.0957 | 0.0078 |

表 4-1: 表中数据节选自数据库 [178]，对应方程 4-13中的$Ce^{2+}$离子分波形状因子参数,这里我们假定采用稀疏磁杂质近似并且忽略离子价态涨落。

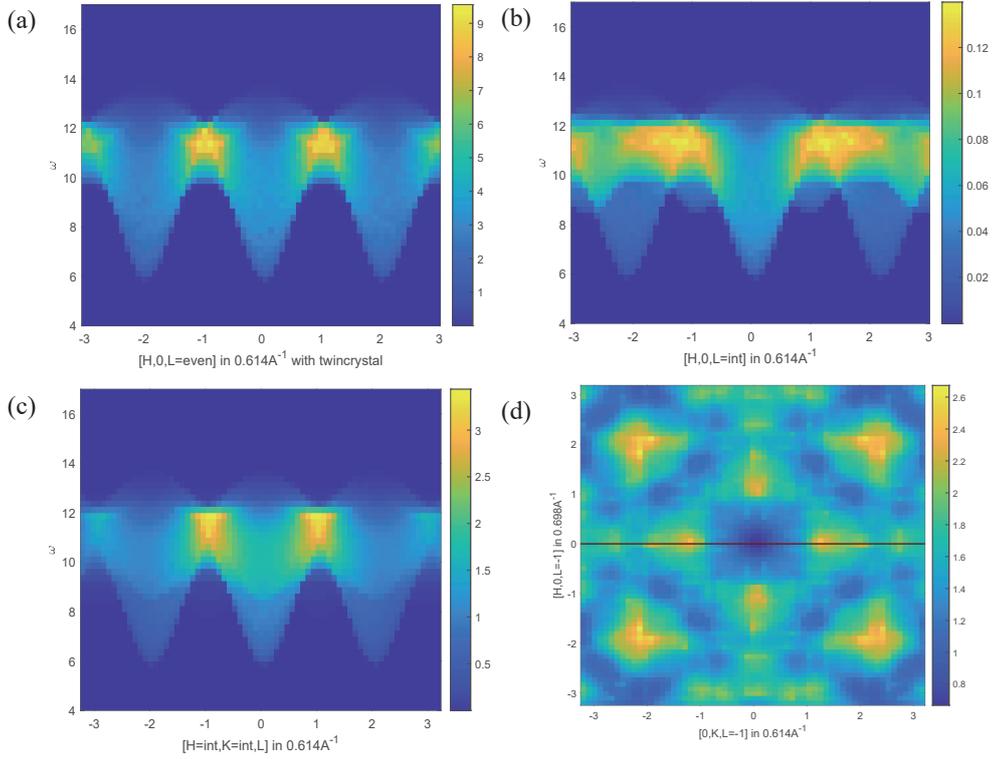

图 4-6: 修正自旋波理论给出的自旋关联函数，$J_1 = 4.1667, J_2 = -2.5, J_3 = 3.3333$，单位meV。(a)表示$S(q_x, \omega)$，并对所有$q_z = 2\pi n, n \in even$求和。对整数$q_z$求和的依据是实验中测量$S(q_x, \omega), S(q_y, \omega)$时中子束沿着c轴即z方向，需要考虑z方向所有截面的总贡献。(b)则表示$S(q_x, \omega)$，对所有$q_z = 2\pi n, n \in Z$求和，相比(a)得到的自旋激子类信号有所展宽。因为假定存在孪晶效应，因此$S(q_y, \omega)$和$S(q_x, \omega)$分布一致。图(c)显示$S(q_z, \omega)$结果，在$L = \pm 2$处相比实验有一定谱权重丢失，没有很好地体现各向异性。图(d)显示$S(q_x, q_y, \omega = const)$，呈现出近似$C_4$对称的强度分布，即xy平面内存在孪晶效应。所有频率(能量)单位均为meV，结果大体和INS实验定性吻合。

在上述近似之后，下面我们给出基于MSWT的自旋关联函数的不同截面图如图 4-6所示，可以看出除了$S(q_z, \omega)$之外，其他截面基本与实验结果符合 4-3。有关$S(q_z, \omega)$的解释有待未来进一步研究，$S(q_z, \omega)$在另一组磁交换参数下能较好符合实验，结果如 4-7所示。

物理图像上，受到高能中部分子(parton)模型的启示，可以把上述磁激发看成

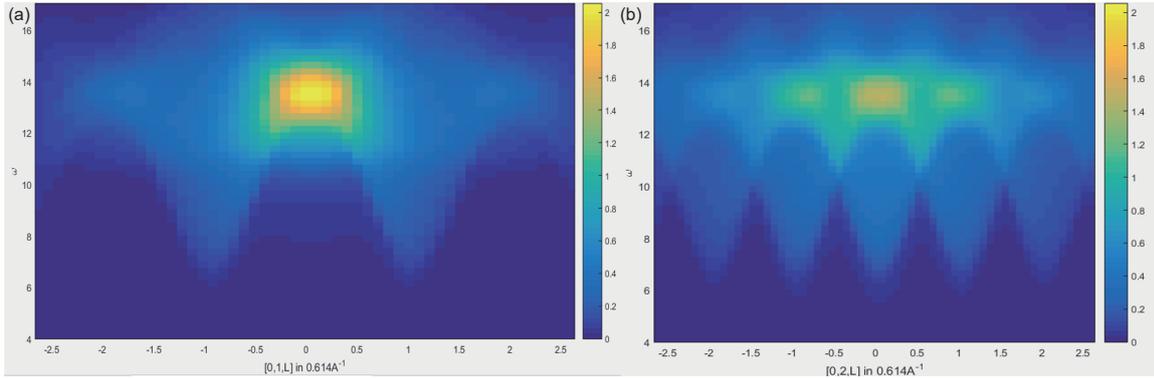

图 4-7: 另一组参数给出的$S(q_z, \omega)$, $J_1 = 20$, $J_2 = -12$, $J_3 = 5$, 单位meV, 可以看出如果不考虑孪晶, 只考虑整数$q_y$的自旋关联总贡献, 能更好地符合实验 4-3的结果。 （关于图片质量详见最后的特别声明）

自旋激子的信号 [179, 180, 181], 如图 4-8。巡游电子和杂质局域电子通过部分子规范场(部分子在这里看作一种规范玻色子) [182, 183]束缚在一起形成激子类似物, 称为自旋激子, 这种点状激发就是自旋激子的磁激发特征。这种部分子其实和高温超导中的投影方法类似 [170], 相当于强关联极限下为了消除双占据态引入的辅助场, 这种部分子成为自旋激子形成的胶水。这类自旋激子束缚态或者说近藤自旋单态需要特定的能量才能破坏, 破坏之后的准粒子散射满足特定的动量守恒, 因此在自旋关联函数只能呈现一些分立点状信号, 而不能呈现单个准粒子的线状连续激发。

　　本节小结, 我们通过MSWT模拟了$CeFe_2Al_{10}$中的反常磁激发, 能定性上解释实验的结果, 然而MSWT仍然需要先验假定杂质上处于磁无序状态。自旋交换参数需要唯象拟合调整, 有关$S(q_z, \omega)$结果的各向异性仍然不能很好地和其他截面的强度分布很好地在同一参数和孪晶假设下自洽化, 巡游电子以及杂化在其中扮演的角色也不能很好地体现。因此下一节我们考虑直接从安德森模型以及dmft出发去考虑$CeFe_2Al_{10}$的反常磁激发谱, 从更微观的电子格点模型考虑自旋关联函数。

## 4.4 安德森杂化模型小尺寸精确对角化求解和DFT+DMFT模拟

　　上一节我们先验地假定所有格点自旋z分量的期望为零, 然后从自旋模型的角度看磁激发谱, 本节我们将从更微观的电子格点模型 4-1出发考虑磁激发, 而非先验引入磁性无序假设。同时我们考虑直接求解多体安德森模型, 而非之前定性论证

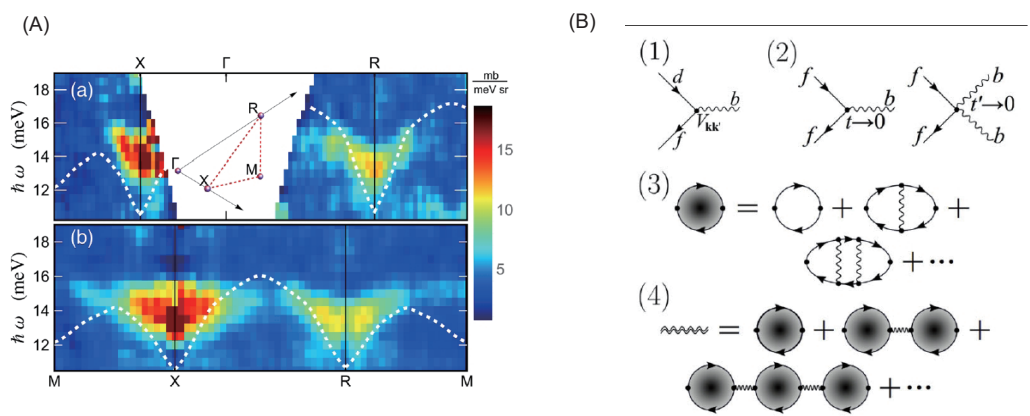

图 4-8: (A)为以典型拓扑近藤绝缘体$SmB_6$为例的自旋激子激发谱,只有点状激发而没有明显的类似自旋波谱权重。图(B)为巡游电子d和局域电子f(重费米子)在部分子b协助下形成自旋激子的物理图像示意图。原图节选自[181]。子图(1)表示自旋激子的形成,(2)则表示杂化导致的巡游电子寿命衰减,并且抑制f电子之间的散射。(3)中费曼图表示所有的费米子线满足的自能自洽关系。(4)表示玻色子线(即部分子)满足极化自洽关系。

中考虑HFA的平均场解 4-3。为了方便求解自旋关联函数,我们考虑求解虚时方向的离散无穷小转移矩阵$T = \exp(-\Delta\tau(H - \mu N))$。其中哈密顿量是我们之前提到的安德森杂化模型 4-2,需要说明的是由于哈密顿量 4-2中的每一项并不对易,非对易造成的误差是关于虚时步长$\Delta\tau$的高阶小量。而因子化之后的每个指数算符的展开可以利用费米子算符反对易性质进行化简,这一操作并不带来额外误差。虚时方向的分解称为Trotter-Suzuki分解,原则上可以通过更为对称的高阶分解减小误差,简单起见仅采用一阶分解。一阶Trotter-Suzuki分解下的转移矩阵化简如下。

$$T = \exp(-\Delta\tau(H - \mu N)) \approx \exp(-\Delta\tau H)\exp(\Delta\tau\mu N) + O(\Delta\tau^2),$$

$$\exp(-\Delta\tau H) \approx \exp(-\Delta\tau H_{hopping})\exp(-\Delta\tau H_{hubbard})\exp(-\Delta\tau H_{hybrid}) + O(\Delta\tau^2),$$

$$\exp(-\Delta\tau H_{hopping}) = \exp(-\Delta\tau \sum_{\langle ij\rangle,\sigma} t_{ij}c_{i\sigma}^\dagger c_{j\sigma})$$

$$= \prod_{\langle ij\rangle,\sigma} \exp(-\Delta\tau t_{ij}c_{i\sigma}^\dagger c_{j\sigma}) + O(\Delta\tau^2) = \prod_{\langle ij\rangle,\sigma}(1 - \Delta\tau t_{ij}c_{i\sigma}^\dagger c_{j\sigma}) + O(\Delta\tau^2),$$

$$\exp(-\Delta\tau H_{hubbard}) = \exp(-\Delta\tau U \sum_i d_{i\uparrow}^\dagger d_{i\uparrow} d_{i\downarrow}^\dagger d_{i\downarrow}) =$$

$$\prod_i \exp(-\Delta\tau U d_{i\uparrow}^\dagger d_{i\uparrow} d_{i\downarrow}^\dagger d_{i\downarrow}) = \prod_i [1 + (\exp(-\Delta\tau U) - 1)d_{i\uparrow}^\dagger d_{i\uparrow} d_{i\downarrow}^\dagger d_{i\downarrow}] + O(\Delta\tau^2),$$

$$\exp(-\Delta\tau H_{hybrid}) = \exp(-\Delta\tau \sum_{i\sigma} V c_{i\sigma}^\dagger d_{i\sigma}) = \prod_{i,\sigma}(1 - \Delta\tau V c_{i\sigma}^\dagger d_{i\sigma}) + O(\Delta\tau^2),$$

$$\exp(\Delta\tau\mu N) = \exp(\Delta\tau\mu \sum_{i\sigma} c_{i\sigma}^\dagger c_{i\sigma}) = \prod_i [1 + (\exp(\Delta\tau\mu) - 1)c_{i\sigma}^\dagger c_{i\sigma}] + O(\Delta\tau^2).$$

$$(4\text{-}14)$$

上述简化用到了费米子算符的反对易关系和费米子粒子数算符的性质，并忽略杂质能级项。下面我们需要做的是给出费米子算符的特定矩阵表示，同时我们仍然假定每个单胞有4个格点，分别对应单胞内的4个$Ce$原子，每个格点有巡游，局域2个轨道以及2个自旋极化态，而每个费米子态有空和占据2种情况。虽然这个模型是三维格点模型，但我们仍然可以以特定次序给费米子编号，即等效地排列成一维链，应用Jordan-Wigner变换4-15，可以把所有费米子算符表示为系列自旋算符的张量积表示[184]。因此求解多体费米子系统的能谱归结为求解一个非局域相互作用自旋系统的能谱，费米子反对称代数是这一非局域性的根源。

$$c_n = (-1)^{n-1}(\prod_{j=1}^{n-1}\sigma_j^z)\sigma_n^-, \quad c_n^\dagger = \sigma_n^+(-1)^{n-1}(\prod_{j=1}^{n-1}\sigma_j^z),$$

$$\sigma^z = \begin{pmatrix} 1 & 0 \\ 0 & -1 \end{pmatrix}, \sigma^- = \begin{pmatrix} 0 & 0 \\ 1 & 0 \end{pmatrix}, \sigma^+ = \begin{pmatrix} 0 & 1 \\ 0 & 0 \end{pmatrix}.$$

$$(4\text{-}15)$$

无穷小时间下的转移矩阵的谱可以用精确对角化(ED)求解，由于体系的希尔

伯特空间会随着尺寸指数增长，即存在指数墙，因此求出转移矩阵所有本征值代价是高昂的，也不实际可行。所幸的是通常人们一般只关注近零温的量子基态以及低激发态的性质，而根据基态熵面积律判据,基态和低激发态求解复杂度随着尺寸是多项式增长的(临界点会多出一个对数修正)[185]。因此为求解我们所关心的低能极限下的自旋关联函数，只需要在ED中保留转移矩阵的几个模最大本征值。即:

$$\langle \sigma_i^a(\tau = N_1\Delta\tau)\sigma_j^b(0)\rangle = \frac{Tr(T^{N_2}\sigma_i^a T^{N_1}\sigma_j^b)}{Tr(T^{N_1+N_2})},$$

$$M^{-1}TM = t, \quad T = MtM^{-1}, \tag{4-16}$$

$$\frac{Tr(T^{N_2}\sigma_i^a T^{N_1}\sigma_j^b)}{Tr(T^{N_1+N_2})} = \frac{Tr(\{M^{-1}\sigma_j^b M\}t^{N2}\{M^{-1}\sigma_i^a M\}t^{N1})}{Tr(t^{N_1+N_2})}.$$

其中$T = MtM^{-1}$表示的是转移矩阵的形式对角化，为了求解基态实时间动力学，实际采用复的时间步长$\Delta\tau = \Delta\tau_1 + i\Delta\tau_2, \Delta\tau_1 << \Delta\tau_2$，其中$\Delta\tau_1$是唯象引入的虚时间隔使得关联函数的行为仅由转移矩阵的几个模最大本征值控制。以保留转移矩阵的前$D$个最大本征值$\{\lambda_1 \cdots \lambda_D\}$为例，自旋关联函数表示如下[186]。

$$A = M^{-1}\sigma_i^a M, B = M^{-1}\sigma_j^b M$$

$$\langle \sigma_i^a(\tau = N_1\Delta\tau)\sigma_j^b(\tau = 0)\rangle \sim \frac{\sum_{a,b=1}^{D} A_{ab}B_{ba}(\lambda_a)^{N_2}(\lambda_b)^{N_1}}{\sum_{a=1}^{D}(\lambda_a)^{N_1+N_2}} \sim$$

$$\frac{A_{1,1}B_{1,1} + \sum_{a=2}^{D}(A_{1a}B_{a1}(\frac{\lambda_a}{\lambda_1})^{N_1} + A_{a1}B_{1a}(\frac{\lambda_a}{\lambda_1})^{N_2})}{1 + \sum_{a=2}^{D}(\frac{\lambda_a}{\lambda_1})^{N_1+N_2}} \sim \tag{4-17}$$

$$A_{1,1}B_{1,1} + \sum_{a=2}^{D}(A_{1a}B_{a1}(\frac{\lambda_a}{\lambda_1})^{N_1} + A_{a1}B_{1a}(\frac{\lambda_a}{\lambda_1})^{N_2}).$$

这里$(N_1 + N_2)\Delta\tau$表示总的时间截断，从关联函数的表达式可以看出，当本征值$\lambda_1 > \lambda_2 > \cdots > \lambda_D$没有简并，转移矩阵的较小本征值对关联函数的贡献随着时间是指数衰减的，因此转移矩阵的最大本征值对关联函数起着主导作用，而简并的本征值会产生相当的贡献，转移矩阵本征值简并的参数点对应着相变点。对上述自旋关联函数4-17做傅里叶变换到频率-动量空间，就可以和INS实验测出的散射截面比较。

上述讨论仍然是从比较唯象的安德森模型出发的，即我们无法先验地知道hopping参数以及Hubbard U等的相对大小，只能是依靠调参使得模拟结果尽可能接近INS实验结果。为了和$CeFe_2Al_{10}$这类实际材料联系，从凝聚态的角度看需要用到密度泛函理论和动力学平均场理论的结合(DFT+DMFT)，这种模拟方法成

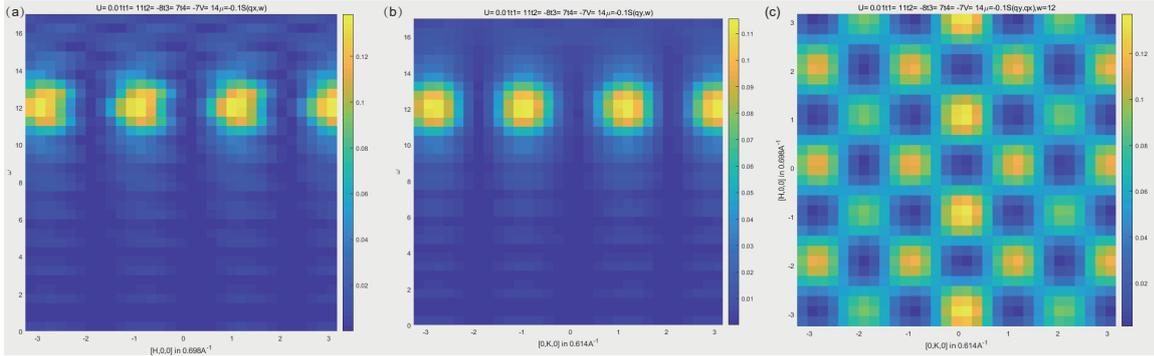

图 4-9: 转移矩阵对角化方法得出的$CeFe_2Al_{10}$安德森杂化模型的自旋关联函数。(a),(b),(c)分别为自旋动力学结构因子分量$S(q_x,\omega)$, $S(q_y,\omega)$, $S(q_x,q_y)$, 能量单位meV, 可以看出在小动量转移下, 自旋关联函数是定性符合实验的, 但对谱权重在大动量下的衰减并不能很好描述。$S(q_z,\omega)$结果与实验仍不符合, 这里没有给出。(关于图片质量详见最后的特别声明)

功地将实际材料计算第一性原理和量子杂质模型联立到一起, 可以统一地处理弱关联金属, 强关联绝缘体和强关联金属。而强关联金属系统是DFT+U无法很好处理的。注意到$CeFe_2Al_{10}$的母体是金属铝, 稀土元素Ce作为杂质原子在其中形成金属间化合物, 因此这个体系实际上应该是更为复杂的含有f电子自旋轨道耦合(SOC)的强关联金属化合物, 原则上除了哈伯德相互作用U, 还需要考虑洪特耦合$J_H$[187]的贡献。

下面我们简要回顾一下DFT+DMFT基本理论和算法框架, 并简述它如何应用在$CeFe_2Al_{10}$这个比较复杂的同时含有SOC以及巡游强关联性质的具体系统。

密度泛函理论的基本依据是基态能以及交换关联势是电子密度的泛函, 即Kohn-Sham定理[170], 根据这个定理可以首先根据元素壳层输入电子密度, 在相应的晶格边界条件下求解Kohn-Sham微分方程的本征值问题, 根据基态波函数更新电子密度。当基态收敛并且电子密度自洽, 则完成了基态能和波函数求解。不过密度泛函理论只能求弱关联材料的基态电子结构, 对于弱关联材料交换关联势近似只是局域电子密度的泛函, 称为局域密度近似(LDA),而当关联变成中等强度, 交换关联势对密度的依赖不再是局域的, 需要考虑密度梯度的修正, 即广义梯度近似(GGA),对于绝缘的中等关联系统, DFT+U也是一种解决方案。然而对于强关联系统例如高温超导体, 近藤化合物, 以及巡游电子强关联系统, 基

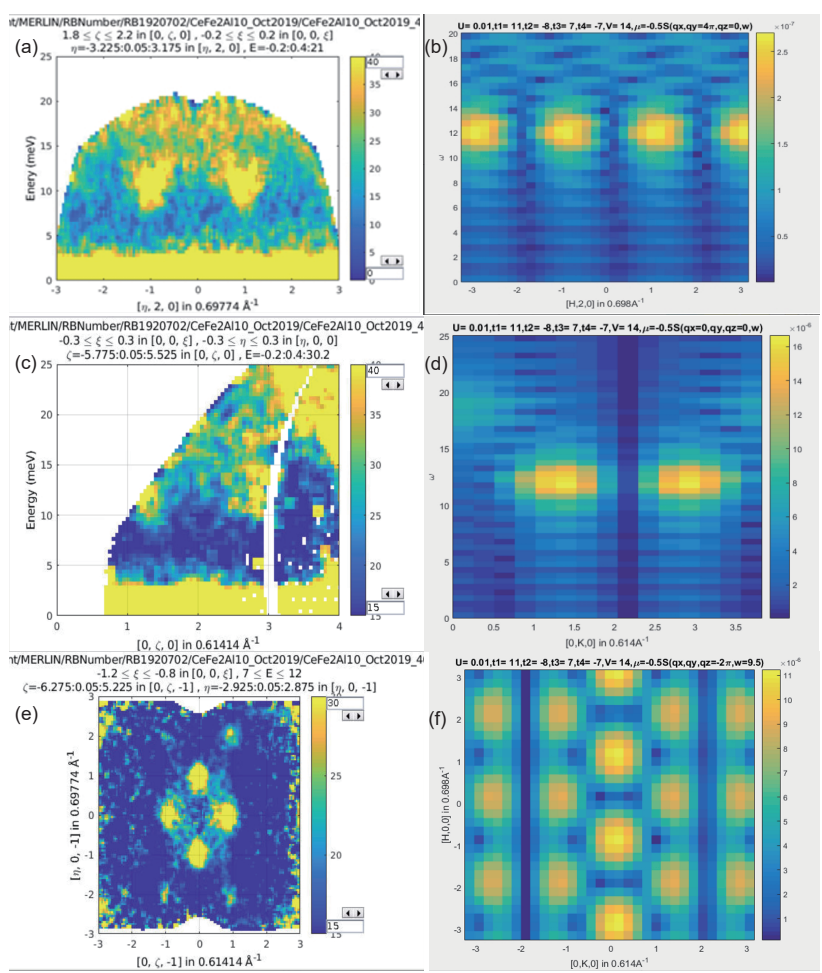

图 4-10: 转移矩阵对角化方法得出的$CeFe_2Al_{10}$安德森杂化模型的自旋关联函数与补充INS实验数据的比较，(a),(c),(e)为$S(q_x,\omega)$, $S(q_y,\omega)$, $S(q_x,q_y)$理论分布，能量单位meV。(b),(d),(f)为对应的INS实验散射强度。实验结果来自DT.Adroja研究组。整体结果定性吻合但大动量处的衰减不足。((a),(c),(e)均是实验合作方邮件中提供的原图，关于图片质量详见最后的特别声明)

于GGA的DFT以及DFT+U均不能很好处理。原因是交换关联势不足以描述系统的所有关联，例如巡游电子给杂质的频率依赖的自能修正，这是静态的交换关联势无法描述的。此时就需要把DFT求解器和DMFT的杂质求解器联合在一起，即DFT+DMFT。首先在DFT步，初始化电子密度得到交换关联势，约化为一个单体电子平均场问题，即动能加上Kohn-Sham势能在晶格周期边界条件下的本征值问题，可以得到一组布洛赫波函数。进行傅里叶变换之后能得到每个杂质原子上的

瓦尼尔函数，用于下一步的DMFT。DMFT的要点是把杂质之间的空间关联映射为虚时方向的关联，即考虑一个杂质或团簇与一个动态更新的巡游电子库(bath)进行耦合，这个电子库是一个动态更新的平均场，DFT+U也是一种特殊的平均场，但其中的关联是静态的，而DMFT中的关联可以具有频率依赖。其相应参数由上一步的DFT给出，即Kohn-Sham方程给出的巡游电子库单粒子格林函数。DMFT步收敛的判据是杂质格林函数和单粒子格林函数的局域瓦尼尔投影数值上相等。迭代过程中这2个格林函数的逆之差就是杂质自能，值得一提的是，DFT中的交换关联势已经考虑了一部分自能修正，因此在构造Kohn-Sham(KS)格林函数局域投影的过程需要扣除重复计数的部分自能即所谓重计数自能(double counting self-energy)，剩下的部分称为剩余自能。杂质格林函数的求解一般要用到精确对角化或连续时间量子蒙特卡洛(CTQMC)求解[188]，前者受限于尺寸和频率分辨率，后者计算复杂度与逆温度成正比，这也是dmft模拟近零温系统的固有困难[163]。杂质自能更新之后，同样会反过来修正电子的能量，化学势产生变化即费米分布发生变化，会生成一个新的电子密度，将其重新代入DFT流程图后可以生成新的基态能和波函数。当基态能收敛以及杂质格林函数和KS格林函数局域投影相同，则整个DFT+DMFT计算完成。流程图如下，选自[187]。

   DFT步为简明起见我们采用LDA，相应的交换关联势对局域电子密度的依赖如下公式 4-18所示[170]。这里采用的能量单位是里德伯(Ry)，$1Ry \approx 13.6eV$。

$$V_{xc} = \begin{cases} -\dfrac{0.4922a_H}{\rho^{2/3}} - \dfrac{0.0374a_H^{-1/2}\rho^{-7/6}}{(1.3334+0.8293a_H^{-1/2}\rho^{-1/6})^2} & , (\rho \le \dfrac{0.2387}{a_H^3}) \\ -\dfrac{0.4922a_H}{\rho^{2/3}} - \dfrac{0.0207}{\rho} + \dfrac{0.004}{a_H\rho^{4/3}} - \dfrac{0.0008}{a_H\rho^{4/3}}\ln(0.6204a_H^{-1}\rho^{-1/3}), & (\rho \ge \dfrac{0.2387}{a_H^3}) \end{cases}$$

$$(4\text{-}18)$$

   对LDA单粒子平均场哈密顿量应用布洛赫定理，可以求出能带和本征波函数。为了方便比较杂质格林函数，考虑把KS格林函数变换到虚时间表象。

$$G_{n,n'}(\tau) = \frac{1}{N}\delta_{n,n'}\sum_{\omega_m,k}[i\omega_m - (\epsilon_{n,k} - \mu + \Delta\Sigma)]^{-1}\exp(i\omega_m\tau),$$

$$\sum_{\omega_m=-\infty}^{\omega_m=+\infty}(i\omega_m - \epsilon)^{-1}\exp(i\omega_m\tau) = (i\pi - \beta\epsilon)^{-1}\beta\exp(i\pi\tau/\beta) \qquad (4\text{-}19)$$

$$F(1, \frac{1}{2} + \frac{i\beta\epsilon}{2\pi}, \frac{3}{2} + \frac{i\beta\epsilon}{2\pi}, \exp(i\frac{2\pi\tau}{\beta})) + h.c.$$

其中$\epsilon_{n,k}$表示费米面以下的能带，$i\omega_m = i(2m+1)\pi/\beta$表示费米子的松

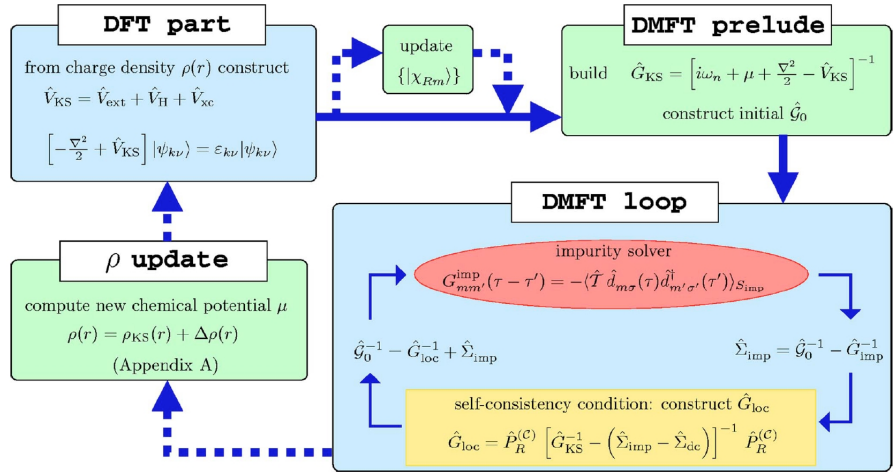

图 4-11: DFT+DMFT迭代流程图。DFT步更新瓦尼尔函数和KS格林函数，代入DMFT步求解局域格林函数，杂质格林函数由杂质求解器给出，两者的逆之差更新杂质自能以及化学势，间接更新电子密度代回DFT步完成一轮循环。DFT+DMFT算法收敛要求DFT基态能收敛以及DMFT中KS局域格林函数收敛到杂质格林函数。原图来自[187]

原(matsubara)频率，$\Delta\Sigma = \Sigma_{imp} - \Sigma_{dc}$是剩余自能，一般而言重计数自能$\Sigma_{dc}$没有一个明确的解析形式，我们一般采用如下的静态极限经验表达式$\Sigma_{dc} = U(n - 1/2) + J(n - 1)/2$。n是杂质占据数，对于Ce杂质，哈伯德相互作用和洪特相互作用分别为$U = 6eV, J = 0.7eV$。数据来自材料数据库网站[189, 190]，同时参考了$CeFe_2Al_{10}$的高温DMFT模拟的计算工作[191]的参数选取。$F(a, b, c, z)$是合流超几何函数。另一方面，对于杂质格林函数的求解，由于实验温度5K，这种低温是CTQMC难以模拟的，因此我们采用ED进行杂质格林函数的求解。与之前提到的虚时转移矩阵表达式4-14不同的是，DMFT问题中转移矩阵一般是虚时依赖的，数值上不好处理，一种简化方案是求解虚时方向的平均转移矩阵进行简化。

$$T(\tau) = \exp(-\Delta S_{imp}) = \exp(-\Delta \tau H_U)\exp(\Delta \tau \mu N)$$

$$\prod_{\tau', m\sigma, m'\sigma'} \exp(\Delta \tau \Delta \tau' d^{\dagger}_{m\sigma} \bar{T}^{\tau}[G_0^{-1}(\tau - \tau')]^{m'\sigma'}_{m\sigma} \bar{T}^{\tau'} d_{m'\sigma'}),$$

$$\bar{T} = [T((N-1)\Delta \tau) \cdots T(\Delta \tau)T(0)]^{1/N}, \quad (4\text{-}20)$$

$$S_{imp} = -\int\int_0^{\beta} d\tau d\tau' \sum_{m,m',\sigma,\sigma'} d^{\dagger}_{m\sigma}(\tau)[G_0^{-1}(\tau - \tau')]^{m'\sigma'}_{m\sigma} d_{m'\sigma'}(\tau') + \int d\tau H_U,$$

$$H_U = \sum_m U_m n_{m,\uparrow} n_{m,\downarrow} + \frac{1}{2}\sum_{mm'\sigma\sigma', m\sigma \neq m'\sigma'} J^{m'\sigma'}_{m\sigma} n_{m,\sigma} n_{m',\sigma'}.$$

其中 $S_{imp}$ 是量子杂质的作用量，包含KS格林函数赋予的动能项，它是虚时关联依赖的。洪特相互作用描述不同轨道自旋构型之间的相互作用。由于泡利不相容原理，需要要求 $m\sigma \neq m'\sigma'$，即参与相互作用的电子不能完全位于相同轨道和相同自旋上。

需要注意的是，方程 4-20和 4-14不一样，前者是一个转移矩阵的自洽关系，原则上需要所有时间片上的转移矩阵 $\{T(\tau)\}$ 共同自洽决定，这里我们做了平均场处理，实际做法是用无相互作用杂质作用量先构造初始转移矩阵，不断迭代直到每个时间片上的转移矩阵收敛。另外虽然ED杂质求解器能相比CTQMC求解器处理更低温的体系，但其能量分辨率相比后者是不足的，两者适用不同的温度，各有优劣。当DFT+DMFT收敛，为简单起见，可以基于随机相近似(RPA)求解自旋关联函数，这与我们关注的长程关联物理是一致的。RPA下的自旋关联函数(线性响应率)如下 [170, 179]。

$$\chi^0_{m\sigma,n\sigma'}(\mathbf{q}, \omega) = \frac{1}{N}\sum_k |A^{m\sigma}(\mathbf{k}+\mathbf{q})|^2 |A^{n\sigma'}(\mathbf{k})|^2 \frac{n_F(E^{m\sigma}(\mathbf{k}+\mathbf{q})) - n_F(E^{n\sigma'}(\mathbf{k}))}{\omega - E^{m\sigma}(\mathbf{k}+\mathbf{q}) + E^{n\sigma'}(\mathbf{k})},$$

$$\chi(\mathbf{q}, \omega) = (I - \tilde{U}\chi^0(\mathbf{q}, \omega))^{-1}\chi^0(\mathbf{q}, \omega),$$

$$\tilde{U} = \frac{1}{2}U_m \delta_{mn}\tau^x + \frac{1}{2}J_{m\sigma,n\sigma'}$$

$$(4\text{-}21)$$

其中 $m, n$ 为能带指标，$A^{m\sigma}(k)$ 为DFT+DMFT自洽完成后的能带m自旋$\sigma$的波函数，表示对自旋关联起贡献的能带投影。注意因为我们在DFT步对于KS方程的解并没有引入自旋极化以及自旋轨道耦合(SOC),因此林德哈（Lindhard）响应率 $\chi^0$ 在这里其实不依赖自旋指标，对2种自旋都是一样的。只有经过

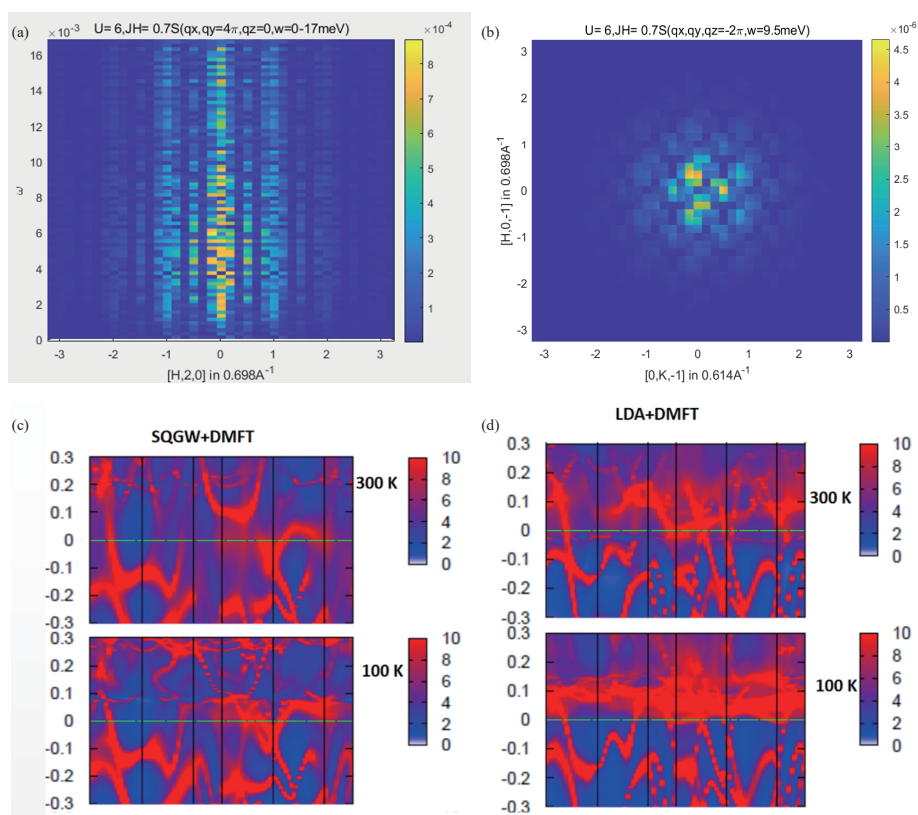

图 4-12: (a)为基于ED求解器的$CeFe_2Al_{10}$自旋关联函数$S(q_x,\omega)$DMFT模拟，能量单位meV，模拟温度5K，(b)为相应$S(q_x,q_y)$自旋关联函数在等能面$\omega=9.5meV$上的分布。然而受限于dmft中ED杂质求解的能量分辨率，结果不如前面的MSWT和安德森杂质有效模型的预测。(c),(d)分别是英国国王大学Mark van Schilfgaarde研究组[189]基于QSGW+DMFT（准粒子自洽GW），以及LDA+DMFT得到的$CeFe_2Al_{10}$系统在100K,300K下的单粒子谱权重，可以看到2种方法在费米面附近仍然有不小的准电子谱权重，即系统更偏向一个金属态而不是半金属态，而基于CTQMC的dmft计算再计算更低的温度将非常困难，这也是传统DMFT模拟此类低温系统单粒子激发谱，磁性激发的固有业界难题。

杂质的自旋相关散射$\tilde{U}$才体现出对自旋的依赖性。如图 4-12为基于ED求解器DMFT的$CeFe_2Al_{10}$自旋关联函数，动量有较好分辨率，然而能量分辨率比较低。基于自然轨道重整化群方法[163]与DMFT的结合的方案将留待后续研究，可能同时解决分辨率和低温困难。

此外，当DFT+DMFT迭代收敛，可以基于涨落耗散定理求出费米面附近的

粒子平均数，证实$CeFe_2Al_{10}$是一个半导体[170]。

$$\langle c^\dagger c\rangle_{\omega=\mu+i\eta} = -\frac{1}{\pi}\sum_{k,m}\mathbf{Im}(G_{KS}(\mathbf{k},m)) = -\frac{1}{\pi}\sum_{k,m}\mathbf{Im}\{\omega - (E_{\mathbf{k},m} - \mu + \Delta\Sigma_{\mathbf{k},m})\}^{-1}|_{\omega=\mu+i\delta}.$$

(4-22)

这里我们需要对频率加一个小的虚部来表示响应是推迟的，让态密度的$\delta$函数分布产生一个有限的展宽，同时剩余自能的虚部也会给予准粒子有限的寿命。基于方程4-22我们可以估算出$CeFe_2Al_{10}$费米面附近电子密度为$1.13\times10^9 - 4.03\times10^{10}cm^{-3}$，和硅中的载流子密度量级相近，因此验证了$CeFe_2Al_{10}$确实是半导体。此外可以对比图 4-12中英国国王大学Mark van Schilfgaarde研究组的DMFT单粒子谱权重结果，在100K-300K下SQGW+DMFT以及LDA+DMFT在费米面附近均给出金属态，说明$CeFe_2Al_{10}$需要在更温才能转变成半金属态，100K的DMFT模拟从解释5K反常磁激发的角度看可能已经不再可靠。

小结，本节基于转移矩阵方法以及DFT+DMFT从更微观的巡游-杂质相互作用模型出发模拟了$CeFe_2Al_{10}$的自旋激发，其中我们并不需要先验地假设每个杂质上是磁无序的。理论上也可以处理其他的磁无序和磁有序相。然而受限于尺寸效应和ED杂质求解器的能量分辨率，简单的转移矩阵法没有很好地复现自旋关联函数在大动量下的衰减，而DFT+DMFT虽然估算出了材料属于半导体，但自旋激发谱的能量分辨率不足，杂质集团也受到尺寸限制，从而并不能很好地符合实验测出的磁激发谱。更准确的基于CTQMC杂质求解器的$CeFe_2Al_{10}$ DFT+DMFT模拟将由Adroja，Dr. Luo等学者(本人未参与此项工作，本文结果与该工作内容无重合)给出[192]，敬请关注。

## 4.5 一般晶体材料的高能第一性原理-量子电动力学与格点规范的应用

之前的几节我们以$CeFe_2Al_{10}$为例讨论了从凝聚态角度怎么处理强关联系统，本节我们从高能物理角度看强关联晶体材料，相当于考虑一个比哈伯德模型，安德森杂质模型以及密度泛函理论更为微观的模型，即从还原论(计算)的角度看凝聚态材料多体系统，相比文小刚等学者的演生论视角是逆向的[193]。基本思想如下，和高能物理中的格点规范第一性原理类似，我们需要在晶格的周期边界条件下自洽求解真实的物质场(即电子)和规范场(电磁场)，这样的偏微分方程的本征解对应于系统作用量或者拉格朗日量变分运动方程的解。在此基础上，可以分别作微扰和非微

扰层面的修正,可以得到各种场构型涨落对特定关联函数的影响。这种方法原则上不需要一些先验唯象参数,例如原子的赝势,紧束缚跃迁积分,哈伯德和洪特参数等。通常只需要给出晶体结构和原子坐标就能够自洽进行,因此该理论上不仅限于如$CeFe_2Al_{10}$的这类特定材料,可以推广到任意结构已知的晶体材料。基于微扰的方案原则上可以得出关联函数的所有量子电动力学(QED)的单圈图修正,而基于非微扰的格点规范(LGT)方案则具有更大的可扩展性,可以处理强耦合,非微扰的情况,并且能通过抽样更多的参数来考虑更多元激发以及无序,掺杂对关联函数的修正影响,例如对原子位置构型在平衡位置附近抽样能考虑声子修正,在保证电荷守恒情况下对离子实原子序数抽样能表示掺杂,电荷无序的修正。本节提出了如上更微观层面的QED第一性原理在凝聚态系统的应用架构,并提出了当把材料中传递库伦作用的电磁场当成准经典场,关联函数的蒙特卡洛抽样会大大简化。唯一的难点是如何在抽样过程中保持电磁场这一规范场的构型仍然处于同一规范,即格点场论的规范固定问题。本节设计了一种可行固定规范算法,相比高能领域LGT的固定规范算法更为简便。本节仅提供一个较为宏观的架构,如何在$CeFe_2Al_{10}$这一具体材料中应用留待未来研究。

本节的主要动机如下,凝聚态第一性原理为模拟实验中的可观测量,仍需要一些经验参数或唯象参数,正如前面基于凝聚态框架下的$CeFe_2Al_{10}$系统模拟,我们大致能看到凝聚态框架的困境。那么一个自然的问题是能否直接从高能物理的第一性原理出发模拟凝聚态中的相对低能长波的普适现象,而不需要调太多的唯象参数呢?我们知道凝聚态多体系统的难点主要在于电子的相互作用,在传统关联电子模型中,电子相互作用的媒介一般都是作了唯象化处理,例如电磁相互作用中的光子,晶格振动中的声子等,这些玻色激发被当成准静态的背景,在这个背景下我们可以求出电子的本征能量和本征态,一般我们没有考虑新的电子分布对这些玻色型背景场有怎样的修正和反作用。而在另一个领域,即光子晶体等人工超材料,光学微腔系统中,人们主要关注这类材料的光学性质[194],一般求解的本征方程是电磁场的麦克斯韦(Maxwell)方程,参与光子之间相互作用或者说调制光子相互作用的是电介质,它的性质也是材料电子决定的,光子晶体中电子参与的媒介作用则是一定程度上做了唯象处理,没有考虑电磁场对其中电介质材料电子的反作用。那么一个很自然的想法就是能不能同时联立求解实际材料中的物质场(费米子波函数)以及传递相互作用电磁场,使得它们同时自洽收敛呢?作者认为这个框架是理论可行

的, 方案具体如下。

我们知道固体中的电子低速极限下服从薛定谔方程, 当考虑自旋的时候服从泡利方程, 而更一般的满足相对论协变性的情况是服从狄拉克方程, 前二者可以通过狄拉克方程的低速极限以及保留狄拉克旋量大分量得到 [195, 196]。考虑到 $CeFe_2Al_{10}$ 这类材料具有SOC等相对论效应, 因此我们可以一开始就从狄拉克拉氏量(密度)出发, 综合处理材料中相对论或非相对论性极限, 以及可能存在自旋极化, 自旋轨道耦合(SOC)的情况。狄拉克场和电磁场满足如下拉格朗日量:

$$L_{Dirac} = \sum_{\mathbf{k},n} \bar{\psi}_{\mathbf{k},n} \gamma^\mu (i\partial_\mu - eA_\mu)\psi_{\mathbf{k},n} - m\bar{\psi}_{\mathbf{k},n}\psi_{\mathbf{k},n} - \frac{1}{4}F_{\mu\nu}F^{\mu\nu}. \tag{4-23}$$

其中 $\gamma^\mu$ 为狄拉克矩阵, 满足伽马代数 $\{\gamma^\mu, \gamma^\nu\} = 2g^{\mu\nu}$, 可以根据实际需要选取不同的表示。$A_\mu$ 为四维矢势, 表示电磁场的势, $F_{\mu\nu} = \partial_\mu A_\nu - \partial_\nu A_\mu$ 为电磁场场强张量。$m$ 为电子的静质量, 单位已经取为 $c = \hbar = 1$, 即真空光速和普朗克常数都为1。$\psi_{\mathbf{k},n}$ 为用动量和能带标记的狄拉克旋量, $\bar{\psi} = \psi^\dagger \gamma^0$ 表示伴随旋量。在非相对论极限下可以约化为泡利模型, 相应的旋量约化为2分量旋量 [196], 具体如下。

$$H_{Pauli} = \frac{(\mathbf{p}+e\mathbf{A})^2}{2m} - e\phi + \frac{e}{4m^2}\vec{\sigma} \cdot [(\mathbf{p}+e\mathbf{A}) \times (\nabla\phi)]$$
$$-i\frac{e}{4m^2}(\nabla\phi) \cdot (\mathbf{p}+e\mathbf{A}) - \frac{e^2}{4m^2}\nabla^2\phi, \tag{4-24}$$
$$L_{Pauli} = \sum_{\mathbf{k},n} \psi^\dagger_{\mathbf{k},n}(i\partial_t - H_{Pauli})\psi_{\mathbf{k},n}.$$

从中可以看到SOC等修正项, 它们来自相对论效应, 被电子静质量压低, 标量势和矢量势的地位也不再对称, 为了形式统一简洁, 以下都采用狄拉克拉氏量进行论证。电子和电磁场所满足的运动方程由如下拉氏量的变分决定。

$$\frac{d}{dt}\left(\frac{\partial L_{Dirac}}{\partial(\partial_t\bar{\psi}_{\mathbf{k},n})}\right) - \frac{\partial L_{Dirac}}{\partial(\bar{\psi}_{\mathbf{k},n})} = 0,$$
$$\partial_\mu\left(\frac{\partial L_{Dirac}}{\partial(\partial_\mu A_\nu)}\right) - \frac{\partial L_{Dirac}}{\partial A_\nu} = 0, \tag{4-25}$$
$$\gamma^\mu(i\partial_\mu - eA_\mu)\psi_{\mathbf{k},n} - m\psi_{\mathbf{k},n} = E_{\mathbf{k},n}\psi_{\mathbf{k},n},$$
$$\partial_\mu F^{\mu\nu} = -e\sum_{\mathbf{k},n}\bar{\psi}_{\mathbf{k},n}\gamma^\nu\psi_{\mathbf{k},n} = J^\nu.$$

即得到著名的狄拉克方程和麦克斯韦方程，其中$J^\nu$为相对论协变的守恒流，即电荷密度与电流，可以由诺特定理导出[195]，我们对于实际材料第一性原理的关键就是如何求解上面2个联立偏微分方程组。对于单电子旋量场，我们采用量子化的处理，在晶格周期边界条件下(玻恩-卡曼边界条件)[170]求解能带$E_{\mathbf{k},n}$（扣除静质能），本征能量对动量的依赖同样也可以通过布洛赫定理引入。对于电磁场，不同于量子光学问题，为了简化我们把电磁场当成是半经典的对象，这是合理的，因为光子的量子效应只在单光子光源，存在外加光学谐振腔[197, 198]等光子数很少的情况才有明显体现。同时我们假定电荷电流，以及电磁势的分布在鞍点构型下和晶格是一致的，即满足和晶格相同的对称性和周期性。基于以上假设，我们可以先初始化电荷分布和电流分布(例如均匀静电荷，无电流)，求解出相应的本征能量和旋量场波函数。再进一步考虑如下的费米分布和系综平均的守恒流。

$$N_C = \sum_{\mathbf{k},n} \frac{1}{\exp[\beta(E_{\mathbf{k},n}-\mu)]+1},$$
$$J^\nu(\mathbf{r}) = -e \sum_{\mathbf{k},n} \frac{\bar{\psi}_{\mathbf{k},n}(\mathbf{r})\gamma^\nu\psi_{\mathbf{k},n}(\mathbf{r})}{\exp[\beta(E_{\mathbf{k},n}-\mu)]+1}.$$
(4-26)

其中$N_C$为一个单胞中的电子总数，由单胞内所有原子的初始壳层决定，包括非价电子。$\mu$为化学势或费米能级，其数值由该粒子数守恒条件4-26决定。守恒流实际是几乎只有费米面以下的电子才起到贡献。根据已经熟知的电动力学[136]知识，可以把经典的电磁势写成电荷电流的积分形式。

$$\phi(\mathbf{r}) = \frac{1}{4\pi\epsilon_0} \sum_{r'\in V'} \frac{\rho(\mathbf{r}')}{|\mathbf{r}-\mathbf{r}'|} + \frac{1}{4\pi\epsilon_0} \sum_{j\in V'} \frac{Z_j e}{|\mathbf{r}-\mathbf{R}_j|},$$
$$\mathbf{A}(\mathbf{r}) = \frac{\mu_0}{4\pi} \sum_{r'\in V'} \frac{\mathbf{J}(\mathbf{r}')}{|\mathbf{r}-\mathbf{r}'|}.$$
(4-27)

其中$\mathbf{r}'$标记所有离散点电子波函数的坐标，j标记所有离子实的位置，静电标量势由所有电子和离子实提供，磁矢势由电子产生的电流提供，这里忽略粒子实的运动对电流的贡献，采用了绝热近似，$V'$是一个足够大的空间使得求出的中心区电磁场收敛。需要注意并澄清的是，方程4-27其实是与晶格同周期的电磁势的一种截断近似，其特点是便于迭代，并没有破坏电子密度以及电磁场具有晶格周期的基本假设。更严格的处理应该是同时对电子部分和电磁场部分应用布洛赫定理求解一个更大的方程组，其中电子和电磁场可以分别用独立的单粒子动量进行标记，相关更

严格的处理方式将留待未来研究。综上联立方程 4-25 4-26 4-27，原则上可以通过迭代等算法使得波函数和晶格尺度的电磁场同时收敛，就能得出多体相互作用修正下的电子波函数。其实麦克斯韦方程和薛定谔方程的联立求解已经有学者做了相应工作[199]，并发展了成熟的高精度保辛差分算法，不过主要集中在超材料电磁方程，电磁多尺度物理场建模领域。而近年来发展的基于QED的DFT主要应用于量子光学，腔电动力学以及以腔为媒介的光子诱导配对的超导[197, 200, 201, 198]等领域，直接在(没有外加光场以及谐振腔的)晶体材料凝聚态系统求解狄拉克-麦克斯韦联立方程组的工作仍然鲜有出现。

可能有读者会产生疑问，上述方程仍然是单电子水平的，它如何刻画多体系统中的相互作用，特别是复杂的体现多粒子统计性质的交换关联相互作用，下面给出一个简明的论证。考虑2个电子之间的绝热交换过程，根据法拉第定律，交换会形成一个电流圈，随之产生一个磁通。

$$\psi_1(\mathbf{r}_1)\psi_2(\mathbf{r}_2) \to \psi_1(\mathbf{r}_1)\exp(-ie\int_{\mathbf{r}_1,C_1}^{\mathbf{r}_2}\mathbf{A}\cdot d\mathbf{r})\psi_2(\mathbf{r}_2)\exp(-ie\int_{\mathbf{r}_2,C_2}^{\mathbf{r}_1}\mathbf{A}\cdot d\mathbf{r})$$

$$= \psi_1(\mathbf{r}_1)\psi_2(\mathbf{r}_2)\exp(-ie(\int_{\mathbf{r}_1,C_1}^{\mathbf{r}_2}+\int_{\mathbf{r}_2,C_2}^{\mathbf{r}_1})\mathbf{A}\cdot d\mathbf{r}) = \psi_1(\mathbf{r}_1)\psi_2(\mathbf{r}_2)\exp(-ie\oint_C\mathbf{A}\cdot d\mathbf{r})$$

$$= \psi_1(\mathbf{r}_1)\psi_2(\mathbf{r}_2)\exp(-i\varphi_0).$$

$$(4\text{-}28)$$

接下来的问题是怎么确定绝热交换产生的相位 $\varphi_0$，一种方案是直接运用所谓的索末菲半经典量子化条件[202]，$\oint_C(p+eA)dq = (n+1/2)h = (n+1/2)2\pi, n \in Z$。因为绝热下机械动量的贡献可以忽略，因此交换过程产生的电磁场贡献的磁通应该是 $\pi$ 的奇数倍，使得2个电子的直积波函数交换后多出一个负号，因此我们有理由推论，只要在方程 4-25中考虑了完整的电磁场，就能描述多体交换相互作用。目前的DFT第一性原理模拟大多只考虑静电势静电荷以及赝势的贡献，把磁矢势和电流的贡献考虑进去的工作仍然比较少，主要集中在分数量子霍尔效应(FQHE)，阿贝尔任意子的密度泛函理论[203, 204]。因此我们有理由推测，描述完整的多体系统需要同时考虑标量势和磁矢势。截止目前，我们已经得到了自洽求解电子旋量波函数和相应电磁场的算法，原则上当迭代自洽以后，我们就可以求出路径积分(配分函数)的鞍点构型，但光有这些信息对于关联函数的求解一般还是不够的，还需要考虑相对鞍点构型(即上述偏微分方程的解)附近各种涨落的贡献[193]，下面两小

节将分别介绍如何从微扰和非微扰角度处理涨落对可观测量即关联函数的贡献。

### 4.5.1 微扰角度考虑涨落:费曼图自洽方法

承接上一节内容,本小节从微扰角度考虑涨落对关联函数的贡献,从QED角度看,任何单圈水平的微观涨落或者说散射过程都是由如下3种基本图构成的,即费米子自能修正,玻色子极化修正,散射顶角修正(包括发射顶角和吸收顶角)[195],在每个顶点处我们施加能动量守恒条件。在前人的工作中,通常研究者都是集中考虑其中一种图形占主导,例如随机相近似(RPA)主要考虑长程涨落,通常用于靠近相变点的情况,以及长波低能理论,其中媒介玻色子的极化占主导,最终玻色子传播子由戴森(Dyson)方程给出[205]。另一种主要考虑短程涨落以及硬核散射,以顶点修正为主导,称为贝特-萨皮特(Bethe-Salpeter)近似,常用于短波光子与物质的相互作用,例如文献[206]。由于我们不能先验的认为一个一般晶体材料性质中是长程还是短程涨落占主导,因此我们需要考虑上述的所有图可能的贡献。

根据 4-13,我们可以很容易地列出如下的费曼图自洽关系式。

$$
\begin{aligned}
-\Sigma(\mathbf{k}) &= \int \frac{d^4 q}{(2\pi)^4} \Gamma^-(\mathbf{k}, \mathbf{k}-\mathbf{q}; \mathbf{q}) W(\mathbf{q}) \Gamma^+(\mathbf{k}-\mathbf{q}, \mathbf{k}; \mathbf{q}) G(\mathbf{k}-\mathbf{q}), \\
\Pi(\mathbf{q}) &= \int \frac{d^4 q}{(2\pi)^4} \Gamma^+(\mathbf{k}, \mathbf{k}+\mathbf{q}; \mathbf{q}) G(\mathbf{k}+\mathbf{q}) \Gamma^-(\mathbf{k}+\mathbf{q}, \mathbf{k}; \mathbf{q}) G(\mathbf{k}), \\
-\Gamma^+(\mathbf{k}, \mathbf{k}+\mathbf{q}; \mathbf{q}) &= \int \frac{d^4 p}{(2\pi)^4} \Gamma^+(\mathbf{k}-\mathbf{p}, \mathbf{k}+\mathbf{q}-\mathbf{p}; \mathbf{q}) G(\mathbf{k}+\mathbf{q}-\mathbf{p}) \\
& \quad \Gamma^+(\mathbf{k}+\mathbf{q}, \mathbf{k}+\mathbf{q}-\mathbf{p}; \mathbf{p}) W(\mathbf{p}) \Gamma^-(\mathbf{k}-\mathbf{p}, \mathbf{k}; \mathbf{p}) G(\mathbf{k}-\mathbf{p}), \\
-\Gamma^-(\mathbf{k}, \mathbf{k}-\mathbf{q}; \mathbf{q}) &= \int \frac{d^4 p}{(2\pi)^4} \Gamma^+(\mathbf{k}-\mathbf{p}, \mathbf{k}-\mathbf{q}-\mathbf{p}; \mathbf{q}) G(\mathbf{k}-\mathbf{q}-\mathbf{p}) \\
& \quad \Gamma^+(\mathbf{k}-\mathbf{q}, \mathbf{k}-\mathbf{q}-\mathbf{p}; \mathbf{p}) W(\mathbf{p}) \Gamma^-(\mathbf{k}-\mathbf{p}, \mathbf{k}; \mathbf{p}) G(\mathbf{k}-\mathbf{p}).
\end{aligned}
\tag{4-29}
$$

即把所有内圈动量积分得到基本费曼图,方程 4-29中的顶角其实可以统一为一个$\Gamma$,反射或吸收取决于玻色子动量$q$的方向,分开只是图像上稍微方便理解。同样费米子和玻色子的格林函数也有相应的修正,即自能和极化修正。

$$
\begin{aligned}
G(\mathbf{k}) &\to G(\mathbf{k})[I + \Sigma(\mathbf{k}) G(\mathbf{k})], \\
W(\mathbf{q}) &\to W(\mathbf{q})[I + \Pi(\mathbf{q}) W(\mathbf{q})].
\end{aligned}
\tag{4-30}
$$

从方程 4-30可以明显看出,当只考虑费米子和玻色子的无穷多次自能和极化修正,按照等比求和(RPA)就能得到著名的戴森方程,可以基于发散条件简单判别相

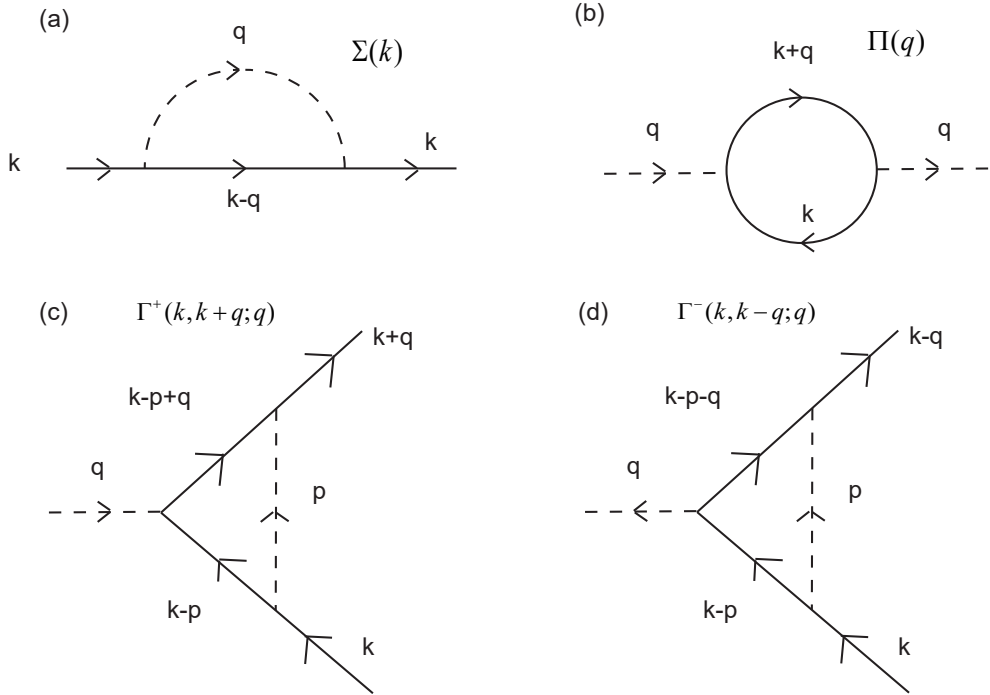

图 4-13: 基本的类QED单圈费曼图, (a)表示费米子自能, (b)表示媒介玻色子的极化, 实际材料系统中玻色子可以是光子, 声子, 以及其他涨落元激发等。(c),(d)分别表示吸收和发射顶角, 在每个顶点都需要满足能动量守恒。

变点和微扰失效的参数区域。

$$G(\mathbf{k}) = G_0(\mathbf{k})[I - \Sigma(\mathbf{k})G_0(\mathbf{k})]^{-1},$$
$$W(\mathbf{q}) = W_0(\mathbf{q})[I - \Pi(\mathbf{q})W_0(\mathbf{q})]^{-1}. \tag{4-31}$$

然而, 从更一般的角度考虑, 应当同时联立方程 4-29 4-30, 输入特定初值的$G_0, W_0, \Gamma_0$然后进行顺序迭代, 直到所有的图形收敛即完成自洽。对于QED或者电磁, 库伦相互作用。我们有$W_0(\mathbf{q}) = \frac{-ig_{\mu\nu}}{\mathbf{q}^2+i\epsilon}$, $\Gamma_0(\mathbf{k}_1, \mathbf{q}; \mathbf{k}_2) = -ie\gamma^\mu\delta(\mathbf{k}_2 = \mathbf{k}_1 + \mathbf{q})$, $G_0(\mathbf{k}) = (\omega - (E_{\mathbf{k},n} - \mu))^{-1}$ [195]。其中$G_0(\mathbf{k})$需要用到方程 4-25中电磁场下狄拉克方程的本征值$E_{\mathbf{k},n}$和系统化学势$\mu$。对于声子, 初值则更为复杂, 声子单体非相互作用格林函数为$W_0(\mathbf{q}) = -\frac{\omega_\mathbf{q}^2}{\omega^2+\omega_\mathbf{q}^2}$; 其中$\omega_\mathbf{q}$为声子色散 [207], 而电声耦合顶角的初值则需要对原子位置求微分进而求解另一个本征值问题得到 [170], 类似BCS问题中的形变势近似。以下有几点需要注意, 首先, 所有的图可能是一个矩阵取值的函数, 相

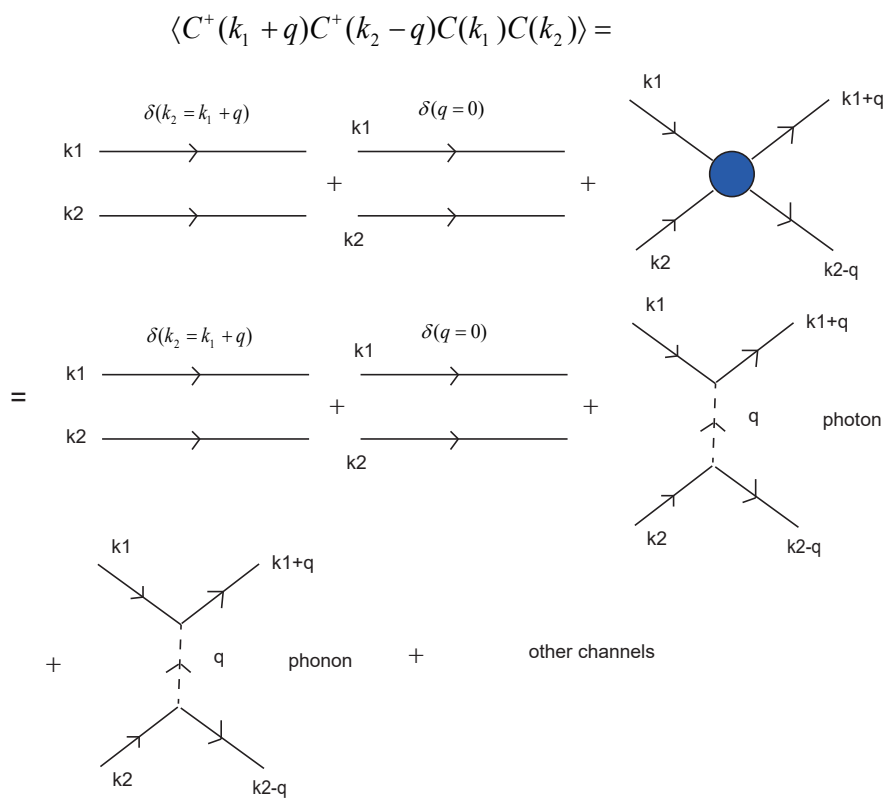

图 4-14: 4费米子关联函数费曼图。对于非相互作用的情况，可以通过维克(Wick)定理解耦为2个自由的费米子格林函数，而相互作用情况则会出现不可约的图(体现为不可约顶点)，树图水平可以通过一些缀饰的中间玻色子实现费米子之间的散射，例如光子媒介的电磁相互作用，声子媒介的BCS电子配对作用等。高温超导等问题中还需要考虑其他涨落等贡献的散射道。

应的乘法推广为矩阵乘法，因此每个因子的顺序不能随意变化。其次，系统中的玻色子可能有多种，例如光子，声子，不同的玻色子之间也会相互修正，例如顶点图中的内线和外线玻色子可能是不同的。最后，当所有的图迭代收敛，原则上我们只精确到了圈图的单圈近似，而没有考虑更高圈的涨落，因为我们考虑的所有图里并没有设计2个及以上内线玻色子交叉且互不可分的情况，因此上述费曼图自洽方法只考虑了部分涨落，没有考虑所有的高阶涨落，而且我们也很难先验地证明上述迭代方案的收敛性，这产生了下一小节引入涨落的非微扰方法的需要。

下面我们假定所有的基本图已经收敛，因此我们可以用它们来表示关联函数，即一般的4费米子期望值，如图4-14所示。这样的关联函数一般可以分解为不

同散射道的求和，例如无散射的自由费米子格林函数，根据维克定理自由费米子的4点函数期望有2种不同的缩并 [195]，以及光子，声子等为媒介的散射通道。声子为媒介的散射可能导致系统出现常规超导。以上为基于变分方法的自洽费曼图求解关联函数算法，它原则上能考虑所有单圈图微扰，并能推广到任意晶体系统，然而上述方法并没有超出微扰框架，收敛性难以预知和控制，下面一节我们将讨论更为一般的非微扰算法来考虑涨落。

### 4.5.2 非微扰角度考虑涨落:格点规范变分经典蒙特卡罗算法

本小节从非微扰角度讨论涨落对关联函数的贡献，一个很自然的想法是采用统计物理的蒙特卡洛方法抽样近似求解关联函数或路径积分，由于这里还涉及到规范场即电磁场的构型，因此本质也是格点规范问题。基本想法就是基于我们之前偏微分方程 4-25求出的场构型附近，人为生成偏离鞍点构型的其他构型，然后根据Metropolis算法确定是否接受新的场构型，和数值求解配分函数和统计物理期望值类似。需要澄清的是，我们这里把旋量场和电磁场当成准经典场，采用经典蒙特卡洛，和需要考虑费米子符号问题的量子蒙特卡洛(QMC)不同，更多详见 [208]。

首先我们简要介绍一下格点规范理论(LGT)，格点规范最初源自研究量子色动力学(QCD)的夸克禁闭，渐进自由等问题 [209, 210]。由于QCD的非微扰特性，基于蒙特卡洛的LGT几乎是第一性原理研究QCD的唯一数值方法，即QCD的第一性原理。LGT近年来也被应用于研究凝聚态中的演生物态，特别是具有一定演生对称性，拓扑性质的物态，在这些物态中，低能自由度受到一些约束，当我们将这些约束解耦，相当于引入规范场与这些描述自由度的场发生耦合，也称为对称规范化(gauging the symmetry)。例如凝聚态中类比夸克禁闭的去禁闭量子临界点(DQCP) [211],旋子(spinon)禁闭 [212]，Z2拓扑序 [213]，U(1)自旋液体 [214]，分形子(fracton)拓扑序 [215]等。同时LGT近期在张量网络和量子模拟等领域也广受关注，例如考虑规范场与费米子物态耦合的投影纠缠对态(PEPS) [216]，Gross-Neveu模型的张量网络点正规化和纠缠 [217]，玻色-哈伯德量子模拟器上观测规范不变性 [218]，Toric code格点规范的量子模拟 [219]等。因此LGT在凝聚态的应用正在迅速发展并且前景巨大。

回到我们的问题，我们待求解的是类似如下的关联函数。

$$\langle J^\mu(\mathbf{r}, t) J^\nu(\mathbf{r} = 0, t = 0) \rangle = \sum_{\psi, \mathbf{A}}$$

$$\frac{\exp(-S[\psi, \mathbf{A}], +\infty, t) J^\mu(\mathbf{r}, t) \exp(-S[\psi, \mathbf{A}], t, 0) J^\nu(\mathbf{r} = 0, t = 0) \exp(-S[\psi, \mathbf{A}], 0, -\infty)}{\exp(-S[\psi, \mathbf{A}], +\infty, -\infty)}.$$

$$(4\text{-}32)$$

其中 $J^\mu = L^\mu + S^\mu$ 为轨道角动量和自旋角动量的和，即总角动量。$S[\psi, \mathbf{A}]$ 为狄拉克拉格朗日量 4-23 对应的作用量。求和 $\sum_{\psi, \mathbf{A}}$ 是对所有满足要求的旋量场，规范场的泛函求和或泛函积分。对于旋量场，原则上抽样的构型中振幅和相位可以任意涨落，对于规范场，则需要满足规范固定条件，否则会出现重复抽样的构型，在物理上产生冗余，这个我们后面将着重讨论。初始构型我们选取鞍点构型，即方程 4-25 的解，它们是配分函数中贡献最大的一项。除了 $\psi, \mathbf{A}$，我们还可以额外对离子实位置进行抽样求和，原则上经过晶格弛豫之后，在离子实平衡位置蒙特卡洛抽样，就相当于考虑声子修正 [220]，即 $\mathbf{R}_i = \mathbf{R}_{i,0} + \delta \mathbf{R}_i$。此外，我们还可以考虑掺杂无序的影响，例如 $N_C = N_{C,0} + \delta n = \sum_i (Z_i + \delta Z_i)$，即假定相对原来的单胞总电子数有一个偏差，相当于在保证电荷守恒的情况下，对离子实电荷数进行抽样。因此上述方法可以推广到对其他元激发相关参数的抽样，考虑这些参数涨落对关联函数的贡献，这种思想是普适可扩展的。最后我们对时空关联函数 4-32 做傅里叶变换，就能得到谱仪通常探测的关联函数。

接下来我们重点讨论 U(1) 规范场(电磁场)$\mathbf{A}$ 的抽样问题以及规范固定算法。由于规范场本身具有冗余性，即等价规范变换之后的场构型仍然描述同一个物理系统，因此如果我们在抽样中如果输入的两个构型属于同一个规范，则会出现重复抽样，传播子中可能会出现零模导致奇异性。因此我们在蒙特卡洛抽样(至少基于变分的蒙特卡洛)时需要保证得到的规范场构型还是同一规范下的，即满足规范固定条件，更多细节可以参考作者的知乎博文 [221]。这里先简要回顾一下高能领域中的 LGT 算法以及规范固定算法。解析上最常用的是所谓法捷耶夫-波波夫鬼场(Faddeev-Popov ghost field)辅助场方法 [195]，数值上用的更多的则是寻找哈尔测度(Haar measure)直接求解路径积分，这种方法不需要固定规范 [210]，但很遗憾本文采用的变分的方法，不能直接应用这种方法，固定规范仍然是必要的。另外一种思路是转化为优化问题，例如寻找最大分岔树 [222],应用模拟退火求解等价的优化问题 [210]，以及随机规范固定算法 [223]。然而对于 3 维欧式空间中的正交晶格上 U(1) 经典规范场，如果控制为库伦规范，可以考虑如下数值上更为简便的算法(独立于以

上规范固定方案)。需要声明的是和量子的规范场不同，经典规范场可以放在格点上，而量子规范场定义在价键(link)上。

考虑如下紧致规范场联络的无穷小规范变换：

$$A_g^\mu = g^{-1}(A^\mu + \frac{i}{e}\partial^\mu)g, g \in G,$$

$$g = \exp(-i\alpha) = \exp(-i\alpha_a T^a) \approx (I - i\alpha_a T^a),$$

$$(A_\mu)_g \approx A_\mu + \frac{1}{e}T^a\partial_\mu\alpha_a - i\alpha_a[A_\mu, T^a] = A_\mu + \frac{1}{e}D_\mu\alpha, \qquad (4\text{-}33)$$

其中G为对应规范群(考虑连续李群)，g为其中距离单位元无穷小的元素,相位参数的协变导数定义为$D_\mu\alpha = \partial_\mu\alpha - ie[A_\mu, \alpha]$。从中我们就可以立刻看到，如果要抽样的新构型不跑出现在所选的规范，即固定规范，则需要相位参数满足测地线方程$D_\mu\alpha = 0$。下面为简明只考虑阿贝尔规范场的情况。对于一个给定格点，类比电磁场的相位变换，考虑把周围价键上的联络吸收进去，有$\exp(-i\alpha) = \exp(-ie\int_{S^{d-1}} A_\mu ds^\mu) = \exp(-ie\int_{S^d} dA)$。最后一步利用了高斯定理，因此可以看到相位$\alpha_n = e\int_{S^d} dA + 2\pi n, n \in Z$。测地线条件相当于在每个点满足$D_\mu\alpha_n = eD_\mu(dA) + 2\pi D_\mu(n(x)) = 0$。其中$n(x)$是磁单极子密度，是规范场紧致性导致的，在这里我们不考虑磁单极子的出现。对于阿贝尔规范场，相当于$D_\mu(dA) = \partial_\mu(dA) = 0$，即$dA = const$，对于3+1d的情况，就是联络的散度为无时空依赖的常数即$\partial_\mu A^\mu = const$，通常教科书中的洛伦兹规范是$\partial_\mu A^\mu = 0$，库伦规范是$\bigtriangledown \cdot \mathbf{A} = 0$ [136]。这其实是固定规范的充分非必要条件，从数值的意义上，仅满足这个条件虽然规范固定，但原则上无法得到同一规范下的场构型。然而以下简明起见，仍然考虑3d空间固定为库伦规范。要求3d空间中散度一致趋于0，考虑镶嵌到高一维并构造如下控制方程。

$$\partial_t(\bigtriangledown \cdot \mathbf{A}) + \beta(\bigtriangledown \cdot \mathbf{A}) = 0, \beta > 0. \qquad (4\text{-}34)$$

这里只要给出随机的初始构型和边界条件，这个方程的不动点就是满足规范条件的，但现在仅凭这个方程 4-34 并不能唯一确定规范势，因此考虑再增加约束方

程。定义:

$$X = \partial_x A^x, Y = \partial_y A^y, Z = \partial_z A^z,$$

$$\partial_t X + \frac{\beta}{3}(X + Y + Z) = 0,$$

$$\partial_t Y + \frac{\beta}{3}(X + Y + Z) = 0, \tag{4-35}$$

$$\partial_t Z + \frac{\beta}{3}(X + Y + Z) = 0.$$

只要上述3个方程同时成立,规范固定条件自动成立,但需要注意,数值上处理的是差分方程组而非微分方程组,前者对参数选择,收敛性将有更严格要求。

$$X_{n+1} = -(\frac{\beta}{3} - 1)X_n - \frac{\beta}{3}Y_n - \frac{\beta}{3}Z_n,$$

$$Y_{n+1} = -\frac{\beta}{3}X_n - (\frac{\beta}{3} - 1)Y_n - \frac{\beta}{3}Z_n, \tag{4-36}$$

$$Z_{n+1} = -\frac{\beta}{3}X_n - \frac{\beta}{3}Y_n - (\frac{\beta}{3} - 1)Z_n.$$

选择合适的参数$\beta$进行迭代,最终$X_\infty + Y_\infty + Z_\infty$就是所求的零散度构型,接下来讨论$\beta$取值,方程 4-36其实严格可解,通项如下。

$$X_n = \frac{2 + (1-\beta)^{n-1}}{3}X_1 + \frac{(1-\beta)^{n-1} - 1}{3}Y_1 + \frac{(1-\beta)^{n-1} - 1}{3}Z_1,$$

$$Y_n = \frac{(1-\beta)^{n-1} - 1}{3}X_1 + \frac{2 + (1-\beta)^{n-1}}{3}Y_1 + \frac{(1-\beta)^{n-1} - 1}{3}Z_1, \tag{4-37}$$

$$Z_n = \frac{(1-\beta)^{n-1} - 1}{3}X_1 + \frac{(1-\beta)^{n-1} - 1}{3}Y_1 + \frac{2 + (1-\beta)^{n-1}}{3}Z_1.$$

显然迭代收敛条件为$|1 - \beta| < 1$,$\beta$越接近1收敛越快,我们也可以直接看出方程的不动点。

$$X_\infty = \frac{2}{3}X_1 - \frac{1}{3}Y_1 - \frac{1}{3}Z_1,$$

$$Y_\infty = -\frac{1}{3}X_1 + \frac{2}{3}Y_1 - \frac{1}{3}Z_1, \tag{4-38}$$

$$Z_\infty = -\frac{1}{3}X_1 - \frac{1}{3}Y_1 + \frac{2}{3}Z_1.$$

显然不动点严格满足零散度条件,对这种严格可解的情况,只要按照上述不动点表达式对初始构型作一次迭代就能完成固定规范。当然对于更一般的没有严格解的情况,迭代是通用方法,但要检验其收敛性。最后通过积分我们可以得到满足库

伦规范的场构型。

$$A^x(x,y,z) = A^x(0,y,z) + \int_0^x dx' X_\infty(x',y,z),$$
$$A^y(x,y,z) = A^y(x,0,z) + \int_0^y dy' Y_\infty(x,y',z),$$
$$A^z(x,y,z) = A^z(x,y,0) + \int_0^z dz' Z_\infty(x,y,z').$$

(4-39)

可以看到，规范势的x分量可以差一个任意的关于y,z的函数，余类推，数值上由随机生成的初值决定。我们可以基于之前求出的鞍点场构型 4-25，加上一些随机涨落作为初值，虽然初始构型不一定满足规范固定条件，但最后会趋于零散度，以上即为经典U(1)规范场的抽样固定规范算法，原则上当然还可以引入一个表示规范固定的δ函数，然后用高斯函数洛伦兹函数等去近似它，但这样做并不够精确，而且这样抽样的构型中其实有很多并不满足规范固定条件的构型需要舍弃，即抽样的效率比较低，本文提出的直接构造满足规范固定条件的算法正好能解决这一问题。这套算法应用于类似$CeFe_2Al_{10}$等强关联实际材料将留待未来研究。

小结本节内容，受高能物理启示，我们从比哈伯德模型，以及密度泛函模型更为微观的量子电动力学模型出发考虑强关联晶体系统的关联函数，对于这样一个相互作用多体量子场论。可以通过边界条件同时求解电子和电磁场的变分方程，在此基础上考虑涨落对关联函数的贡献。从微扰角度考虑涨落，可以从基本QED费曼图出发，考虑它们之间的自洽修正关系，根据变分解作为初值代入，原则上可以考虑所有单圈修正涨落对关联函数的贡献。从非微扰角度看，该问题是一个多体版本的格点规范(LGT)问题，在半经典近似下，在变分鞍点解附近对所有电子旋量场和电磁场中满足规范固定条件的构型进行抽样求和原则上就能求出非微扰关联函数，对于库伦规范或洛伦兹规范的规范固定，本节找到了一种直接构造的办法生成满足规范固定条件的场构型，并且该算法框架可以推广到对更多系统参数抽样来考虑其他修正效应的情况，例如声子修正对应原子位置抽样，掺杂对应粒子数和原子序数抽样，具有高度可扩展性。并且这种算法不局限于特定材料，理论上只要给定单胞内初始原子位置就能进行，避免过多唯象参数的引入。

## 4.6 本章小结

本章分别对传统凝聚态唯象模型和高能物理启示的第一性原理模型进行比较，求解同一个巡游磁无序强关联系统$CeFe_2Al_{10}$。传统凝聚态方法层面，包括修正自

旋波理论，转移矩阵精确对角化方法，以及LDA+DMFT方法，它们能在一定程度上解释$CeFe_2Al_{10}$的磁无序和激发谱，其中自旋激子图像一定程度借鉴了高能物理中的规范禁闭理论。但上述方案中为了和中子散射的实验相吻合，需要调节许多唯象参数以及预设比较多的先验假设，这些参数因材料而异，因此具有一定局限性。另一方面，受到高能第一性原理启示，我们发展出一套同时自洽求解微观电子(旋量)波函数以及微观电磁场的算法，原则上能通过微扰和非微扰的办法计算任意关联函数，微扰的自洽费曼图方案考虑了所有单圈修正，而非微扰的格点规范方法原则上能考虑电子密度涨落，电磁场规范涨落，声子涨落等多种自由度对关联函数的修正，具有高度可扩展性，可以直接移植到其他材料，同时也解决了实际阿贝尔电磁场抽样过程中比较微妙的规范固定问题，然而实际操作，计算的复杂程度和需要处理的自由度数就要远大于唯象的凝聚态模型。因此本章的结论是对于实际材料，凝聚态方法和高能方法是互补的，取决于我们需要微观处理到什么程度以及唯象的程度，根据研究对象的实际需要选择算法。对于$CeFe_2Al_{10}$磁无序及自旋激子磁激发的实验和理论更完善的解释，敬请关注即将出现的新工作 [192]。

# 第5章 总结与展望

本论文着重讨论了高能物理以及量子信息在凝聚态物理中应用的几个例子，它们都为传统凝聚态问题从演生论和还原论提供了新的视角。

首先，对于实空间晶格结构存在应变，扭转，弛豫的情况，受到广义相对论和弯曲时空qft的启示，相应的变形都能等效为弯曲时空，当无变形的低能有效理论是一个已知理论，例如单层石墨烯半填充的狄拉克理论，则变形后的理论相当于原始费米子对有效弯曲时空或者坐标架序参量的响应。这种理论不需要转角公度，不需要具有平移对称性，因此理论上适用于任意转角。即这套理论是一种统一描述小转角和准晶转角的候选者。同样基于这套几何理论可以预言RBG的性质，这是传统的类BM模型做不到的，RBG也作为一种新的非平衡弗洛凯系统，具有更为丰富的拓扑性质。相应的开放问题例如RBG中是否存在时间晶体，时间准晶相等，同样值得学界未来进行进一步研究。当弯曲时空能描述扭转摩尔体系，那么摩尔体系或一般的应变体系能否反过来模拟宇宙真实的弯曲，引力效应等，这样的逆问题也是值得深思的。

其次，我们从多体角度考虑TBCB中可能的FCI相，最终在$1.608°$对应的手征极限，导带$\nu = 1/5$填充数值观测到10重简并基态，它们受到多体能隙保护。同时受到量子信息中纠缠谱启示，我们将其应用在FCI相候选者的准空穴计数上，计数结果显示MATBCB的低能近简并基态准空穴计数满足Halperin自旋单态的广义泡利原理，这和高能物理中的色单态是类似的，从而证明了MATBCB中高陈数FCI相的存在性。同时未来基于高能中的生成函数方法的启示，也可能对摩尔系统这类问题的准量子化学模拟(特别是dmrg模拟)中生成数量巨大的相互作用构型提供便利。另外，如何实现其中的非阿贝尔FCI相，tMoTe2, R5G等体系是否能实现相应的高陈数零场FCI相(即FQAHE)，FCI准空穴计数能否模拟真实的高能色单态等也是值得未来深入研究的方向。

最后从演生论的角度看，任何在局域希尔伯特空间存在约束，例如强关联导

致双占据的排除，磁性阻挫等很多情况下，我们都可以通过引入一个辅助场来实现对这一约束的解耦，相应的代价是这个辅助场有时会呈现出类似规范场的行为，例如高温超导的投影方法，磁性阻挫问题的演生格点规范。在第四章中类似$CeFe_2Al_{10}$这类近藤半金属其中的自旋激子图像需要用到部分子辅助场，其在约束解耦过程中也充当了类似规范玻色子的作用，充当杂化电子空穴配对胶水。而从还原论的角度看，多体QED是比紧束缚模型，安德森模型等更为微观的模型。它的重整化低能应该给出和凝聚态唯象模型一致的结果，不同的是其需要的先验参数大大减少，类比光子晶体的电磁场求解，多体QED中联立求解电子场和电磁场应该能统一描述材料的性质，并且更具有普适性，适用于晶体态的绝缘体，导体以及半金属等。相应的格点规范问题也可以直接应用高能的方法，通过固定固定规范后采用准经典蒙特卡洛算法求解各种物理可观测关联函数，并能通过对各种参数抽样考虑不同涨落对关联函数的修正。其中的固定规范算法怎么推广到非阿贝尔规范场，底流形拓扑非平庸的情况以及考虑磁单极子的情况仍然是十分有趣的开放问题。

从上述例子我们可以发现高能物理，量子信息等领域与凝聚态之间有着许多共同语言和工具，未来在这些领域之间我们坚信一定会有更多的联系桥梁，并且这些领域必将在相互启示中共同取得全新的进展。

# 参考文献

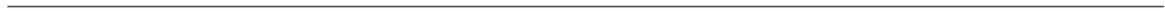

# 附录A 量子几何迹条件取等和波函数全纯性的等价证明

本附录简要证明量子几何迹条件取等和波函数全纯性的等价性，为了简单起见只考虑阿贝尔情况，将态矢量分解成实部虚部$|u\rangle = |u_1\rangle + i|u_2\rangle$，复变函数柯西黎曼条件表示为：$|\partial_{k_x} u_1\rangle = |\partial_{k_y} u_2\rangle$, $|\partial_{k_y} u_1\rangle = -|\partial_{k_x} u_2\rangle$。先看量子度规的迹。

$$
\begin{aligned}
Tr(g) &= \langle\partial_{k_x} u|\partial_{k_x} u\rangle + \langle\partial_{k_y} u|\partial_{k_y} u\rangle - \langle\partial_{k_x} u|u\rangle\langle u|\partial_{k_x} u\rangle - \langle\partial_{k_y} u|u\rangle\langle u|\partial_{k_y} u\rangle \\
&= \langle\partial_{k_x} u_1|\partial_{k_y} u_2\rangle - \langle\partial_{k_y} u_1|\partial_{k_x} u_2\rangle - \langle\partial_{k_x} u_2|\partial_{k_y} u_1\rangle + \langle\partial_{k_y} u_2|\partial_{k_x} u_1\rangle \\
&\quad -[((\langle u_1|\partial_{k_x} u_1\rangle + \langle u_2|\partial_{k_x} u_2\rangle)^2 + (\langle u_1|\partial_{k_x} u_2\rangle - \langle u_2|\partial_{k_x} u_1\rangle)^2] \\
&\quad -[((\langle u_1|\partial_{k_y} u_1\rangle + \langle u_2|\partial_{k_y} u_2\rangle)^2 + (\langle u_1|\partial_{k_y} u_2\rangle - \langle u_2|\partial_{k_y} u_1\rangle)^2] \\
&= \langle\partial_{k_x} u_1|\partial_{k_y} u_2\rangle - \langle\partial_{k_y} u_1|\partial_{k_x} u_2\rangle - \langle\partial_{k_x} u_2|\partial_{k_y} u_1\rangle + \langle\partial_{k_y} u_2|\partial_{k_x} u_1\rangle \\
&\quad -2[(\langle u_1|\partial_{k_y} u_2\rangle - \langle u_2|\partial_{k_y} u_1\rangle)^2 + (\langle u_1|\partial_{k_x} u_2\rangle - \langle u_2|\partial_{k_x} u_1\rangle)^2].
\end{aligned}
\tag{1}
$$

其中每个等号都用柯西黎曼条件代换，再看QGT的虚部反对称组合的虚部即贝里曲率。

$$
F = \mathbf{Im}\{\langle\partial_{k_x} u|(I - |u\rangle\langle u|)|\partial_{k_y} u\rangle - \langle\partial_{k_y} u|(I - |u\rangle\langle u|)|\partial_{k_x} u\rangle\}.
\tag{2}
$$

显然方程 1 2两者的二次项都是$\langle\partial_{k_x} u_1|\partial_{k_y} u_2\rangle - \langle\partial_{k_y} u_1|\partial_{k_x} u_2\rangle - \langle\partial_{k_x} u_2|\partial_{k_y} u_1\rangle + \langle\partial_{k_y} u_2|\partial_{k_x} u_1\rangle$。因此下面主要比较四次项，考虑其中的一半：

$$
\begin{aligned}
&(\langle\partial_{k_x} u_1| - i\langle\partial_{k_x} u_2|)(|u_1\rangle + i|u_2\rangle)(\langle u_1| - i\langle u_2|)(|\partial_{k_y} u_1\rangle + i|\partial_{k_y} u_2\rangle) \\
&= [((\langle\partial_{k_x} u_1|u_1\rangle + \langle\partial_{k_x} u_2|u_2\rangle) + i(\langle\partial_{k_x} u_1|u_2\rangle - \langle\partial_{k_x} u_2|u_1\rangle)] \\
&\quad [((\langle u_1|\partial_{k_y} u_1\rangle + \langle u_2|\partial_{k_y} u_2\rangle) + i(\langle u_1|\partial_{k_y} u_2\rangle - \langle u_2|\partial_{k_y} u_1\rangle)]
\end{aligned}
\tag{3}
$$

考虑方程 3 的虚部并代入柯西黎曼条件。

$$
\begin{aligned}
&(\langle\partial_{k_x}u_1|u_2\rangle - \langle\partial_{k_x}u_2|u_1\rangle)(\langle u_1|\partial_{k_y}u_1\rangle + \langle u_2|\partial_{k_y}u_2\rangle)+ \\
&(\langle\partial_{k_x}u_1|u_1\rangle + \langle\partial_{k_x}u_2|u_2\rangle)(\langle u_1|\partial_{k_y}u_2\rangle - \langle u_2|\partial_{k_y}u_1\rangle) \\
&= (\langle\partial_{k_x}u_1|u_2\rangle - \langle\partial_{k_x}u_2|u_1\rangle)(-\langle u_1|\partial_{k_x}u_2\rangle + \langle u_2|\partial_{k_x}u_1\rangle)+ \\
&(\langle\partial_{k_y}u_2|u_1\rangle - \langle\partial_{k_y}u_1|u_2\rangle)(\langle u_1|\partial_{k_y}u_2\rangle - \langle u_2|\partial_{k_y}u_1\rangle)
\end{aligned}
\tag{4}
$$

因为 $|u_1\rangle, |u_2\rangle$ 均为实矢量，另外一半四次项给出的结果也相同。即 $Tr(g) = |F|$ 等价于 $|\partial_{k_x}u_1\rangle = |\partial_{k_y}u_2\rangle$, $|\partial_{k_y}u_1\rangle = -|\partial_{k_x}u_2\rangle$。

# 附录B 单层棋盘格霍尔丹-哈伯德模型的PES校正

本附录校正文献[93]的PES，其中的霍尔丹-哈伯德(Haldane-Hubbard)模型对应的$\nu = 1/3$的$C = 1$拓扑平带和$\frac{1}{3}$填充最低朗道能级(LLL)具有精确对应，准空穴计数也符合$(1,3)_{C=1}$广义泡利原理，即拓扑简并基态为3重简并，PES能隙以下的态严格服从解析解$N_x N_y \frac{(N_x N_y - 2N_A - 1)!}{(N_x N_y - 3N_A)!(N_A!)}$。与原文的区别是我们校正中采用的相互作用强度是文献[93]中的10倍。

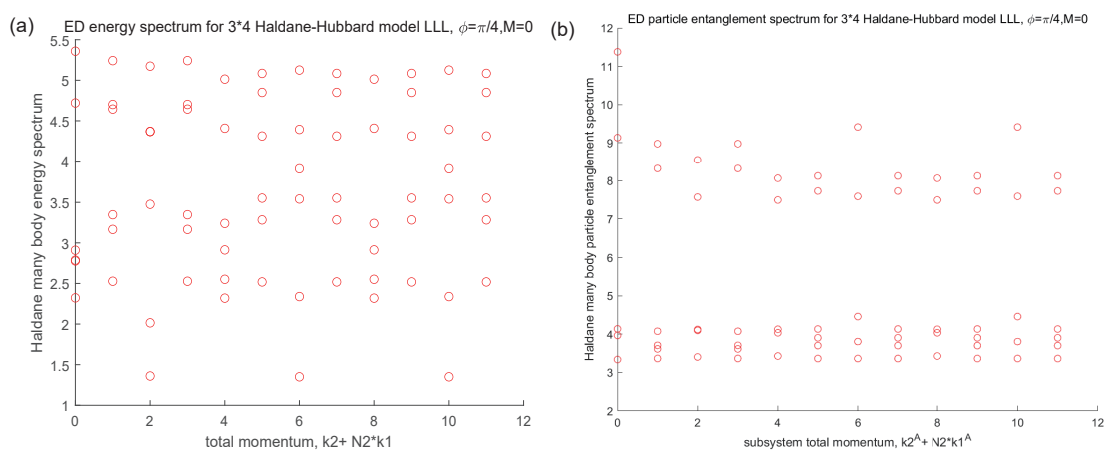

图1: 如图(a)(b)为单层棋盘格霍尔丹-哈伯德在次近邻相位$\phi = \pi/4$下的多体能谱和PES，动量空间尺寸$3 \times 4$。从(a)中可以看到近3重基态简并，(b)中$\xi = 6$为大致的PES能隙位置，能隙以下的态为42个，符合正文提到的$(1,3)_{C=1}$广义泡利原理，其中$K^A = 0, 2 \cdots 10$的区域有3个态，$K^A = 1, 3 \cdots 11$的区域有4个态，一共42个准空穴态。

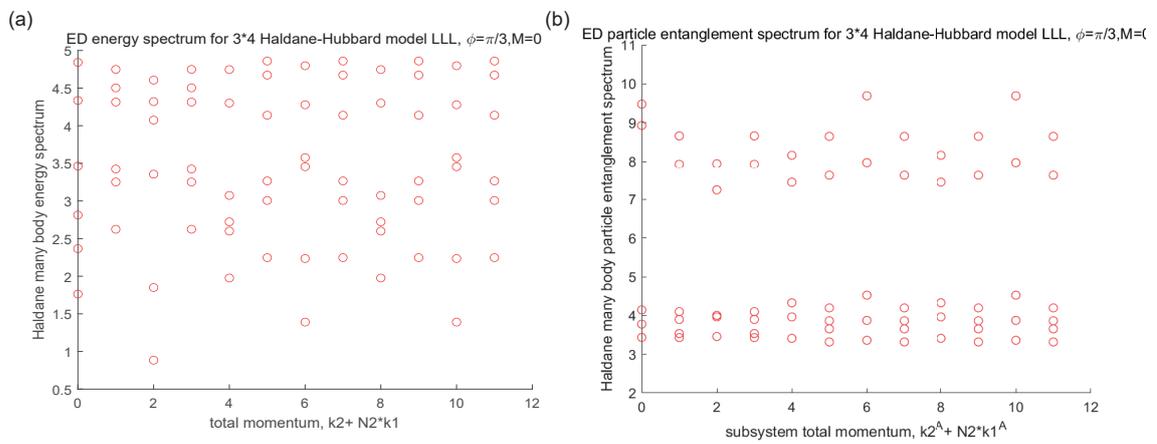

图 2: 如图(a)(b)为相同尺寸和参数下的$\phi = \pi/3$下的多体能谱和PES，可以看到多体谱的三重简并性没有$\phi = \pi/4$那样精确，但PES仍然给出和 1一样的准空穴计数。

# 附录C 其他填充数和尺寸下的数值结果

同时根据文献[128]预言，$\nu = 1/6$的$C = 2$系统具有平移对称破缺的FCI相，遗憾的是现有的$4 \times 6$尺寸，手征极限参数下并没有看到FCI简并基态候选者，即磁通泵浦并没有完全打开能隙。有可能需要像RnG体系那样通过HFA修正来打开能隙。同时平移对称性破缺可能使PES计数的广义泡利原理不再成立，平移对称性和内部color自由度的复合对称性一般保持，因此平移对称性破缺会伴随color内部自由度也破缺，此时可能需要采用更一般的赝势方法[150]，将留待以后工作。

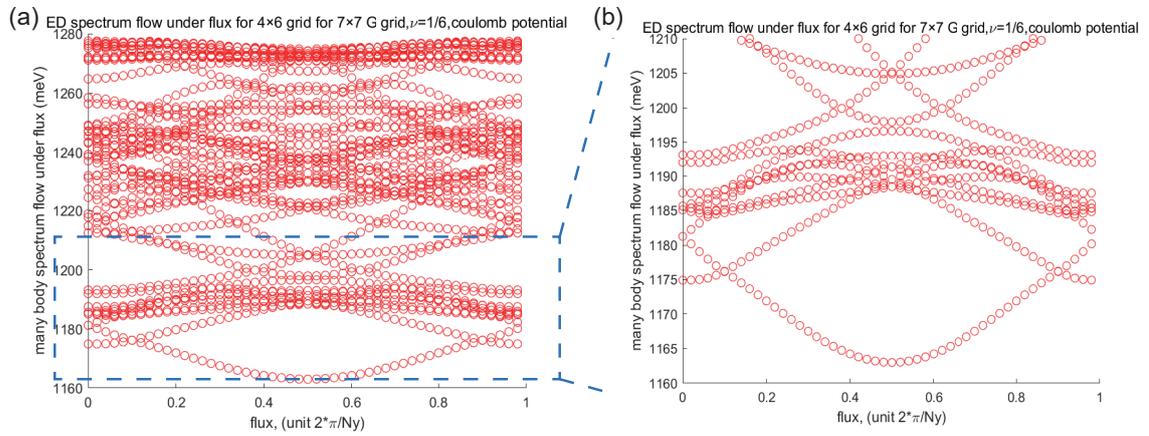

图 3: 如图(a),(b)为MATBCB在$\nu = 1/6$填充下的$4 \times 6$动量空间离散格点密度下的磁通泵浦，其中(b)为(a)图的低能放大图，根据文献[128]，$\nu = 1/6$的$C = 2$系统具有平移对称破缺的FCI相，遗憾的是现有尺寸当前参数下并没有看到FCI简并基态候选者，即磁通泵浦并没有完全打开能隙。

# 在学期间发表论文情况

## 与学位论文相关的学术论文

- **Jia-Zheng Ma**, Trinanjan Datta， Dao-Xin Yao: Effective curved space-time geometric theory of generic-twist-angle graphene with application to a rotating bilayer configuration *Physical Review B.105.245102* (2022)

  （已发表，与本学位论文第二章相关）

- **Jia-Zheng Ma**, Rui-Zhen Huang, Guo-Yi Zhu, Ji-Yao Chen, Dao-Xin Yao: Fractional Chern insulator candidate on a twisted bilayer checkerboard lattice *Physical Review B.110.165142* (2024)

  （已发表，与本学位论文第三章相关）

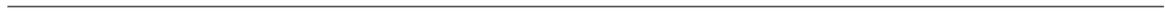

# 特别声明

本声明回应第一轮审稿中个别评审人提出的质疑，以及笔者对该评审的一些看法。首先笔者对该提出拒稿，本轮不予答辩的评审人的观点是部分认同的，上轮稿件中特别第二章确实规范性，格式等仍需要打磨。但仅凭格式规范问题，因为莫须有的创新性不突出否定全文创新性，按照上一版论文中第二章的印象分给全文打分，评价章节之间关联不紧密等原因就剥夺笔者本轮答辩资格。笔者认为这样的评审有失公平，这一点也可以从其他评审人的评分佐证，该评审人与其他评审人的打分存在较大差异，评语和另一位打85分但评级为A的评审人有相当的重叠，但却打出了59分的不合格分数，且并不能给出物理上合理的拒稿理由。总体评分为CAA，笔者原本有申诉机会，但碍于发起申诉可能对学院学校声誉产生负面影响，最终当时没有选择申诉，半年后的现在，笔者在本声明中逐一回应并申诉。

1.评审人认为本文"创新点凝练不足，且多次出现"重复出"某某结果的表述，很难辨认出哪些部分是作者的原创性工作。"这样否认全文创新性的论断是莫须有的。创新点的表述也许确实需要进一步凝练，但在上一版论文中，第一个工作中明显提到创新点是"提出了全新的经典U(1)规范场的规范固定算法"，第二个工作"创新地提出TBG的弯曲时空视角的一般转角理论，并首先推广到转动双层石墨烯体系这种前人理论无法预言的系统"。第三个工作"创新性地探索了C=2的高陈数分数陈绝缘体候选者，它具有超出最低朗道能级对应的新物理。"上述创新点都已经表述地十分清楚，TBG的弯曲时空理论工作已在prb发表，TBCB FCI候选者工作近期也被prb接受，说明至少在prb水平，该评审人的判断是存在一定失误的。如果该评审人仍然以分辨不出创新点为由认为本文创新度不足，那么该评审人是否真的对摩尔电子学，QED第一性原理的前沿有充分的了解或调研呢？笔者认为也要打个问号。

2. "第一章的文献引用过少，缺乏充分介绍，写作多处不规范。"这点笔者认为大致中肯，将增加特别是实验方面的文献。

3. "缺乏对研究背景和方法的系统介绍，2-4章读起来更像是各自独立的期刊论文，而非构成一篇博士论文的三个紧密联系的章节。" 本文题目说得很清楚，高能和量子信息视角的强关联系统理论。提供的是一种统一的视角，方法论，无论强关联系统演生出怎样的新奇现象，其实都可以从规范场理论，有效的时空理论，多体纠缠这样的统一视角进行考虑。狭义上来看，2-4章当然可以认为是毫无联系，但这不见得是件坏事。正是看似毫无联系，才体现了这套思想方法的普适性。这也十分符合复旦大学某位万老师提出的"量子信息，拓扑物态与演生时空量子模拟"的大方针，笔者也认为要对某个学科知识进行深化变革，正是需要这种跨界交叉研究的精神，这也是一个与时俱进的理论工作者应有的野心，笔者认为这是一种自信。

4. "论文的语言问题较多，很多语句的表述流畅性较差或过于口语化，多处公式缺少标点符号。" 这点大致中肯，特别是原版中第二章的部分，将重新打磨。

5. "原稿第2章更像是一个半成品工作，一些图片如2-6上的文字清晰度过低，几乎无法辨认。还有很多图片（如2-10）似乎是直接从绘图软件中截图出来的。" 2-6清晰度本身是没问题的，只是文字大小需要调整。其他图主要选自笔者的早年的研究笔记，其质量确实达不到sci论文发表的水平，但实际背后有原因。图2-10的实验数据图来自笔者和实验合作方的私人通讯，实验图即合作方通过Merlin中子谱仪得出的原始处理结果，笔者不便进一步处理。原稿第二章工作其实是笔者早期和实验学者的合作工作，做了2年后，没有获得实验合作者的认可，最后沦为废案，后续同研究组的高年级博士接手后又做了3年（笔者后续没有参与），仍然没有让各方都满意的结果，相当于该项目烂尾。笔者在项目初期积累的部分原始数据，代码和原图也因为新旧设备更替部分丢失，因为当时也没有发表的打算。但同时笔者后续也通过借鉴高能领域的方法对这个问题有了新的想法，独立提出了基于格点规范的新算法，跳出了具体材料的限制。受限于计算资源与$CeFe_2Al_{10}$系统复杂度，最终只能完成整个框架的其中一个小零件，即规范固定算法，未来整个架构的实现也许是需要许多大研究组的协作，并且应该先在较为简单的体系例如过渡金属单质中予以验证，才能应用于$CeFe_2Al_{10}$这类复杂系统。原始数据，代码，图像笔者会尽量寻回，但实在无法寻回时笔者将不会继续投入时间，毕竟在一个5年都无法完成的项目上继续浪费时间，对于上升期的博士生来说无论如何都是赔本的买卖。尽管原稿第2章可以说是一项失败的研究，但笔者认为博士论文不应该只写成功，失败的经历也同样重要，可以让后续的想要进入这一领域的研究者少走弯路，

以及不用去重复发现一些东西。以成败看论文同样有悖国家"破五唯"的大方针。因为部分图片质量低扣分，笔者可以理解，但仅凭部分图片质量低直接打不及格，笔者认为这是不公平的，是舍本逐末。如果按照该评审人的逻辑，只要这项工作没有完全做完，就可以否定全文的创新价值直接拒稿，那么笔者将永远无法通过外审，当然笔者也无话可说，这个博士学位不要也罢。

6. "参考文献格式不统一，多篇引文（如PR系列论文）卷期号不全。" 这源于毕业论文和pr系列latex模版的差异，同样的bib放在pr系列latex是能编译出正常的卷期号的。笔者将尽量进行修正，如无法更改，只能保持原状。

上述即为该评审人的拒稿理由，笔者认为如果评审人能找出论文物理上的根本错误，缺陷，或者说明论文在物理上的创新度确实达不到博士要求，那么笔者会对这个拒稿表示接受。然而仅凭上述原因，笔者认为这样的拒稿有失公平。因此如果在接下来流程中，仍然有对本文有异议者，对本文有所批评，认为该学位论文达不到博士水平，请麻烦多花一些时间，找到物理上合情合理的理由，谢谢。笔者尊重所有对本文提出合理批评的同行。最后，笔者认为，真金不怕火炼，因此也欢迎教育部以及业内专家对本文进行检查，笔者将坚决捍卫自己的理论。本声明与笔者的导师，学院，学校无关，笔者将承担一切责任。

附：如果对本文有异议者仍然不太清楚什么是"从物理上给出合理的批评理由"，笔者可以在这里给出一个例子。

本篇论文在理论方法以及创新性上存在较大问题，理由如下。

1. 第二章中采用的TBG弯曲时空理论，其实是一个3+1d的Dirac理论，而一般石墨烯问题中采用的是2+1d 演生Dirac理论，这里作者有误导读者的嫌疑。其次，本章提出的理论缺乏一个从格点模型(例如紧束缚模型)出发的微观理论，作者并没有证明这个理论是UV完备自洽的，2个mini valley也没有同时出现Dirac cone，不能实现一般转角下的inter valley coupling。建议作者采用完全动量空间的紧束缚模型，考虑一般转角下固定截断动量半径范围内所有紧束缚项的贡献，不要以所谓的弯曲时空理论作为噱头进行炒作。

2. 第三章的TBCB的FCI候选者，实际是系列工作[Phys.Rev.X.1.021014(2011)], [Phys.Rev.B.87.205137(2013)], [Phys.Rev.Research.4.043151(2022)] 的简单缝合，并没有提出什么新的东西。其中两篇论文是近十年前的工作，因此第三章的时效性新颖性存疑，作者提出的理论无法考虑存在平移对称破缺的情况，以及可能的CDW与FCI共存的情况，其中的FCI相也是老生常谈的abelian FCI相，创新性较为欠缺。此外如果按照作者的计数理论，PES的准空穴计数，对于某些尺寸，某些子系统粒子数(例如$5 \times 6$尺寸，$N_A = 2, \nu = 1/5$)，这套理论给出的计数似乎并不是一个整数，这也是这个理论可能有缺陷的地方。建议作者采用更一般的赝势方法，DMRG方法综合考虑FCI相与CDW的竞争，并研究更有价值的non abelian FCI相。

3. 第四章基于QED求解重费米子化合物$CeFe_2Al_{10}$的激发谱。从输运实验可以看出，这个近藤化合物的杂化能隙很小，换算成温度只有$15K$。而作者提出的所谓多体QED理论的能标要远高于这个杂化能标，如果作者不能论证多体QED在重整化下会流向这样一个低能近藤杂化有效理论，那么有理由相信，作者的这套多体QED算法只是没有意义的数值游戏，甚至是一个民科式的理论。建议作者采用自然轨道重整化群结合DMFT的方法[2209.14178]或者基于投影蒙特卡罗的杂质求解器，求解这个近藤化合物的近零温性质，而不要妄想用QED高能理论来全盘代替相当不同的DFT+DMFT理论。

以上即为对本学位论文的批评理由。

# 致谢

行文于此，感慨万千，六年的硕博生涯一晃而过。回顾往昔，家人，老师，朋友，同行都每时每刻给予我莫大的帮助，在此一并表示衷心的感谢。

首先我要感谢我的硕博导师姚道新教授，在他的带领下我有幸在研究生刚入学时期就接触到了强关联量子多体的前沿领域。姚老师非常鼓励学生进行多个前沿的独立自主研究，重视不同领域之间的联系，在姚老师的鼓舞下我选修了量子场论，广义相对论，自学了微分几何等看上去和凝聚态物理并不直接相关的课程，为未来的研究打下坚实的基础。另外姚老师对学生要求也很严格，主张学生集中在重大科学问题，学界疑难问题上，甘坐冷板凳，肯下苦功夫。趁着年轻有创造力，不需要操心教学任务和项目基金申报的时候潜心解决业界难题，而非灌水应付了事。姚老师经常教导学生，做理论不能和实验脱节，理论要能解释实验，并且在做理论过程中需要思考可能的实验实现，甚至预言新的实验，这些教诲令我受用至今。同时姚老师也非常慷慨，我也得以多次受姚老师资助参加学术会议，向国内外先进的同行学习交流，我也多次借助姚老师的人脉得到了许多合作的机会。姚老师对我计算资源上的支持更是不遗余力，帮助我完成了许多大规模的计算任务，再次对他表示衷心感谢。

其次，我要感谢所有给予我启示的同学，同行，合作者。第一个工作中，首先感谢李光杰博士，叶鹏老师对本工作的启发和独到见解。感谢合作者Trinanjan Datta教授对该项研究的计划安排，他在论文修改方面和提升论文的创新性和重要性方面提出了很多宝贵意见。微分几何以及标架变换方面，黄泽敏，任杰，Jaakko Nissinen，Matthew Horner等老师，同行提供了很多有价值的讨论。拓扑相的物理方面，感谢严忠波老师的许多帮助和有价值的讨论。

第二个工作中，首先我想感谢陈继尧，朱国毅，黄瑞珍三位合作老师，将这个工作引导向正确的方向，几位老师对数值计算严谨的态度让我收获良多。工作期间，Jun-Kai Dong(生僻字无法显示)，Daniel Eric Parker，李昕昕，蔡家麒，潘高